\documentclass[a4paper,fleqn]{cas-dc}

\usepackage[authoryear]{natbib}

\usepackage{caption}
\usepackage{subcaption}

\usepackage[pages=all]{background}

\backgroundsetup{
 scale=6,                  
 color=black!30,             
 angle=45,                 
 opacity=0.3,              
 contents={Preprint – Under Review},   
}

\begin{document}
\let\WriteBookmarks\relax
\def\floatpagepagefraction{1}
\def\textpagefraction{.001}
\shorttitle{Regularized Machine Learning for System Identification of Ship Free-Running Manoeuvres from CFD-Based Synthetic Data} 
\shortauthors{R.F.~Suárez~et~al.}

\title [mode = title]{Regularized Machine Learning for System Identification of Ship Free-Running Manoeuvres from CFD-Based Synthetic Data:\\ A Comparative Study}                      



\author[]{R.F.~Suárez}[auid=000,bioid=1]
\ead{ricardo.suarez.fernandez@tuhh.de}

\credit{Conceptualization, Formal analysis, Investigation, Methodology, Resources, Visualization, Writing – original draft, review and editing} 


\affiliation[]{%
  organization={Hamburg University of Technology (TUHH)},%
  addressline={Institute for Fluid Dynamics and Ship~Theory~(M8)},%
  addressline={Am~Schwarzenberg-Campus~4~(C)},%
  postcode={21073},%
  city={Hamburg},%
  country={Germany}%
}

\author{J.C.~Berndt}[orcid=0009-0009-3984-2807]
\cormark[1]
\ead{j.berndt@tuhh.de}

\credit{Conceptualization, Formal analysis, Investigation, Visualization, Methodology, Validation, Writing – original draft, review and editing} 

\author{M.~Abdel-Maksoud}[]

\credit{Investigation, Funding acquisition, Supervision, Writing - review and editing}
\ead{m.abdel-maksoud@tuhh.de}

\cortext[cor1]{Corresponding author}


\begin{abstract}
This study investigates supervised machine learning techniques for identifying ship hydrodynamic coefficients from CFD-generated data from free-running simulations. Specifically, ordinary least squares and regularized regression methods are applied to Abkowitz-type manoeuvring models. Training and validation datasets are derived from URANS simulations of zig-zag and turning circle manoeuvres, which are validated against experimental benchmark data. The analysis evaluates the effects of coefficient set size, minimum training length required for predictive model training, and manoeuvre combinations on model performance. Results demonstrate the suitability of large-angle zig-zag manoeuvres for hydrodynamic system identification, provided that multicollinearity is addressed through appropriate coefficient selection, regression models, or input data variability. Larger coefficient sets offer greater model flexibility for variable conditions but are more prone to multicollinearity. Regularized regression techniques effectively mitigate multicollinearity and notably enhance prediction accuracy, as does incorporating more diverse manoeuvring data. Among tested models, Ridge regression provided the best compromise between computational efficiency and prediction accuracy.
\end{abstract}


\maketitle

\section{Introduction}
\label{subsec:introduction}

Enhanced manoeuvrability is a core requirement for both ship design and operational safety. Its significance was formalised by the \citet{IMO2002} (IMO) through resolution MSC.137(76), which established the Standards for Ship Manoeuvrability. Mathematical models are widely applied to predict manoeuvring performance during ship design, with their predictive fidelity determined largely by the underlying hydrodynamic coefficients

The \cite{IMO2005} summarises four primary methods for determining these coefficients. The first is based on captive model tests, such as the Oblique towing (OTT), rotating arm (RAT), circular motion (CMT) or planar motion Mechanism (PMM) tests, which is the most established approach. The second relies on empirical formulas or databases (see \cite{sutulo_mathematical_2011}). The third employs theoretical or computational methods, where Computational Fluid Dynamics (CFD) is used to conduct virtual captive tests, like PMM simulations. The fourth is System Identification (SI), which determines coefficients by analysing data from free-running model tests or full-scale trajectories. Once obtained, these coefficients are fed into a mathematical model used in system-based simulations to predict the ship's dynamic response and performance parameters, often faster than real time.

However, these conventional approaches each present notable limitations. CFD-based methods, whether simulating captive tests or free-running manoeuvres, remain computationally expensive and time-consuming, making them impractical for testing a large variety of conditions during the early design phases. Experimental free-running and captive model tests suffer from similar preparation and execution demands, restricting their use beyond controlled studies. Full-scale trials, although indispensable for validation, are confined to the final stages of ship development. These constraints underscore the need for approaches like SI, which offer an alternative pathway to efficiently obtain reliable manoeuvring models, particularly during preliminary design and for operational analysis.

Given these constraints, SI from limited input data has provided a rigorous route for parameter estimation in ship manoeuvring. Early studies in ship steering dynamics introduced model-reference and probabilistic approaches, including maximum-likelihood and recursive schemes \cite{astromIdentificationShipSteering1976,kallstromExperiencesSystemIdentification1979}. Subsequent advances incorporated filtering and equation-error methods, notably the Kalman filter and its non-linear variants \cite{Hwang1980}, which offered recursive state estimation in the presence of noise and helped popularise least-squares (LSQ) formulations for parameter identification. Despite their widespread use, Kalman filtering and LSQ-based methods remain constrained in manoeuvring applications by practical limitations: sensitivity to initial values, ill-conditioning of the normal equations, multicollinearity among regressors, simultaneous parameter drift, and cancellation phenomena \citep{bonci_2015}.

To address these issues, machine-learning estimators with explicit regularisation have been introduced in parametric identification settings and are naturally expressed as quadratic programming. The above limitations have been documented for Abkowitz-type models, where correlated motion variables and interacting terms may enable error-compensating fits that undermine out-of-sample performance \citep{bonci_2015,wang_zou_2018}. Early applications by \citet{luoParametricIdentificationShip2009,zhangIdentificationAbkowitzModel2011a} employed support vector regression (SVR) for manoeuvring models; subsequent studies proposed measures targeting parameter identifiability and drift reduction, including simplified structures and sample reconstruction via difference methods \citep{luoParameterIdentifiabilityShip2016}. Subsequent work employed PSO and SVR for parameter identification, enabling optimal selection of structural and regularisation parameters while using difference-based sample reconstruction to diminish multicollinearity and simultaneous drift \citep{luoPSOSVM2016}. Hybrid schemes then combined SVR with optimisation-based hyperparameter selection to mitigate noise sensitivity and collinearity \citep{banShipTrackRegression2017,xuRealTimeParameterEstimation2018}. More recent refinements broadened the approach via SVR variants and complementary Bayesian or derivative-based SI frameworks \citep{wangIdentificationShipManoeuvring2019a,xueHydrodynamicParameterIdentification2020,alexanderssonSystemIdentificationVessel2022}. 

A recent contribution by \cite{chillcce_data-driven_2023} presents a data-driven SI workflow that estimates Abkowitz-type hydrodynamic coefficients from measured free-running model-test (EFD) time series and validates manoeuvre predictions against benchmark data from the SIMMAN 2008 workshop \citep{simman2008}. The identification proceeds by solving the equations of motion directly, i.e., without inverting the mass matrix or integrating the equations in time, while enforcing physics-based bounds on the coefficients and using a recently developed Euler-equation approach to determine zero-frequency added masses \citep{el_moctar_2022}. The study adopts a reduced-parameter Abkowitz formulation originally proposed by \cite{Mucha2017}, whose coefficients identified from RANS-based captive simulations--serve as a comparative baseline for validation and consistency checks. This methodology has been further systematised and extended in subsequent work, including a doctoral thesis that consolidates direct and indirect SI, added-mass computation via Euler equations, and verification-validation procedures for shallow and deep water conditions \cite{chillcce_phd_2023}.

The present study employs a model-based SI framework, wherein the non-linear mathematical model is formulated using Taylor-series expansions of the hydrodynamic forces and moments, based on the approach proposed by \cite{Abkowitz1964Hy5}. Within this framework, a constrained Quadratic Programming (QP) estimator is employed to identify hydrodynamic coefficients for the MOERI KVLCC2 benchmark using a Least-Squares (LSQ) baseline and three regularised machine learning (ML) models: Lasso, Ridge and SVR. This study pursues four main objectives:

\begin{enumerate}
\item \textit{Minimum training dataset length per manoeuvre:} Determine the minimum data-length requirement per input manoeuvre using the number of overshoot cycles in ZZ 35$^\circ$/35$^\circ$ (PS) needed to obtain stable, accurate coefficient estimates suitable for re-integration and cross-manoeuvre prediction.
\item \textit{Estimator comparison under a unified QP formulation:} Conduct a head-to-head evaluation of LSQ, Lasso, Ridge, and SVR on common training and testing manoeuvres, quantifying robustness and generalisation beyond the training set. In addition, the workflow remains reproducible and solver-agnostic under the unified QP formulation, with systematic hyperparameter tuning for the regularised estimators.
\item \textit{Analysis and mitigation of identification instabilities:} Evaluate and address simultaneous ill-conditioning, cancellation phenomena, and multicollinearity by combining: (a) per-predictor diagnostics via the Variance Inflation Factor (VIF) and component-wise force and moment decompositions; (b) dimensional reduction of the manoeuvring model formulation given in \cite{lewis_PNA_1988} to a reduced coefficient set; (c) regularised estimators (Lasso, Ridge, SVR) within the unified QP setting; and (d) manoeuvre-combination strategies in CFD to enrich excitation. The main goal is to obtain stable, physically interpretable coefficients and to prevent error-cancelling fits resulting in bad prediction performance for off-training manoeuvres.
\item \textit{Minimum number of CFD-generated manoeuvres:} Identify the minimal set of different manoeuvres required to train and validate the estimators with reliable out-of-sample prediction, and validate the resulting CFD manoeuvres against benchmark data for free-running tests (e.g. SIMMAN 2014: ZZ 20$^\circ$/20$^\circ$ (PS) and TC 35$^\circ$ (SB)) \cite{simman2014}, followed by benchmarking of the SI outcomes on the validated datasets.
\end{enumerate}

\section{Methods}
\label{subsec:methods}

The section outlines the methodological framework used to predict ship manoeuvring performance, which integrates high-fidelity numerical data with advanced statistical modelling. The approach is fundamentally structured into two parts: data generation and SI.

The Abkowitz Manoeuvring Model provides an well established approach for defining the set of hydrodynamic coefficients required. A robust system identification begins with CFD simulations that generate the requisite training and testing datasets, the processing of which is detailed in section~\ref{subsec:data_prep}. The core of the SI process involves solving a QP problem to fit the input data. This framework facilitates a comparison between the established LSQ regression model and several regularized ML models, specifically Lasso, Ridge, and SVR (\ref{subsec:reg_ML_methods}). Finally, the section addresses model stability by examining multicollinearity diagnostics (\ref{subsec:vif}) and concludes with the definition of the test case (\ref{subsec:test_case}) used to demonstrate the capability of the developed framework.

\subsection{Abkowitz Manoeuvring model}
\label{subsec:abkowitz_model}

\noindent The Abkowitz model is formulated as a system of ordinary differential equations (ODEs) in time domain. Within this framework, the surrounding flow field is not explicitly resolved in space, hence memory effects such as wake evolution or vortex history are not taken directly into account. Instead, their net influence is absorbed into the hydrodynamic coefficients, which summarise unsteady and nonlinear flow phenomena.

\noindent Following the  Abkowitz model formulated in Principles of Naval Architecture, Vol.~III \cite{lewis_PNA_1988}, the model describes ship manoeuvring as a three-degrees-of-freedom (3-DOF) problem in the horizontal plane. A ship-fixed frame is used with the $x$-axis pointing forward, the $y$-axis to port, and the $z$-axis downwards. The kinematic variables are the surge velocity $u$, sway velocity $v$, and yaw rate $r$, while $\delta$ denotes the rudder angle. With ship mass $m$, yaw moment of inertia $I_{z}$, and longitudinal centre of gravity (CG) coordinate $x_{G}$, the equations of motion are
\begingroup\small
\begin{equation}
\begin{aligned}
m(\dot{u} - vr - x_{G}r^2) &= X, \\
m(\dot{v}+ ur + x_{G}\dot{r}) &= Y, \\
I_{z}\dot{r} + mx_{G}(\dot{v} + ur) &= N,
\end{aligned}
\label{eq:motion_eq}
\end{equation}
\endgroup
\noindent where $X$, $Y$, and $N$ are the surge and sway forces and the yaw moment due to hydrodynamic effects, respectively.

As part of the data pre-processing, all kinematic and dynamic signals are non-dimensionalised using the Prime System I as in \citet{chillcce_data-driven_2023}, and centralised with respect to the reference condition. This mitigates scale disparities, reduces unit dependence, and improves the conditioning of the regression problems used to identify hydrodynamic coefficients. It also facilitates consistent comparison across model-scale and full-scale datasets and stabilises subsequent sensitivity analyses (e.g., VIF and regularisation paths).

In the following, primed symbols denote non-\linebreak dimensional variables, defined according to the prime system where non-dimensional values are computed by $L=L_{pp}$ the reference length and $U$ the reference speed. Perturbations from the reference condition $(u_{0}=U,\, v_{0}=0,\, r_{0}=0,\, \delta_{0}=0)$ are non-dimensionalised as: 
\begin{equation}
\begin{aligned}
u' &= \frac{u}{U},\quad
v' = \frac{v}{U},\quad
r' = \frac{L\, r}{U},\\[3pt]
\dot u' &= \frac{\dot u}{U^{2}/L},\quad
\dot v' = \frac{\dot v}{U^{2}/L},\quad
\dot r' = \frac{\dot r}{U^{2}/L^{2}},
\end{aligned}
\label{eq:prime_vars}
\end{equation}
with ship advance speed $U$ corresponding to the forward speed of the ship before the manoeuvre is initialized and the ship's length between the perpendiculars $L = L_\mathrm{pp}$.

\noindent The corresponding non-dimensional mass, inertia, forces and moment terms are:
\begingroup\small
\begin{equation}
\begin{split}
m'=\frac{m}{\rho L^{3}},\quad
I_{z}'=\frac{I_{z}}{\rho L^{5}},\\
X'=\frac{X}{\rho L^{2}U^{2}},\quad
Y'=\frac{Y}{\rho L^{2}U^{2}},\quad
N'=\frac{N}{\rho L^{3}U^{2}},
\end{split}
\label{eq:prime_forces}
\end{equation}
\endgroup

\noindent where $\rho$ denotes the water density.

For readability, the subsequent expansions are written in dimensional form without prime symbols.

The hydrodynamic forces and the yaw moment are represented as a truncated Taylor series expansion (up to third order) in $(u,v,r)$ and $\delta$, with the surge velocity deviation from the initial value $\Delta u=u-U_0$:
\begin{align}
X &= X_0 + X_u \Delta u + X_{uu} \Delta u^2 + X_{uuu} \Delta u^3 \nonumber\\
&\quad + X_{vv} v^2 + X_{rr} r^2 + X_{\delta\delta} \delta^2 \nonumber\\
&\quad + X_{vvu} v^2 \Delta u + X_{rru} r^2 \Delta u + X_{\delta\delta u} \delta^2 \Delta u \nonumber\\
&\quad + X_{vr} v r + X_{v\delta} v \delta + X_{r\delta} r \delta + X_{vru} v r \Delta u \nonumber\\
&\quad + X_{v\delta u} v \delta \Delta u + X_{r\delta u} r \delta \Delta u + X_{\dot{u}} \dot{u}
\label{eq:X_Abkowitz}
\end{align}

\begin{align}
Y &= Y_0 + Y_u \Delta u + Y_{uu} \Delta u^2 + Y_v v \nonumber\\
&\quad + Y_{vvv} v^3 + Y_{vrr} v r^2 + Y_{v\delta\delta} v \delta^2 + Y_{vu} v \Delta u \nonumber\\
&\quad + Y_{vuu} v \Delta u^2 + Y_r r + Y_{rrr} r^3 + Y_{rvv} r v^2 \nonumber\\
&\quad + Y_{r\delta\delta} r \delta^2 + Y_{ru} r \Delta u + Y_{ruu} r \Delta u^2 + Y_\delta \delta \nonumber\\
&\quad + Y_{\delta\delta\delta} \delta^3 + Y_{\delta vv} \delta v^2 + Y_{\delta rr} \delta r^2 + Y_{\delta u} \delta \Delta u \nonumber\\
&\quad + Y_{\delta uu} \delta \Delta u^2 + Y_{vr\delta} v r \delta + Y_{\dot{v}} \dot{v} + Y_{\dot{r}} \dot{r}
\label{eq:Y_Abkowitz}
\end{align}

\begin{align}
N &= N_0 + N_u \Delta u + N_{uu} \Delta u^2 + N_v v \nonumber\\
&\quad + N_{vvv} v^3 + N_{vrr} v r^2 + N_{v\delta\delta} v \delta^2 + N_{vu} v \Delta u \nonumber\\
&\quad + N_{vuu} v \Delta u^2 + N_r r + N_{rrr} r^3 + N_{vvr} v r^2 \nonumber\\
&\quad + N_{r\delta\delta} r \delta^2 + N_{ru} r \Delta u + N_{ruu} r \Delta u^2 + N_\delta \delta \nonumber\\
&\quad + N_{\delta\delta\delta} \delta^3 + N_{\delta vv} \delta v^2 + N_{\delta rr} \delta r^2 + N_{u\delta} \delta \Delta u \nonumber\\
&\quad + N_{\delta uu} \delta \Delta u^2 + N_{vr\delta} v r \delta + N_{\dot{v}} \dot{v} + N_{\dot{r}} \dot{r}
\label{eq:N_Abkowitz}
\end{align}

\noindent The coefficients $X_{(\cdot)}$, $Y_{(\cdot)}$ and $N_{(\cdot)}$ are the hydrodynamic coefficients to be identified. They quantify the nonlinear interactions among surge, sway, yaw, and rudder motions, providing a polynomial approximation of the underlying ship hydrodynamics. Equations (\ref{eq:X_Abkowitz}--\ref{eq:N_Abkowitz}) hold in dimensional or non-dimensional form; in this study, the non-dimensional form is used generally.

\subsection{CFD Simulations}
\label{subsec:CFD_setup}

The flow simulations in this work were performed using Siemens Simcenter STAR-CCM+ version 2023.2 Build 18.02.008 on the TUHH High-Performance Cluster. The fluid field is discretized by the Navier-Stokes equations, which describe the conservation of mass and momentum 
\begin{align}
    \frac{\partial}{\partial t} \int_V \rho \, \mathrm{d}V + \int_S \rho (\mathbf{v} \cdot \mathbf{n}) \, \mathrm{d}S = 0 \,,
    \label{equ:NS_cont}
\end{align}

and
\label{equ:NS_momentum}
\begin{equation}
\label{equ:NS_momentum}
\begin{aligned}
&\frac{\partial}{\partial t} \int_V \rho u_i\, \mathrm{d} V + \int_S \rho u_i (\mathbf{v} \cdot \mathbf{n}) \, \mathrm{d}S\\
&= \int_S (\tau_{ij} \mathbf{i}_j - p \mathbf{i}_i) \cdot \mathbf{n} \mathrm{d}S + \int_V \rho \mathbf{g} \cdot \mathbf{i}_i \mathrm{d}V + \int_V \rho \mathbf{q}_i  \mathrm{d}V\;,
\end{aligned}
\end{equation}

with volume $V$ of the control volume enclosed by surface $S$, fluid velocity vector $\mathbf{v}$ with its components $u_i$, surface normal vector $\mathbf{n}$, time $t$, pressure $p$, fluid density $\rho$, components of the viscous stress tensor $\tau_{ij}$ unit vector $\mathbf{i}_j$ in the direction $x_j$ and gravitational acceleration $\mathbf{g}$. 

Equations (\ref{equ:NS_cont}) and (\ref{equ:NS_momentum}) are formulated in Unsteady Rey-nolds-Averaged form (URANS). Closure of these is achieved with Menter's $k$-$\omega$-SST turbulence model \cite{menter_ten_2003}. The water-air interface is resolved using the Volume-of-Fluid (VOF) method from \cite{hirt_volume_1981}. A sharp interface between both phases is achieved with the High-Resolution Interface-Capturing (HRIC) method from \cite{muzaferija_two-fluid_1998} for the discretization of the convective term in the transport equation of the volume fraction. 

Spatial derivatives in the governing equations, as well as the temporal derivatives, are approximated by second-order schemes. To solve the resulting system of linear partial differential equations, the STAR-CCM+ Implicit Unsteady Segregated Flow Solver is used. The solver uses an algebraic multi-grid method with Gauss-Seidel relaxation, V-cylces for pressure and volume fraction and flexible cycles for velocity fields. Under-relaxation factors for velocity and pressure are set to 0.8 and 0.2, respectively. The number of inner iterations per time step is set to 5. Within each iteration, the conservation equations are solved, and pressure and velocity corrections are determined using the SIMPLE method. Finally, the transport equation for the volume fraction is solved. The time step is set to correspond to 1/400\textsuperscript{th} of the ship-length-time $L/U$.

The fluid properties are specified according to representative values employed in physical model testing. Densities of water (Index w) and air (Index a) are $\rho_\mathrm{w} = 1000\,\mathrm{kg/m^3}$ and $\rho_\mathrm{a} = 1.2\,\mathrm{kg/m^3}$ with dynamic viscosities of $\mu_\mathrm{w} = 8.9 \cdot 10^{-4}\,\mathrm{Pa\ s}$ and $\mu_\mathrm{a} = 1.86 \cdot 10^{-5}\,\mathrm{Pa\ s}$.

\begin{figure}[ht!]
    \centering
    \includegraphics[width=0.95\linewidth]{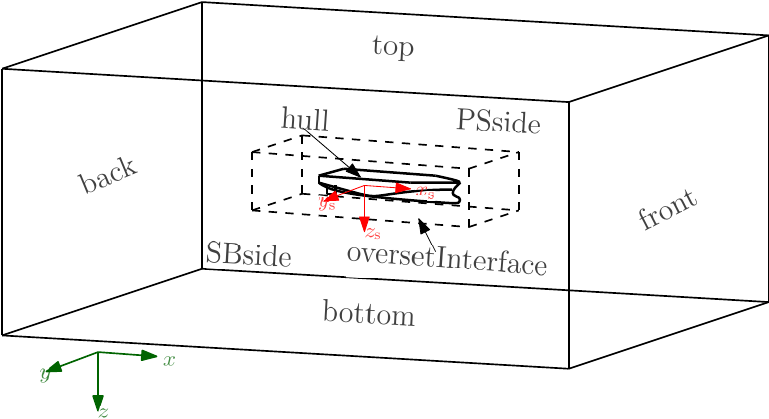}
    \caption{Computational domain geometry and boundary denotation}
    \label{fig:CFD_comp_domain_geom}
\end{figure}

The geometry of the computation domain is depicted in \autoref{fig:CFD_comp_domain_geom}. It extends in the longitudinal direction from $x = -2.56$ to $1.85\,L_\mathrm{pp}$, in the transverse direction from $y = -1.85$ to $1.85\,L_\mathrm{pp}$ and in the vertical direction from $z = 0.91\,L_\mathrm{pp}$ (below the still water level, SWL) to $z = -0.37\,L_\mathrm{pp}$ (above SWL).

To model relative motions between the ship and the background domain as well as rudder motions, the computational domain is separated into different regions: the \textit{overset hull} region, including the vicinity of the hull; the \textit{overset rudder} region, close to the rudder surface; and the \textit{background domain} region for the farfield. During the simulation of the ship manoeuvre, the hull moves in 6 DOF according to the external and fluid forces acting on the hull surface. The background domain follows the motions of the ship in the horizontal plane, i.e.\ translations in $x$- and $y$-direction and rotations around the $z$-axis. The rudder motion corresponds to the ship motion superimposed with a motion of the rudder around it's vertical axis as prescribed by the rudder angle $\delta$.

\begin{figure}[ht!]
    \centering
    \includegraphics[width=0.975\linewidth]{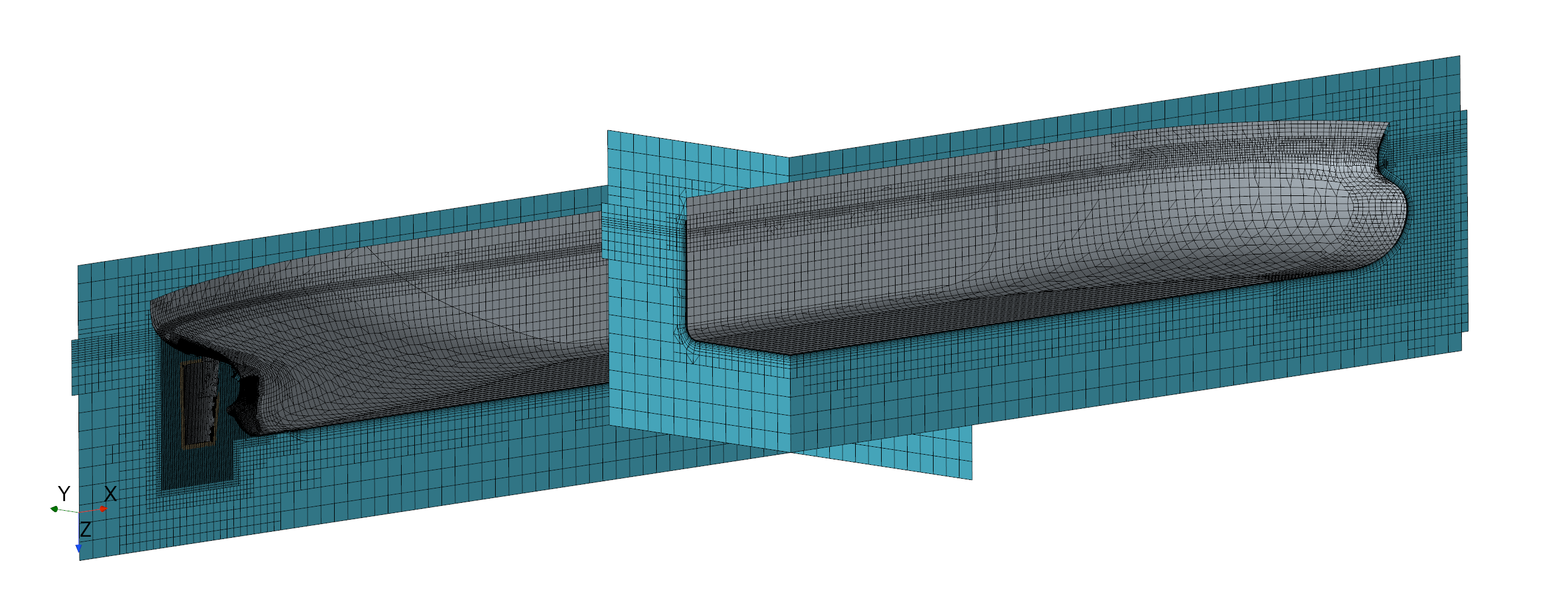}
    \caption{CFD grid of \textit{overset hull} region}
    \label{fig:CFD_hull_grid}
\end{figure}

\begin{figure}[ht!]
    \centering
    \includegraphics[width=0.975\linewidth]{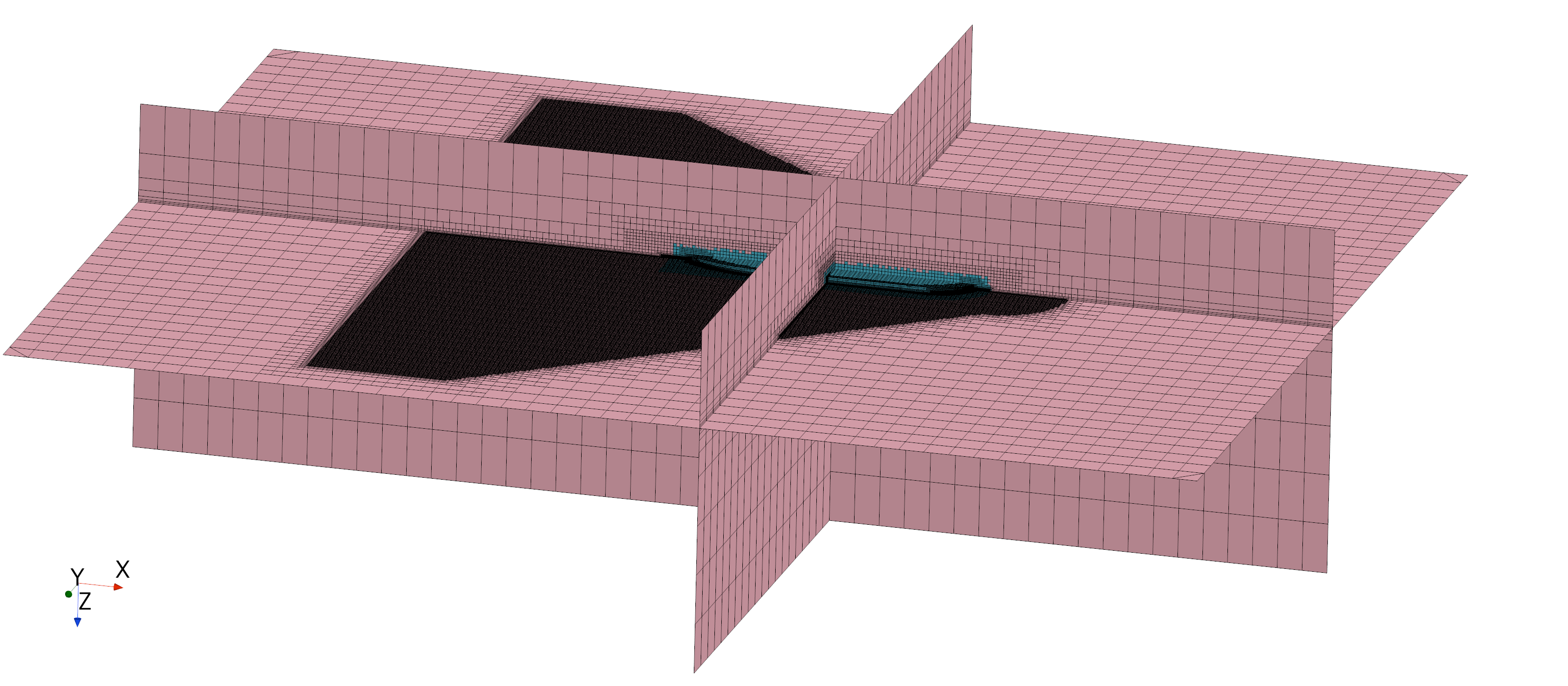}
    \caption{CFD grid of full computational domain}
    \label{fig:CFD_full_grid}
\end{figure}

The numerical grid for spatial discretization of the computational domain is generated with the STAR-CCM+ Trimmed Cell Mesher, which generates unstructured hex-dominant grids with user-specified local refinements to capture complex geometries or specific flow details. The cell size is based on a base cell size corresponding to 1/87\textsuperscript{th} of the ship length $L_\mathrm{pp}$. It coincides with the cell size in the overset interface between \textit{overset hull} region and \textit{background domain} region, and the maximum cell size in the vicinity of the hull surface. To capture specific details of the hull geometry, such as high curvature or characteristic edges, smaller cell sizes down to 3.125\% of the base size are used locally. The cell size close to the rudder surface corresponds to 6.25\% of the base size. The resulting grids are presented in Figures \autoref{fig:CFD_hull_grid} and \autoref{fig:CFD_full_grid}.

Prism layers on the hull surface ensure a non-dimensional wall distance $y^+$ averaged over the submerged part of the hull surface of $30$ to $60$. Wall functions are used to model the near-wall flow. 

At the boundaries in the horizontal direction (\textit{back}, \textit{PSside}, \textit{SBside} and \textit{front} in \autoref{fig:CFD_comp_domain_geom}), a fluid velocity equal to zero and the volume fraction according to the calm water condition are specified. Turbulence intensity is set to 1\% and the turbulent viscosity ratio of 10 is considered. At the upper boundary (\textit{top}), a zero-pressure gradient and ambient pressure are prescribed. The boundary below the hull is considered to be a slip-wall since deep-water conditions are assumed for all cases presented in this work.

\begin{figure}[ht!]
    \centering
    \includegraphics[width=0.975\linewidth]{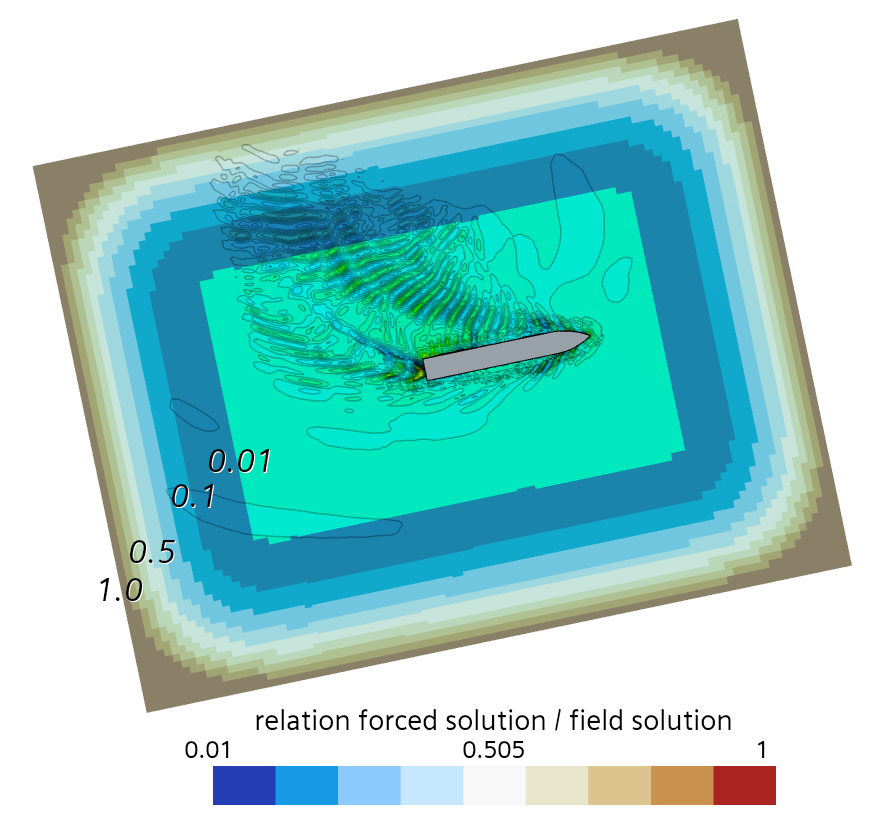}
    \caption{Forcing zone blending for a manoeuvring ship with moving domain}
    \label{fig:CFD_forcing_blending}
\end{figure}

To minimize reflections of waves radiated or diffracted from the ship at the vertical domain boundaries, the solution is gradually forced towards the calm water solution based on the distance to these boundaries. The forcing influences the fluid field with source terms for momentum and volume fraction in the governing equations. Forcing is applied within a distance of one ship length $L_\mathrm{pp}$ to the vertical boundaries with an exponential blending function, as presented in \autoref{fig:CFD_forcing_blending}. The forcing strength was optimized according to the analytical theory developed by \citet{peric_analytical_2018}.

To reduce the computational effort of the simulations, the propeller is modelled with the STAR-CCM+ actuator disk model, which computes the influence of the propeller on the fluid field and the forces on the ship as a function of the propeller rotation rate and the velocity field in the propeller inflow plane. The propeller inflow plane is placed 10\% of the propeller diameter upstream of the propeller centre and has a diameter 10\% larger than the propeller diameter. The actuator disk model calculates the local advance ratio $J = V_\mathrm{inflowplane}/(n\,D)$ with the sampled local velocity $V_\mathrm{inflowplane}$ corrected to account for the propeller-induced velocity. The thrust and torque coefficients $K_\mathrm{T}$, $K_\mathrm{Q}$ are read from the propeller open water curve as a function of $J$. 
Propeller forces are applied as external forces in the rigid body motion solver, and the propeller induced velocities are introduced into the fluid field by source terms in the governing equations. The thrust distribution in the propeller plane is computed according to the Goldstein optimum distribution \cite{goldstein_1929}. The propeller open water characteristics are provided using the open-water diagram data published on the SIMMAN2014 website \cite{simman2014_website_NRMI_curve}.

A numerical self-propulsion test is required to determine a realistic initial condition for starting the ship manoeuvring simulations. In such a test, the ship moves straight ahead with a constant propeller rotation rate $n$ and a neutral rudder angle $\delta_\mathrm{n}$. To define this state, PID controllers are employed to determine the rudder angle $\delta$ and the propeller rotation rate $n$ based on the course and speed deviations from target values, $\Delta \psi (t)$ and $\Delta u (t)$:
\begin{align}
    \delta (t) = K_{\delta,\mathrm{P}} \Delta \psi (t) +   K_{\delta,\mathrm{I}} \int_0^t \Delta \psi (t) \mathrm{d} \tau + K_{\delta,\mathrm{D}} \frac{\mathrm{d} \Delta \psi (t)}{\mathrm{d} t}\,,
\end{align}

\begin{align}
    n (t) = K_{n,\mathrm{P}} \Delta u (t) +   K_{n,\mathrm{I}} \int_0^t \Delta u (t) \mathrm{d} \tau + K_{n,\mathrm{D}} \frac{\mathrm{d} \Delta u (t)}{\mathrm{d} t}\,,
\end{align}
with the PID coefficients $K_{\delta,\mathrm{P}} = 3.0\,\mathrm{}$, $K_{\delta,\mathrm{I}} = 0.1248\,\mathrm{s}^{-1}$, $K_{\delta,\mathrm{D}} = 15.87\,\mathrm{s}$, $K_{n,\mathrm{P}} = 80.0\,\mathrm{m}^{-1}$, $K_{n,\mathrm{I}} = 120.0\,\mathrm{m}^{-1}\mathrm{s}^{-1}$ and $K_{n,\mathrm{D}} = 0.0\,\mathrm{m}^{-1}\mathrm{s}$.

Numerical simulations were performed on the TUHH High-Performance Computing Cluster (HPC) on two nodes, each with 32 CPUs. The simulation time was approximately 3-4 days for a standard manoeuvre.

\subsection{Data Preparation for Regression Methodology}
\label{subsec:data_prep}

This subsection describes the model-based problem framework employed to convert raw time series data from free-running tests into a supervised learning task. Specifically, the identification of hydrodynamic coefficients is formulated as a linear regression model. Based on the Abkowitz model, a feature space is constructed where the target variable, defined as the hydrodynamic force, is a linear combination of the input features. Within this regression framework, the unknown hydrodynamic coefficients constitute the set of parameters to be estimated. This section outlines the data processing pipeline for this regression formulation, focusing on the surge dynamics; however, this applies in the same manner to lateral and yaw motion.

The target vector for the regression model is constructed on a sample by sample basis. For each data instance, the target value $X(\cdot)$ is computed by isolating the known inertial contributions on the right-hand side of the governing equation of motion. This value represents the net hydrodynamic surge force, which is the quantity the model is trained to predict:
\begin{equation}
  X(\Delta u,v,r,\delta,\dot{u}) \;=\; m\,\dot{u} \;-\; m\big(vr + x_G r^{2}\big),
  \label{eq:surge-balance-core}
\end{equation}
The left-hand side now represents the hydrodynamic force in surge direction as a function of the hydrodynamic coefficients $X_{(\cdot)}$, which need to be predicted for constructing a system-based manoeuvring model.

The structure of the linear model is defined by a polynomial basis derived from the kinematic variables. This process generates the feature set used for the regression. Consistent with the Abkowitz model, the hydrodynamic surge force $X$ is expressed as a linear combination of these basic features:
\begin{align}
X - X_{\dot{u}} \dot{u} &= X_0 + X_u \Delta u + X_{uu} \Delta u^2 + X_{uuu} \Delta u^3 \nonumber \\
&\quad + X_{vv} v^2 + X_{rr} r^2 + X_{\delta\delta} \delta^2 \nonumber \\
&\quad + X_{vvu} v^2 \Delta u + X_{rru} r^2 \Delta u + X_{\delta\delta u} \delta^2 \Delta u \nonumber \\
&\quad + X_{vr} v r + X_{v\delta} v \delta + X_{r\delta} r \delta + X_{vru} v r \Delta u \nonumber \\
&\quad + X_{v\delta u} v \delta \Delta u + X_{r\delta u} r \delta \Delta u,
\label{eq:X_Abkowitz_b}
\end{align}

where the hydrodynamic coefficients ($X_u, X_{uu}$, etc.) are the model parameters to be estimated.

The added-mass derivative $X_{\dot{u}}$, which represents a primarily potential fluid effect, can be excluded from the prediction and computed directly using a RANS method, as outlined in \citet{el_moctar_2022}. In contrast, damping and steering coefficients are dominated by complex, viscous-flow phenomena that are most effectively characterized through SI based on experimental or numerical data from free-running manoeuvres.

Therefore, the model is constrained by treating $X_{\dot{u}}$ as a known parameter supplied by additional CFD simulations, which are given in \autoref{tab:cfd_added_mass_results}. The corresponding term, $X_{\dot{u}}$, is transposed to the right-hand side of the equation. This refines the regression's target variable to represent hydrodynamic effects that are not dependent on surge acceleration. This includes not only damping and coupling terms but also, critically, the control or steering forces generated by the rudder angle $\delta$.
\begin{equation}
  X \;=\; \big(m - X_{\dot{u}}\big)\dot{u} - m\big(vr + x_G r^{2}\big).
  \label{eq:surge-linear-core}
\end{equation}

The problem is then formulated in matrix form. The feature matrix, denoted as $A\equiv{A}_X$, is constructed by stacking the basis feature rows from each time sample. The parameter vector, $x\equiv{x}_X$, contains the set of unknown hydrodynamic coefficients to be determined. Finally, the target vector, $b\equiv{b}_X$, is assembled by stacking the corresponding target values computed for each sample using Eq. \eqref{eq:surge-linear-core}.
These components are defined as:

\begin{subequations}\label{eq:design_triplet}
\begin{gather}
A_X \;=\;
\left[
\begin{array}{cccc}
\Delta u & \Delta u^{2} & \cdots & r\,\delta\,\Delta u\\
\vdots   & \vdots       & \ddots & \vdots\\
\vdots   & \vdots       & \ddots & \vdots
\end{array}
\right], \\
x_X \;=\;
\begin{bmatrix}
X_{u}\\
X_{uu}\\
\vdots\\
X_{r\delta u}
\end{bmatrix}, \\
b_X \;=\;
\begin{bmatrix}
\big(m - X_{\dot{u}}\big)\dot{u} - m\big(v r + x_G r^{2}\big)\\
\vdots
\end{bmatrix}.
\end{gather}
\end{subequations}

Stacking all samples produces the linear regression
\begin{equation}
    A x = b\;.
    \label{eq:Ax_equal_b_blocks}
\end{equation}

\subsection{Regularized Machine Learning Models for System Identification}
\label{subsec:reg_ML_methods}

To address the limitations of the standard LSQ approach, this study adopts a Machine Learning (ML) framework. The fundamental difference lies in the optimization objective. Whereas LSQ exclusively seeks to minimize the sum of squared errors (the empirical risk), ML models augment this objective with a regularization term. This principle is known as Empirical Risk Minimization with regularization.

This study evaluates three prominent models that implement this regularized framework: Lasso, Ridge, and SVR. The first two augment the LSQ objective with a penalty on the coefficient vector's magnitude. Lasso regression applies an $\ell_1$ norm penalty depending on the sum of the absolute coefficient values. This is known to induce sparsity by driving some coefficients to exactly zero, thus performing an implicit feature selection. Ridge regression, in contrast, uses an $\ell_2$ norm penalty Euclidean norm that shrinks the coefficients, providing the necessary stability to counteract multicollinearity. SVR operates on a different principle by modifying the loss function itself. It introduces an $\epsilon$-insensitive loss, which creates a margin where errors are not penalized, making the model inherently robust to outliers. Together, this suite of models allows for a systematic comparison of different regularization strategies aimed at improving the robustness and physical consistency of the estimated parameters.

A key advantage of the selected models is that they can all be formulated as convex Quadratic Programming (QP) problems. The convexity of this formulation is a crucial property, as it guarantees that the optimization process converges to a unique, global optimum, precluding the issue of local minima.

Representing these regression techniques within a unified QP framework provides a consistent mathematical basis for a direct comparison of their performance. Furthermore, it ensures that the solutions are solver independent, facilitating reproducibility. For implementation, the baseline LSQ problem is solved using the python standard ordinary least squares library. The regularized and margin-based models are implemented using the python convex programming CVXPY library, a domain-specific language for convex optimization that interfaces with various high-performance solvers. The choice of solver, however, does not alter the mathematical specification of the problem.

As described using the example of the surge balance, the identification uses the linear form in Eq.~\eqref{eq:Ax_equal_b_blocks}.
where $A$ collects the kinematic predictors prescribed by the Abkowitz-type expansion, $x$ contains the unknown hydrodynamic coefficients, and $b$ is a known right-hand side constructed from measured kinematics and known inertial terms (including added mass treated as known constants). By enriching this baseline with regularisation and convex constraints, the identification attains improved robustness to multicollinearity and better physical consistency of the estimated coefficients. The specific LSQ, Lasso, Ridge, and SVR formulations are presented in the next subsection.

\subsubsection{Least-Squares Regression Model}

Ordinary Least Squares LSQ serves as the classical baseline for the parameter estimation task. Its objective function represents the direct minimization of the empirical risk with no regularization. Given the linear relation in Eq.~\eqref{eq:Ax_equal_b_blocks}, the coefficient vector $x \in \mathbb{R}^{p}$ is found by solving:
\begin{equation}
  \min_{x}\; \tfrac{1}{2}\,\|A x - b\|_{2}^{2},
  \label{eq:ols_obj}
\end{equation}
where $\|\cdot\|_{2}$ denotes the Euclidean norm. The leading factor \( \tfrac{1}{2} \) is conventional and does not affect the minimiser, since it scales the objective by a positive constant.

When the matrix $A^{\top}A$ is full rank, this problem has the well-known analytical solution:
\begin{equation}
  x^{\star} = (A^{\top}A)^{-1}A^{\top}b,
  \label{eq:normal_eqs}
\end{equation}
However, in practice, the presence of ill-conditioning and multicollinearity in the feature matrix $A$ often renders this solution numerically unstable, motivating the use of constrained and regularized formulations.

Physical knowledge can be incorporated by imposing hard constraints on the coefficients, which defines the bounded (or constrained) least-squares problem:
\begin{equation}
  \min_{x}\; \tfrac{1}{2}\,\|A x - b\|_{2}^{2}
  \quad \text{s.t.} \quad \mathrm{lb} \leq x \leq \mathrm{ub},
  \label{eq:cls_bounds}
\end{equation}
where the vectors $\mathrm{lb}$ and $\mathrm{ub}$ encode physical limits, such as sign restrictions or stability-motivated bounds.

The bounded problem is solved with SciPy's \texttt{LSQ\_linear}. This constrained LSQ formulation serves as the primary reference against which the regularized models are assessed.

\subsubsection{Lasso Regression Model}

The Lasso (Least Absolute Shrinkage and Selection Operator) model extends the LSQ formulation by augmenting the objective function with an $\ell_1$ penalty on the coefficient vector. This introduces a soft constraint that penalizes model complexity. The optimization problem is:
\begin{equation}
  \min_{x}\; \tfrac{1}{2}\,\|A x - b\|_{2}^{2} \;+\; \lambda\,\|x\|_{\ell_1}
  \quad \text{s.t.} \quad \mathrm{lb} \le x \le \mathrm{ub},
  \label{eq:Lasso_obj}
\end{equation}
where $\|x\|_{\ell_1}=\sum_{j=1}^{p}|x_j|$, $\lambda>0$ controls the regularization strength. The $\mathrm{lb}$ and $\mathrm{ub}$ constraints enforce the same physical restrictions as in the LSQ model.

The defining property of $\ell_1$ regularisation is its ability to induce sparsity. This means it can drive some coefficients to be exactly zero, effectively performing an implicit feature selection. This process yields a more parsimonious model where weak or redundant basis features are excluded, which can significantly improve the interpretability of the hydrodynamic model. By introducing this bias, Lasso reduces the variance of the estimator, often leading to improved predictive performance on unseen data. It is important to note that a coefficient can only be driven to exactly zero if the interval defined by its bounds, $[\mathrm{lb},\mathrm{ub}]$, contains zero.

The optimization problem in Eq.\,\eqref{eq:Lasso_obj} is convex and is implemented using the \texttt{CVXPY} library to ensure reproducibility. Unlike the $\ell_2$ penalty used in Ridge regression, which retains all predictors, the Lasso penalty enforces sparsity and is thus capable of identifying a reduced set of the most influential hydrodynamic coefficients.

\subsubsection{Ridge Regression Model}
\label{subsubsec:Ridge_regression}

Ridge regression also extends the LSQ baseline by adding a penalty term; it uses the squared $\ell_2$ Euclidean norm of the coefficient vector. This form of regularization is particularly effective at stabilizing models in the presence of multicollinearity. The optimization problem is defined as:
\begin{equation}
  \min_{x}\; \tfrac{1}{2}\,\|A x - b\|_{2}^{2} \;+\; \lambda\,\|x\|_{\ell_2}^{2}
  \quad \text{s.t.} \quad \mathrm{lb} \le x \le \mathrm{ub},
  \label{eq:Ridge_obj}
\end{equation}
where $\|x\|_{\ell_2}^{2}=\sum_{j=1}^{p}x_j^{2}$ and $\lambda>0$ control the degree of shrinkage; the $\mathrm{lb}$ and $\mathrm{ub}$ constraints enforce the same physical restrictions as in Eq.~\eqref{eq:cls_bounds}. As mentioned above, the leading factor \( \tfrac{1}{2} \) is conventional and does not affect the minimiser.

The $\ell_2$ penalty continuously shrinks all coefficients towards zero as $\lambda$ increases, but unlike the $\ell_1$ penalty, it does not set them to exactly zero. This penalty renders the objective function strictly convex, which guaranties a unique solution. Furthermore, it improves the numerical conditioning of the problem by effectively adding a positive constant to the diagonal of the $A^{\top}A$ matrix, which directly reduces the variance of the estimated parameters.

When faced with a group of highly correlated features, Ridge tends to distribute the effect among them, shrinking their coefficients together. This "grouping" effect often leads to more physically plausible and stable solutions compared to Lasso, which may arbitrarily select one feature from the group.

The problem in Eq.\,\eqref{eq:Ridge_obj} is formulated in \texttt{CVXPY} with standard convex optimisation solvers, ensuring reproducibility and solver independence.

\subsubsection{Support Vector Regression}

Support Vector Regression (SVR) distinguishes itself from the previous methods by fundamentally altering the loss function rather than augmenting the OLS objective. SVR introduces an $\epsilon$--insensitive loss, where errors smaller than a specified margin $\epsilon$ are completely ignored. The model's goal is to find a function that fits the data within this "tube" while penalizing only the points that fall outside it.

The objective function contains two competing components: an $\ell_2$ regularization term on the coefficient vector, which is identical to that of Ridge regression and penalizes model complexity, and a term that penalizes prediction errors. The primal optimization problem is formulated as:

\begin{align}
\min_{x, \xi} \quad & \tfrac{1}{2} \| x \|_{\ell_2}^2 + C \sum_{i=1}^n \xi_i \label{eq:SVR_primal} \\
\text{s.t.} \quad
& (Ax - b)_i - \epsilon \leq \xi_i, \quad i = 1, \dots, n, \nonumber \\
& -(Ax - b)_i - \epsilon \leq \xi_i, \quad i = 1, \dots, n, \nonumber \\
& \xi_i \geq 0, \quad i = 1, \dots, n, \qquad \mathrm{lb} \leq x \leq \mathrm{ub}. \nonumber
\end{align}

Here, the so-called slack variables $\xi_{i} \ge 0$ measure the magnitude of the prediction error for points outside the $\epsilon- tube$.

The model is governed by two key hyperparameters. The margin $\epsilon$ defines the width of the tube where errors incur no penalty. The hyperparameter $C>0$ controls the trade-off between the regularization term and the error term. A large $C$ places a high penalty on errors, forcing the model to fit the training data more closely, while a small $C$ prioritizes a simpler model with a larger margin, increasing its tolerance for noise. This framework is advantageous for hydrodynamic identification where minor deviations may be acceptable, but large prediction errors must be strongly penalized.

This convex optimization problem is implemented in \texttt{CVXPY}. While the dual formulation of SVR allows for non-linear mappings via the "kernel trick," this work is restricted to the linear primal formulation, which aligns directly with the established regression framework.

\subsection{Multicollinearity diagnostics via Variance Inflation Factor}
\label{subsec:vif}

Multicollinearity refers to strong linear dependence among predictors in the Abkowitz-type design matrix, a situation that arises naturally when polynomial and interaction terms are included. In practice, this leads to an ill--conditioned normal matrix \(A^{\top}A\), such that LSQ becomes variance dominated: small perturbations in the data or model produce large oscillations in individual coefficient estimates, and large but cancelling force and moment components may appear despite a small residual. To overcome this challenge, the Variance Inflation Factor (VIF) is used to quantify predictor-level linear dependence, supporting decisions on regularisation and on the excitation-data design required for identification.

For each regressor \(X_j\), the VIF is defined as
\[
    \mathrm{VIF}_j \;=\; \frac{1}{\,1 - R_j^2\,},
\]
where \(R_j^2\) is obtained by regressing \(X_j\) on all remaining predictors. Values near \(1\) indicate weak linear dependence, whereas large values signal near-collinearity and variance amplification in the corresponding coefficient estimate.

For the purposes of computation, scaling and matrix formulation, let \(X\in\mathbb{R}^{n\times p}\) represent the predictor matrix built from the monomial prefactors of the force and moment expansions (one matrix per degree of freedom). For each column \(X_j\), set \(y=X_j\) and regress it on \(Z=X_{-j}\) using ordinary least squares to obtain \(R_j^2\). When predictors are standardised (zero mean, unit variance) and constant terms are excluded, the correlation matrix \(C=\mathrm{corr}(X)\) yields the equivalent closed forms
\[
\mathrm{VIF}_j \;=\; (C^{-1})_{jj},
\qquad
R_j^2 \;=\; 1 \;-\; \frac{1}{(C^{-1})_{jj}}.
\]

In practice, columns are mean-centred (and typically standardised) before forming interaction terms; constant terms (e.g. \(X_0, Y_0, N_0\)) naturally result in zero correlation ($\mathrm{VIF} = 0$). Because VIF measures structure in the predictors, it depends on the dataset and on the selected coefficient set, not on the identification algorithm. When sway and yaw share identical motion variables, the corresponding VIF fields are identical by construction. Thresholds used as flags (e.g. \(5\)-\(10\)) are heuristic and sensitive to scaling and to the excitation content of the manoeuvres.

Large VIF values correspond to strong linear dependence among predictors, which directly manifests as an ill-conditioned normal matrix and unstable least-squares estimates. This effect is evident for the full set of hydrodynamic coefficients suggested in \cite{lewis_PNA_1988}, where several coupled surge, sway, and yaw terms exhibit VIF levels well above common thresholds, consistent with the unrealistic yet --for the input data-- cancelling force and moment components observed in the simulations (\autoref{fig:forceComps_zz3535ps_diff_coeff_sets}). These high VIFs quantify the same multicollinearity that produces near singular matrices and large variance amplification in the coefficient estimates. In contrast, with the reduced set of coefficients from \citet{Mucha2017} the VIF levels decrease substantially, reflecting a better conditioned regression problem and improved numerical robustness in the identified coefficients (\autoref{fig:vif_values_diff_coeff_sets_X} and \autoref{fig:vif_values_diff_coeff_sets_YN}).

Reducing multicollinearity cannot rely solely on eliminating predictors with high VIF values, since these may correspond to physically relevant terms in the hydrodynamic model. Mitigation can be achieved through dimensional reduction strategies that preserve the dominant physical effects (e.g. adopting the reduced coefficient set from \citet{Mucha2017}) or by enriching the training data with additional manoeuvres that excite different dynamic modes and thus alter the correlation structure of the predictors. Regularisation techniques provide a complementary route: Ridge regression stabilises the inversion of \(A^{\top}A\) by penalising large coefficients, thereby reducing the impact of ill-conditioning without discarding potentially important cross-terms; Lasso regression can improve interpretability by eliminating uninformative terms, but care is required to avoid suppressing terms that are relevant not for the input manoeuvre but other manoeuvres that the hydrodynamic model should capture.

In summary, the VIF analysis complements the regularisation study by quantifying multicollinearity within each degree of freedom. It indicates that the ill-conditioning encountered is predominantly structural originating from polynomial coupling in the Abkowitz expansion and from limited excitation in single-manoeuvre datasets rather than purely numerical. Consequently, a consistent strategy combines reduced coefficient sets, diversified manoeuvre inputs, and Ridge-type regularisation to mitigate multicollinearity and to deliver stable, physically interpretable parameter estimates across identification scenarios.

\subsection{Test Case}
\label{subsec:test_case}

The KRISO Very Large Crude Carrier 2 (KVLCC2) is a benchmark model of a modern tanker ship developed by the Korea Research Institute for Ships \& Ocean Engineering (KRISO, formerly MOERI) as a benchmark model geometry for research and validation of experimental and numerical methods for ship hydrodynamics. It features a typical hull form for ships with high block coefficients and is used for the analysis of ship resistance, sea-keeping and manoeuvring characteristics.

\begin{figure}[ht!]
    \centering
    \includegraphics[width=0.975\linewidth]{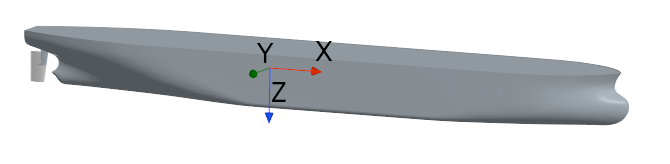}
    \caption{KVLCC2 hull geometry and ship-fixed coordinate system}
    \label{fig:KVLCC2_hull_geom}
\end{figure}

The geometry used for the numerical simulations is presented in \autoref{fig:KVLCC2_hull_geom}. Main Particulars are given in \autoref{tab:KVLCC2_particulars} for the full scale ship and the model scale $\lambda_\mathrm{s} = 1:45.714$, with distance of the centre of gravity from the keel $\mathrm{KG}$, horizontal distance from the midship section forward $\mathrm{LCG}$ and the yaw radius of gyration $k_{zz}$.

\begin{table}[ht!]
    \caption{Main Particulars of the KVLCC2 ship in full scale and model scale}
    \label{tab:KVLCC2_particulars}
    \centering
    \begin{tabular}{>{\raggedright}llrr}
    \hline
    Symbol & Unit & full scale & model scale \\ \hline
    $L_\mathrm{pp}$ & [m] & 320 & 7 \\ 
    $B$ & [m] & 58 & 1.2688 \\ 
    $T$ & [m] & 20.8 & 0.455 \\ 
    $\nabla $ & [m\textsuperscript{3}] & 312622 & 3.2724 \\ 
    KG & [m] & 18.6 & 0.4069 \\ 
    LCG & [m] & 11.136 & 0.2436 \\ 
    $k_\mathrm{zz}/L_\mathrm{pp}$ & [-] & 0.25 & 0.25 \\ 
    \end{tabular}
\end{table}

The numerical flow simulations are performed in model scale, while further investigations are conducted using  non-dimensional variables.
\begin{figure}[ht!]
    \centering
    \includegraphics[width=0.975\linewidth]{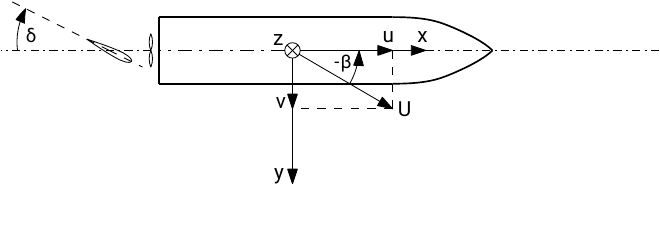}
    \caption{Coordinate system}
    \label{fig:COS}
\end{figure}

\autoref{fig:COS} presents the ship-fixed coordinate system used in the present work. The origin coincides with the centre of gravity of the ship. The $x$-axis points in the ship forward direction, the $y$-axis to the starboard and the $z$-axis downwards in the direction of gravitational acceleration; therefore, the yaw-moment $N$ is positive when turning to starboard. The rudder angle is defined by the mathematically positive rotation around the $z$-axis; i.e. positive if the trailing edge of the rudder is on the port side. The drift angle is defined as $\beta = \mathrm{tan}^{-1}(-v/u)$.

\section{Results}
\label{sec:results}

In section \ref{subsec:cfd_sim_results} the CFD simulation results are summarized and validated using free-running model test data from the SIMMAN 2014 workshop \cite{simman2014}. CFD results are then used to predict hydrodynamic coefficients in Sections \ref{subsec:set_of_coeffs} to \ref{subsec:several_manoeuvre_sets_used}.

The manoeuvring metrics reported in the following sections are selected and defined in accordance with IMO Resolution MSC.137(76) \cite{IMO2002} and the associated explanatory guidance. The key metrics are defined as follows.

For a zig-zag test, the first rudder execute is the application of a specified rudder angle during an initially straight approach. The second execute occurs when the rudder is shifted to the opposite side after the ship reaches a prescribed heading deviation from the original heading. The first overshoot angle (in degrees) is the maximum additional heading deviation beyond the target heading measured after the second execute. Similarly, the second overshoot angle is the maximum additional heading deviation beyond the target heading measured after the subsequent rudder reversal. The time to second execute, $t_{\mathrm{exec2}}$, is the time interval from the first execute to the instant when the heading deviation first reaches the prescribed target. The time to check yaw, $t_{\mathrm{check}}$, is the time interval from the second execute to the first overshoot peak (i.e., the first local extremum of the heading deviation following the second execute). Both time-based metrics are non-dimensionalised using $t^{*}=tU/L$, where $L$ is the ship length and $U$ is the approach speed.

For the turning circle manoeuvre, all metrics are non-dimensionalised by the ship length $L$. The advance at $90^{\circ}$ (AD) is the distance along the initial course from the rudder execute point to the position where the heading deviation first reaches $90^{\circ}$. The transfer at $90^{\circ}$ (TR) is the lateral displacement perpendicular to the initial course at the instant when the heading deviation first reaches $90^{\circ}$. The tactical diameter at $180^{\circ}$ (TD) is the lateral displacement at the instant when the heading deviation first reaches $180^{\circ}$. Finally, the steady turning diameter (SD) is the diameter of the final steady turning circle at a constant rudder angle.

\subsection{CFD Simulation Results}
\label{subsec:cfd_sim_results}

An overview of the CFD simulation cases carried out in the present study is given in \autoref{tab:cfd_simulation_overview}. Reference data from the SIMMAN 2014 workshop \cite{simman2014} is available for free-running model tests for zig-zag 20°/20° (ZZ\,20$^\circ$/20$^\circ$) and turning circle (TC\,35$^\circ$) manoeuvres to port side (PS) and starboard (SB) direction.

\begin{table}[ht!]
    \centering    
    \caption{Overview of CFD ship manoeuvre simulations and case specifiers (ZZ: zig-zag manoeuvre, TC: turning circle, PS: port side, SB: starboard side)}
    \label{tab:cfd_simulation_overview}
    \begin{tabular}{ccc}
        Manoeuvre specifier & Reference data used for validation \\
        \hline
        ZZ 10$^\circ$/10$^\circ$ PS & - \\
        ZZ 20$^\circ$/20$^\circ$ PS & SIMMAN 2014 \\
        ZZ 35$^\circ$/35$^\circ$ PS & - \\
        ZZ 35$^\circ$/35$^\circ$ SB & - \\
        TC 35$^\circ$ PS & SIMMAN 2014 \\
        TC 35$^\circ$ SB & SIMMAN 2014 \\
    \end{tabular}
\end{table}

\begin{figure}[ht!]
    \centering
    \includegraphics[width=1\linewidth]{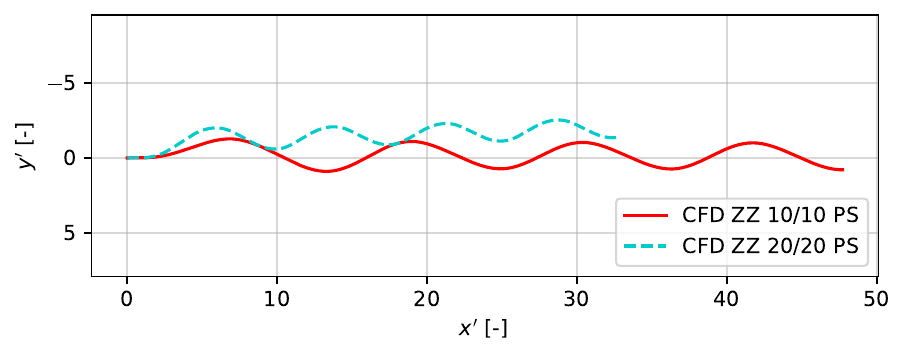}
    \caption{CFD simulation trajectories for the ZZ\,10$^\circ$/10$^\circ$and ZZ\,20$^\circ$/20$^\circ$manoeuvres starting to PS}
    \label{fig:cfd_zz_results}
\end{figure}

\begin{figure}[ht!]
    \centering
    \includegraphics[width=1\linewidth]{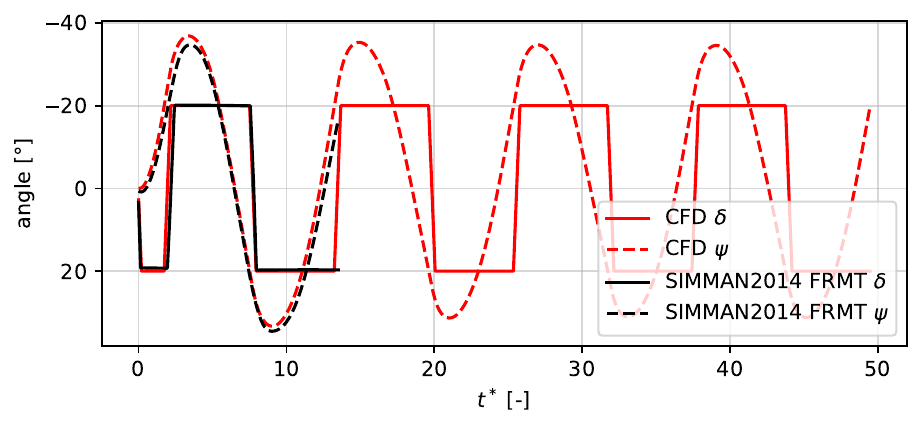}
    \caption{CFD simulation yaw angle results for the ZZ 20/20 manoeuvre starting to PS in comparison with experimental model test (EFD) data from SIMMAN 2014 \citep{simman2014}}
    \label{fig:cfd_zz2020_angles_results}
\end{figure}

\autoref{fig:cfd_zz_results} presents the trajectories of the ZZ\,10$^\circ$/10$^\circ$and ZZ\,20$^\circ$/20$^\circ$manoeuvre with rudder turning initially to PS direction (initial course change to PS). Experimental reference data from SIMMAN 2014 \cite{simman2014} is used for validation of the CFD simulation setup for the time series of rudder and yaw angle during the ZZ\,20$^\circ$/20$^\circ$ manoeuvre and compared to the results from the CFD simulation in \autoref{fig:cfd_zz2020_angles_results}. \autoref{tab:zz2020_valVar} contains a comparison of numerical values of selected validation variables. The agreement between CFD results and experimental reference data is satisfactorily with absolute differences of the validation variables up to 11.6\% for the rudder execution times and up to 19.0\% for the overshoot angles. 

\begin{table}[ht!]
    \centering
    \caption{Comparison of validation variables for the CFD ZZ 20$^\circ$/20$^\circ$ PS manoeuvre to experimental model test (EFD) data from SIMMAN 2014 \cite{simman2014}}
    \label{tab:zz2020_valVar}
    \begin{tabular}{c|c c |c}
        Validation variable & CFD & EFD & difference \\
        \hline
        \hline
        \multicolumn{3}{l}{rudder execution times fullscale [s]}\\
        \hline
        2nd execute $t_2$ & 69.6 & 78.7 & -11.6\%\\ 
        3rd execute $t_3$ & 302.1 & 304.1 & -0.7\%\\
        \hline
        \multicolumn{3}{l}{overshoot angles [°]}\\
        \hline
        Overshoot 1& 16.9 & 14.2 & +19.0\%\\ 
        Overshoot 2& 13.3 & 15.4 &-13.6\% \\
    \end{tabular}
\end{table}

\begin{figure}[ht!]
    \centering
    \includegraphics[width=1\linewidth]{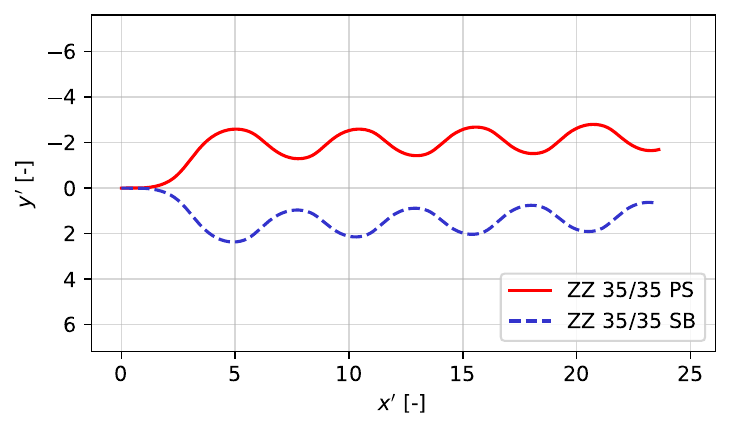}
    \caption{CFD simulation trajectories fr the ZZ\,35$^\circ$/35$^\circ$ manoeuvre starting to PS and SB}
    \label{fig:cfd_zz3535_results}  
\end{figure}

\autoref{fig:cfd_zz3535_results} presents results of the large angle zig-zag manoeuvres ZZ\,35$^\circ$/35$^\circ$ PS and SB. These simulations were performed to generate manoeuvring time series data that contains a lot of dynamic information for the extraction of hydrodynamic coefficients. Both simulations resulted in a trajectories featuring a transverse displacement of the ship at the beginning of the manoeuvre due to the initial turning direction followed by close-to periodic transverse motion and a mean forward motion in $x$-direction. 

\begin{figure}[ht!]
    \centering
    \includegraphics[width=0.8\linewidth]{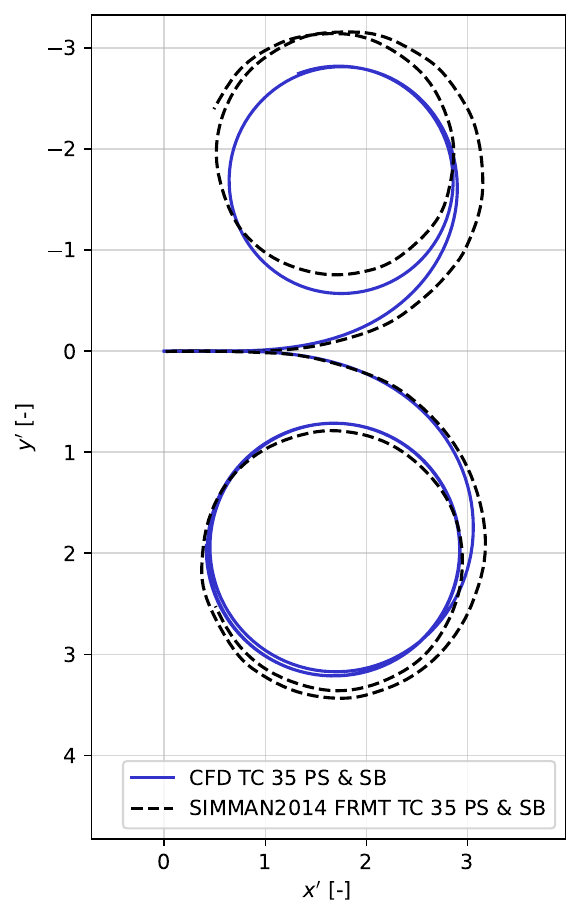}
    \caption{CFD simulation trajectories fr the TC 35$^\circ$ manoeuvre starting to PS and SB}
    \label{fig:cfd_tc_results}
\end{figure}

Trajectories for the turning circle manoeuvres to PS and SB are presented in \autoref{fig:cfd_tc_results} along with the trajectories of free-running model tests from SIMMAN 2014 \cite{simman2014}. Numerical values of selected validation variables are compared in \autoref{tab:tc35ps_valVar} and \ref{tab:tc35sb_valVar}. The calculated values for turning circle diameter, advance and transfer in the CFD simulations were smaller than in the experiments, especially for the PS turning direction, indicating a larger turning ability of the numerical model. Overall absolute differences of the validation variables were about up to 11.5\% for PS and up to 9.8\% for SB turning circle manoeuvres.

\begin{table}[ht!]
    \centering
    \caption{Comparison of validation variables for the CFD TC PS manoeuvre to experimental model test data (EFD) from SIMMAN 2014 \cite{simman2014}}
    \label{tab:tc35ps_valVar}
    \begin{tabular}{c|c c|c}
        Validation variable & CFD & EFD & difference \\
        \hline
        \hline
        \multicolumn{3}{l}{Trajectory dimensions fullscale [m]}\\
        \hline
        Advance AD & 897 & 985 & -8.9\%\\ 
        Transfer TR  & 355 & 399 &-11.0\% \\
        Tactical diameter TD  & 873 & 986 & -11.5\% \\
        Turning diameter D  & 716 & 787 & -9.0\%\\
        \hline
        \multicolumn{3}{l}{Trajectory times fullscale [s]}\\
        \hline
        $T_{90}$ & 155 & 168 & -7.7\%\\
        $T_{180}$ & 331 & 348 & -4.9\%\\
        $T_{360}$ & 763 & 761 & +0.3\%\\
    \end{tabular}
\end{table}

\begin{table}[ht!]
    \centering
    \caption{Comparison of validation variables for the CFD TC SB manoeuvre to experimental model test data (EFD) from SIMMAN 2014 \cite{simman2014}}
    \label{tab:tc35sb_valVar}
    \begin{tabular}{c|c c|c}
        Validation variable & CFD & EFD& difference \\
        \hline
        \hline
        \multicolumn{3}{l}{Trajectory dimensions fullscale [m]}\\
        \hline
        Advance AD & 951 & 989 & -3.8\%\\ 
        Transfer TR  & 395 & 438 & -9.8\%\\
        Tactical diameter TD  & 990 & 1072 & -7.6\% \\
        Turning diameter D  & 808 & 806 & +0.2\%\\
        \hline
        \multicolumn{3}{l}{Trajectory times fullscale [s]}\\
        \hline
        $T_{90}$ & 165 & 176 & -6.3\%\\
        $T_{180}$ & 354 & 371 & -4.6\%\\
        $T_{360}$ & 788 & 803 & -1.9\%\\
    \end{tabular}
\end{table}

The overall agreement between numerical results and experimental reference data is considered to be satisfactory, considering the relatively coarse CFD grid and model simplifications due to the virtual disk model (see \autoref{subsec:CFD_setup}).

Additional CFD simulations had to be carried out to determine acceleration depended coefficients i.e. the added-mass terms as described in section \ref{subsec:data_prep}. The numerical simulation setup followed the procedure defined by \citet{el_moctar_2022} closely and is therefore not described in detail here. The resulting added-mass terms are given in \autoref{tab:cfd_added_mass_results}.

\begin{table}[h!]
    \centering
    \caption{Hydrodynamic added-mass coefficients from CFD simulations; values are non-dimensionalised and have to be multiplied by factor $10^{-3}$}
    \label{tab:cfd_added_mass_results}
    \begin{tabular}{ccccc}
         $X_{\dot{u}}$ & $Y_{\dot{v}}$ & $N_{\dot{r}}$ & $Y_{\dot{r}}$ & $N_{\dot{v}}$\\
         \hline
         $-1.132$ &  $-15.138$ & $-0.803$ & $-0.469$ & $-0.468$ \\
    \end{tabular}
\end{table}

Numerical simulations of free-running manoeuvres took about 30 to 60 hours computation time depending on the computation nodes of the HPC cluster that had been assigned to the specific job.


\subsection{Evaluation of different sets of hydrodynamic coefficients}
\label{subsec:set_of_coeffs}

For system-based manoeuvring simulations employing the Abkowitz model, multiple possibilities exist concerning the hydrodynamic coefficients included for each degree of freedom. In general, employing a larger set of coefficients yields a more complex model capable of capturing the non-linear dependence of hydrodynamic forces on ship motions, provided the additional coefficients are defined in a physically meaningful way. However, predicting a large number of coefficients from free-running ship motion data is challenging due to the increased risk of multicollinearity among combinations of manoeuvring variables, which arises simply because the number of possible variable combinations grows (cf. \citet{bonci_2015, wang_zou_2018}). 

As an example, the Abkowitz model proposed in \cite{lewis_PNA_1988} has a relatively a set of 68 hydrodynamic coefficients. Conversely, \citet{Mucha2017} presents a comparatively small set of 32 coefficients tailored specifically for the KVLCC2 ship, derived through local sensitivity analysis. This section investigates the impact of employing these two different coefficient sets. The hydrodynamic coefficients are predicted by simple least squares fitting to data obtained from a ZZ 35$^\circ$/35$^\circ$ PS manoeuvre. The models' abilities to reproduce both the input manoeuvre and an independent test manoeuvre TC 35$^\circ$ PS, unseen by the prediction algorithm, are evaluated. The predicted manoeuvres are then compared against the CFD simulation results from section \ref{subsec:cfd_sim_results} to assess accuracy.

\begin{figure}[ht!]
    \centering
    \begin{subfigure}[b]{0.49\textwidth}
        \centering
        \includegraphics[width=1\linewidth]{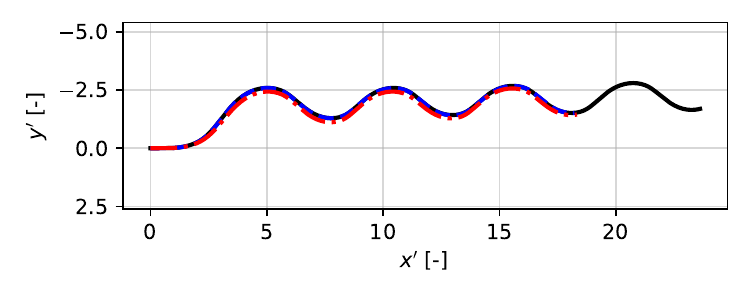}
        \caption{ZZ 35º/35º PS manoeuvre (input manoeuvre)}
    \end{subfigure}

    \begin{subfigure}[b]{0.4\textwidth}
        \centering
        \includegraphics[width=1\linewidth]{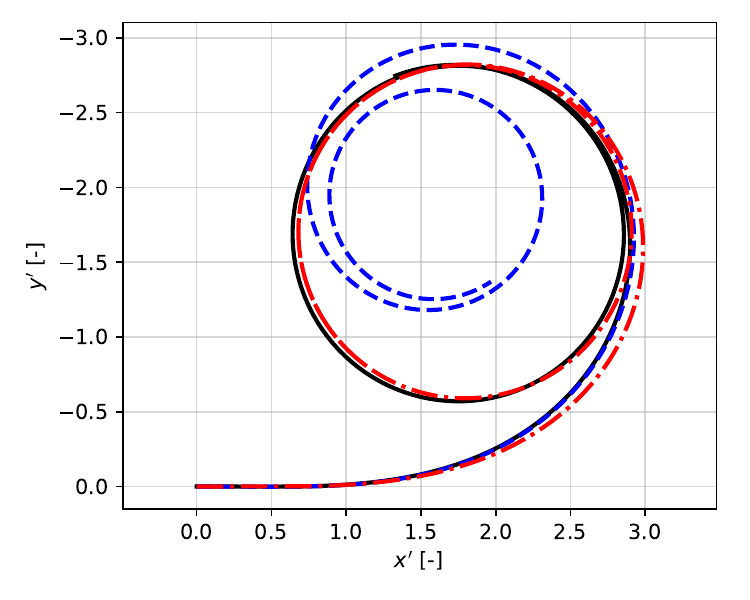}
        \caption{TC 35$^\circ$ PS manoeuvre (test manoeuvre)}
    \end{subfigure}
    \centering
    \includegraphics[width=0.55\linewidth]{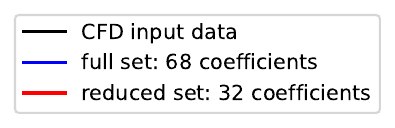}
    \caption{Predicted Trajectories of the input manoeuvre (a) and a test manoeuvre (b) in comparison with CFD simulation data for the full set of coefficients given in \cite{lewis_PNA_1988} and the reduced set, suggested for the KVLCC2 by \citet{Mucha2017}}
    \label{fig:trajets_diff_coeff_sets}
\end{figure}

\begin{figure*}[ht!]
    \centering
    \begin{tabular*}{\textwidth}[t]{cc}
        \includegraphics[width=0.49\linewidth, trim={10 10 10 10},clip]{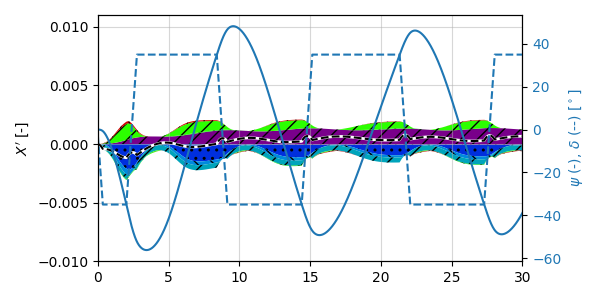} & \includegraphics[width=0.49\linewidth, trim={10 10 10 10},clip]{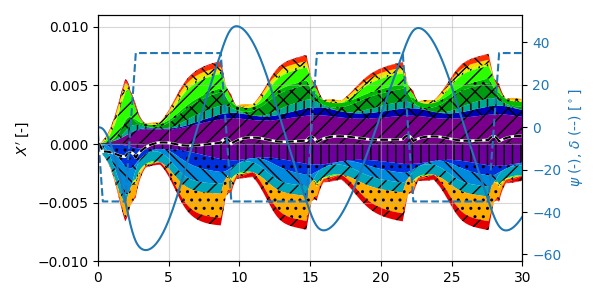} \\
        \raisebox{10pt}{\includegraphics[width=0.3\linewidth, trim={10 10 10 10},clip]{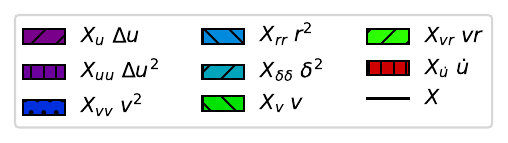}} & \includegraphics[width=0.4\linewidth, trim={10 10 10 10},clip]{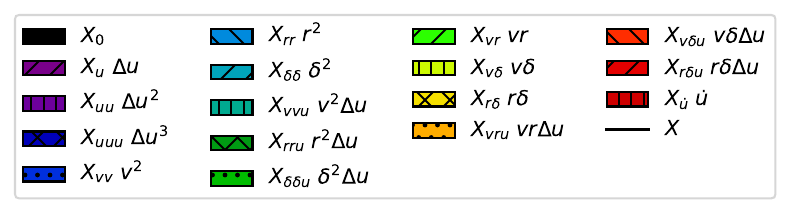} \\
        \includegraphics[width=0.49\linewidth, trim={10 10 10 10},clip]{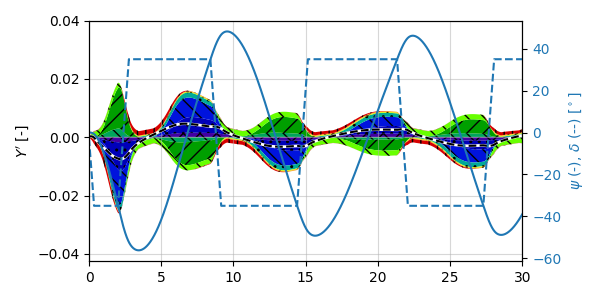} & \includegraphics[width=0.49\linewidth, trim={10 10 10 10},clip]{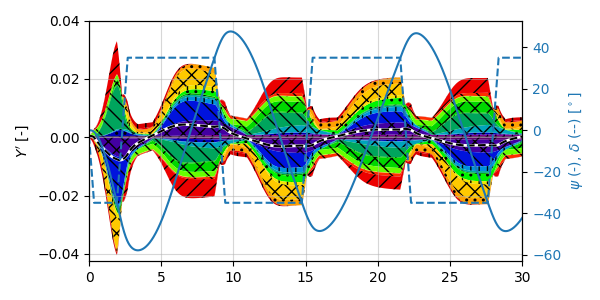} \\
        \raisebox{10pt}{\includegraphics[width=0.3\linewidth, trim={10 10 10 10},clip]{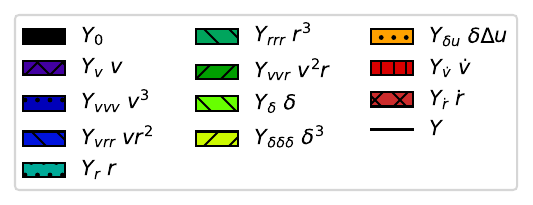}} & \includegraphics[width=0.4\linewidth, trim={10 10 10 10},clip]{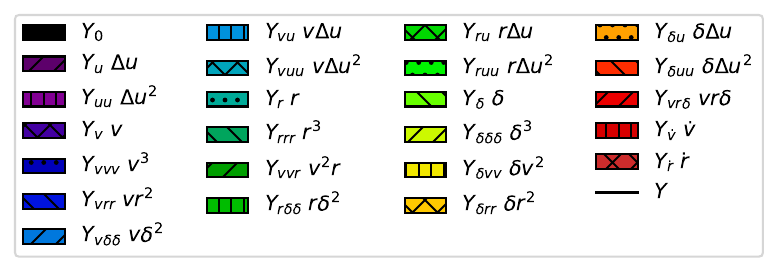} \\
        \includegraphics[width=0.49\linewidth, trim={10 10 10 10},clip]{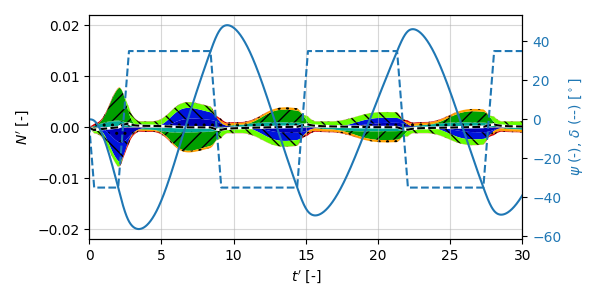} & \includegraphics[width=0.49\linewidth, trim={10 10 10 10},clip]{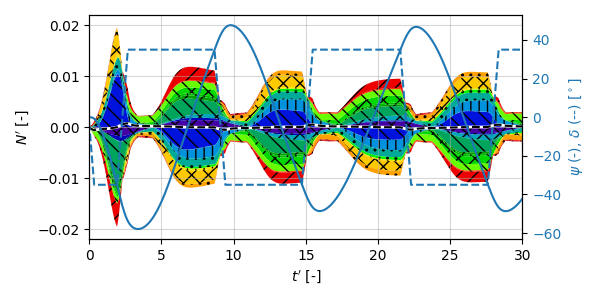}  \\
        \raisebox{10pt}{\includegraphics[width=0.3\linewidth, trim={10 10 10 10},clip]{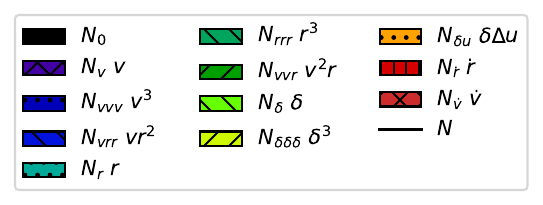}} & \includegraphics[width=0.4\linewidth, trim={10 10 10 10},clip]{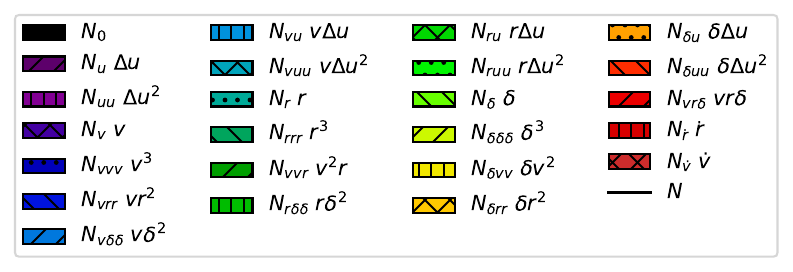} \\
        \begin{subfigure}[b]{0.45\textwidth}\caption{Reduced set of 32 coefficients} \end{subfigure} & \begin{subfigure}[b]{0.45\textwidth}\caption{Full set of 68 coefficients} \end{subfigure} \\
    \end{tabular*}
    \caption{Resulting non-dimensional hydrodynamic force components per hydrodynamic coefficient during the ZZ 35$^\circ$/35$^\circ$ PS input manoeuvre for surge, way and yaw direction (from top to bottom) for the reduced set (left) and the full set of hydrodynamic coefficients (right)}
    \label{fig:forceComps_zz3535ps_diff_coeff_sets}
\end{figure*}

\begin{figure*}[ht!]
    \centering
    \begin{tabular*}{\textwidth}[t]{cc}
        \includegraphics[width=0.49\linewidth, trim={10 10 10 10},clip]{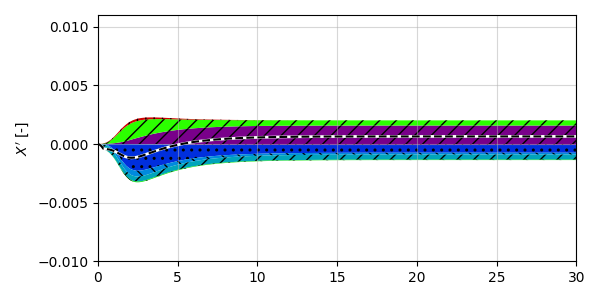} & \includegraphics[width=0.49\linewidth, trim={10 10 10 10},clip]{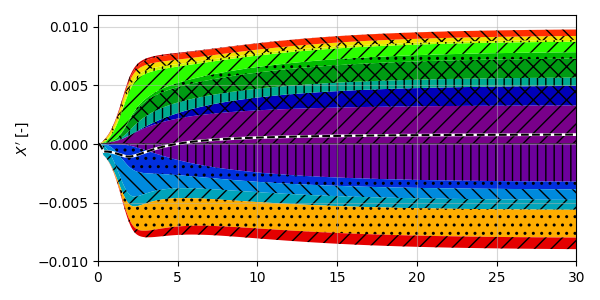} \\
        \raisebox{10pt}{\includegraphics[width=0.3\linewidth, trim={10 10 10 10},clip]{figs/results_coeffsets/99a_Mucha_legend_X.pdf}} & \includegraphics[width=0.4\linewidth, trim={10 10 10 10},clip]{figs/results_coeffsets/99a_PNA_legend_X.pdf} \\
        \includegraphics[width=0.49\linewidth, trim={10 10 10 10},clip]{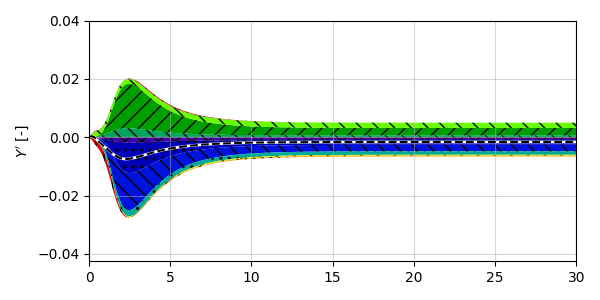} & \includegraphics[width=0.49\linewidth, trim={10 10 10 10},clip]{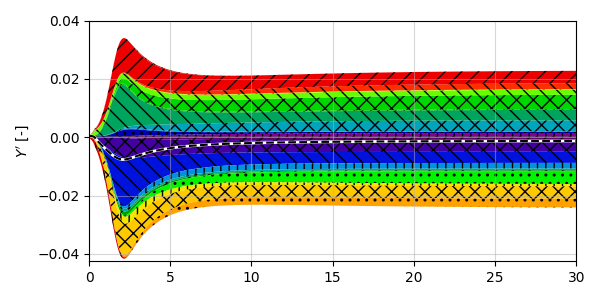} \\
        \raisebox{10pt}{\includegraphics[width=0.3\linewidth, trim={10 10 10 10},clip]{figs/results_coeffsets/99a_Mucha_legend_Y.pdf}} & \includegraphics[width=0.4\linewidth, trim={10 10 10 10},clip]{figs/results_coeffsets/99a_PNA_legend_Y.pdf} \\
        \includegraphics[width=0.49\linewidth, trim={10 10 10 10},clip]{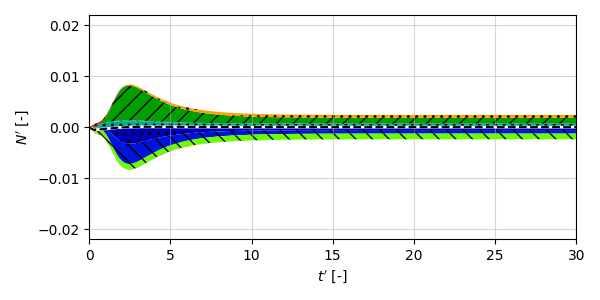} & \includegraphics[width=0.49\linewidth, trim={10 10 10 10},clip]{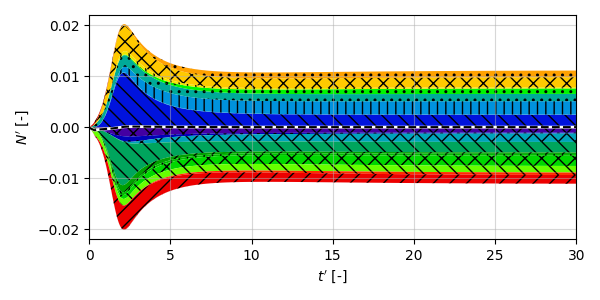} \\
        \raisebox{10pt}{\includegraphics[width=0.3\linewidth, trim={10 10 10 10},clip]{figs/results_coeffsets/99a_Mucha_legend_N.pdf}} & \includegraphics[width=0.4\linewidth, trim={10 10 10 10},clip]{figs/results_coeffsets/99a_PNA_legend_N.pdf} \\
        \begin{subfigure}[b]{0.45\textwidth}\caption{Reduced set of 32 coefficients} \end{subfigure} & \begin{subfigure}[b]{0.45\textwidth}\caption{Full set of 68 coefficients} \end{subfigure} \\
    \end{tabular*}
    \caption{Resulting non-dimensional force components of the hydrodynamic force per hydrodynamic coefficient during the TC 35$^\circ$ PS test manoeuvre, surge, sway and yaw direction (from top to bottom),  reduced set (left) and full set of hydrodynamic coefficients (right)}
    \label{fig:forceComps_tc35ps_diff_coeff_sets}
\end{figure*}

\autoref{fig:trajets_diff_coeff_sets} compares the predicted trajectories obtained using the larger set of 68 and the smaller set of 32 hydrodynamic coefficients. As expected, both models reproduce the input manoeuvre ZZ 35$^\circ$/35$^\circ$ PS very accurately. The larger set of coefficients achieves an almost perfect fit slightly better than the reduced set reflecting its greater complexity and flexibility. However, when simulating the TC 35$^\circ$ PS test manoeuvre, the larger coefficient set exhibits unacceptable deviations from the reference data, whereas the smaller set predicts the test manoeuvre with much higher fidelity.

The generally superior performance of the smaller coefficient set can be explained by examining the hydrodynamic force components shown in \autoref{fig:forceComps_zz3535ps_diff_coeff_sets} and \autoref{fig:forceComps_tc35ps_diff_coeff_sets} as a function of time, corresponding to the input and test manoeuvres, respectively. Although the total hydrodynamic forces (depicted as black dashed lines in \autoref{fig:forceComps_zz3535ps_diff_coeff_sets}) for the input manoeuvre are nearly identical for both models since both were calibrated to fit this data the individual force components in the larger coefficient set are unrealistically large. This behaviour stems from multicollinearity, where two or more combinations of manoeuvring variables counterbalance each other, allowing certain hydrodynamic coefficients to assume excessively large values without noticeably affecting the overall sum of force components. Consequently, while the input manoeuvre is matched almost perfectly despite these inflated components (cf. \autoref{fig:trajets_diff_coeff_sets} (a)), a different manoeuvre, such as the turning circle, where the variable combinations do not cancel out similarly, is predicted poorly by the larger set (cf. \autoref{fig:trajets_diff_coeff_sets} (b), dashed blue line).

\begin{figure}[ht!]
    \centering
    \includegraphics[width=0.9\linewidth, trim={0 0 0 0},clip]{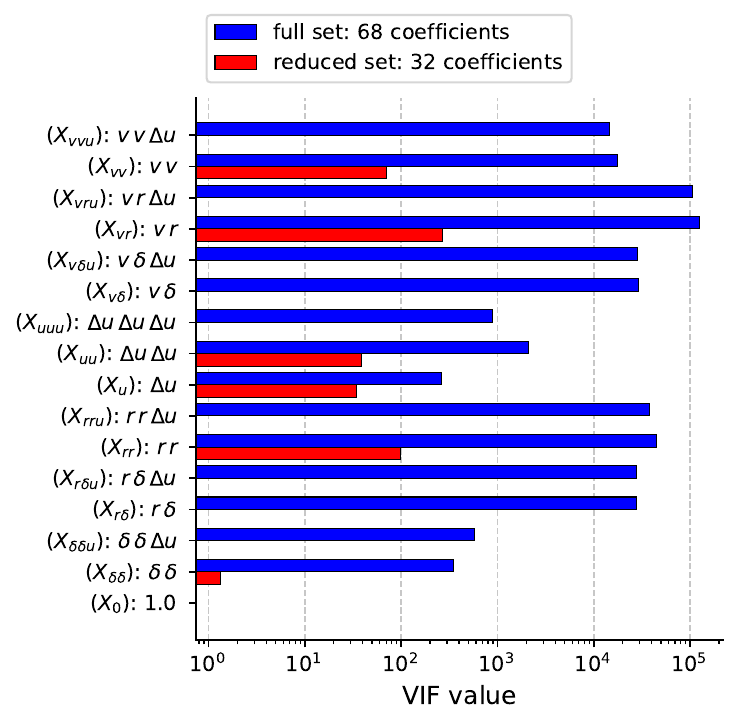}
    \caption{VIF values for surge coefficients ($X$) for CFD simulation data for the full set of coefficients given in \cite{lewis_PNA_1988} and the reduced set, suggested by \citet{Mucha2017}}
    \label{fig:vif_values_diff_coeff_sets_X}
\end{figure}

\begin{figure}[ht!]
    \centering 
    \includegraphics[width=0.9\linewidth, trim={0 0 0 0},clip]{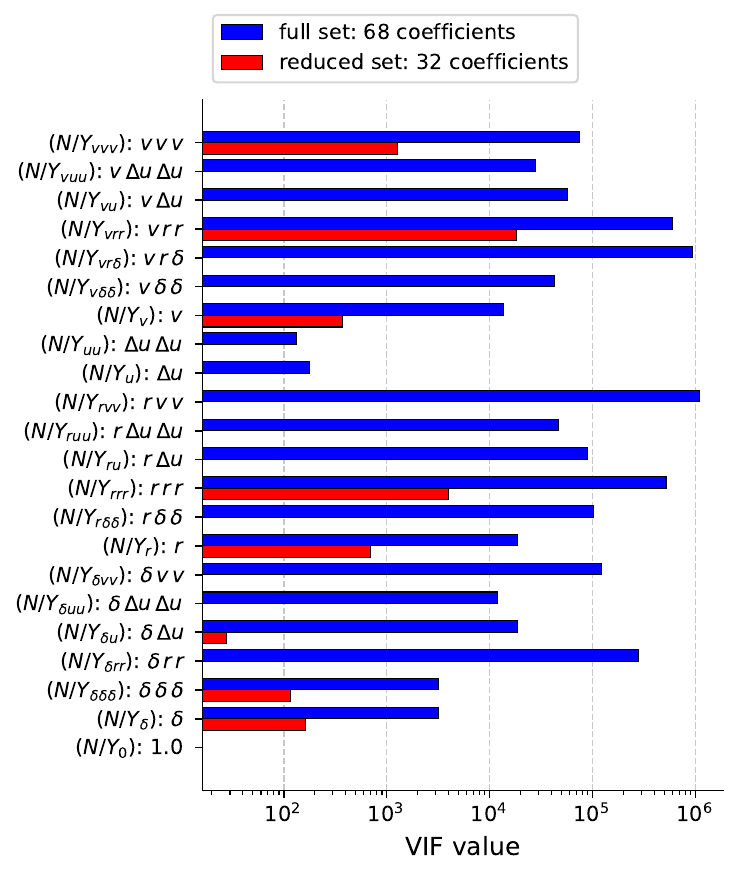}
    \caption{As \autoref{fig:vif_values_diff_coeff_sets_X} but for sway and yaw coefficients ($Y$/$N$)}
    \label{fig:vif_values_diff_coeff_sets_YN}
\end{figure}

In contrast, the smaller coefficient set, identified via local sensitivity analysis, mitigates the issue of multicollinearity and resultant cancellation errors. Therefore, when employing a straightforward least squares fitting approach, this reduced set of coefficients is more robust and better suited for predicting manoeuvres beyond the calibration data from one input manoeuvre.

Additionally, it should be noted that the turning circle predicted using the full set of coefficients does not reach a steady turning trajectory, i.e. the instantaneous turning radius is still changing even after two consecutive turns. This is also reflected in the force component time series in \autoref{fig:forceComps_tc35ps_diff_coeff_sets} (b), where the surge components (top right panel) are not converged to a constant value within the a simulation time of $t^\prime < 25$. Time series of forward speed that will be presented in section \ref{subsec:model_performance} consequently report continuous speed loss in contrast to the CFD results and theoretical considerations.

A statistical analysis via VIF value computation of the input data according to section \ref{subsec:vif} provides additional insight on the cause of the deviations. A VIF analysis is performed for both sets of coefficients to assess the the multicollinearity included in the data. VIF values for the different combinations of motion variables are reported in \autoref{fig:vif_values_diff_coeff_sets_X} and \autoref{fig:vif_values_diff_coeff_sets_YN} on a logarithmic scale for the different degrees of freedom. Values for sway and yaw direction are presented in a single figure because since they are the same due to the same motion variables included in the sets of coefficients. For the reduced set of coefficients values are set to zero if the corresponding coefficients are not included in the set. The resulting VIF values are several orders of magnitude smaller for motion variables used by the smaller coefficient set, making the SI significantly less susceptible to cancellation errors. 

\subsection{Influence of Training‑Set Size on Model Fidelity}
\label{subsec:data-requirements}
The amount of data required for a stable and accurate manoeuvring model was evaluated through numerical investigations based on the ZZ $35^{\circ}/35^{\circ}$ PS manoeuvre. Hydrodynamic coefficients from the reduced set presented in section~\ref{subsec:set_of_coeffs} are predicted using the LSQ model. Starting from the full data record (eight overshoots, OS), the input time series was progressively truncated down to a minimum of time series until 1\,OS. Using time series shorter than 1\,OS is not feasible, since the associated hydrodynamic coefficients generated numerical overflow when re-integrated in the manoeuvring simulator even when trying to reproduce the input data.

\begin{figure}[ht!]
    \centering
    \begin{subfigure}[b]{0.49\textwidth}
        \centering
        \includegraphics[width=0.98\linewidth, trim={0 20 0 0},clip]{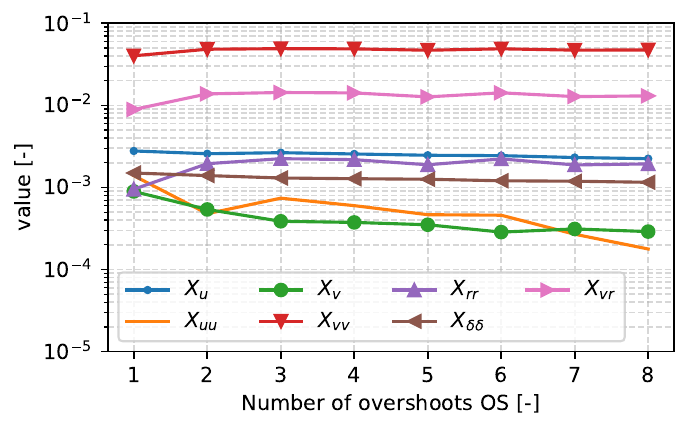}
        \caption{surge direction}
    \end{subfigure}

    \begin{subfigure}[b]{0.49\textwidth}
        \centering
        \includegraphics[width=.98\linewidth, trim={0 20 0 0},clip]{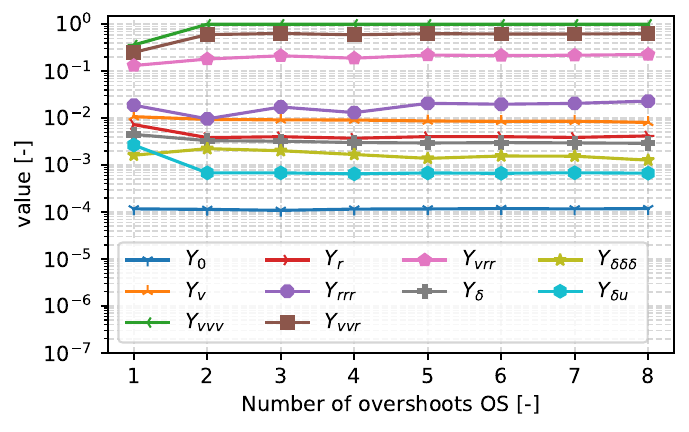}
        \caption{yaw direction}
    \end{subfigure}

    \begin{subfigure}[b]{0.49\textwidth}
        \centering
        \includegraphics[width=.98\linewidth, trim={0 0 0 0},clip]{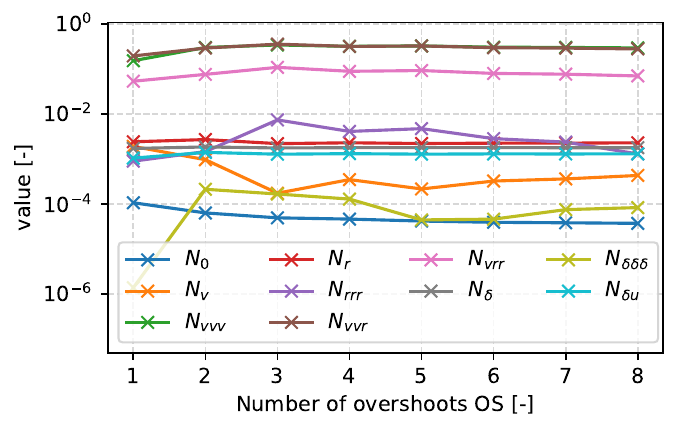}
        \caption{sway direction}
    \end{subfigure}
    
    \caption{Hydrodynamic coefficients predicted from the ZZ\,35$^\circ$/35$^\circ$ PS manoeuvre as a function of the time series length used (until 1 to 8 overshoots)}
    \label{fig:coeffs_diff_dataset_length}
\end{figure}

\autoref{fig:coeffs_diff_dataset_length} presents the identified coefficients for surge ($X$), sway ($Y$), and yaw ($N$) as a function of input length. A logarithmic scale highlights variations over several orders of magnitude for 1-3\,OS in many coefficients. For $\geq 4$\,OS, most coefficients converge. Terms that remain non-converged for $\geq 4$\,OS (e.g.\ $X_{uu}$, $Y_{\delta\delta\delta}$, $N_{rrr}$, $N_{\delta\delta\delta}$) have negligible contribution during the input manoeuvre due to either small coefficient magnitude or weak excitation; this is consistent with their barely noticeable component contributions in \autoref{fig:forceComps_zz3535ps_diff_coeff_sets} (left). Consequently, the prediction of these coefficients is less accurate, due to their relatively small excitation in the input data, which is not of concern as long as other predicted manoeuvre do not depend on these coefficients in a greater manner than the input manoeuvre.

\begin{figure}[ht!]
    \centering
    \begin{subfigure}[b]{0.49\textwidth}
        \centering
        \includegraphics[width=1\linewidth]{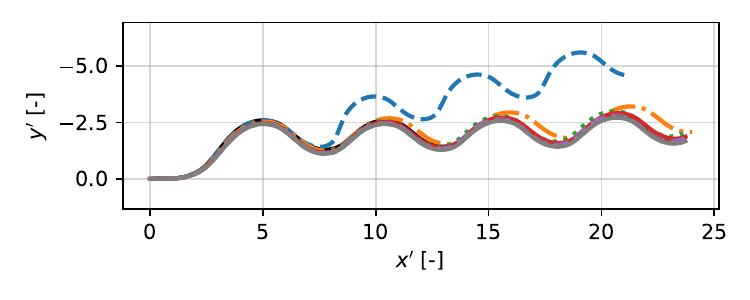}
        \caption{ZZ 35º/35º PS manoeuvre (input manoeuvre)}
    \end{subfigure}

    \begin{subfigure}[b]{0.49\textwidth}
        \centering
        \includegraphics[width=1\linewidth]{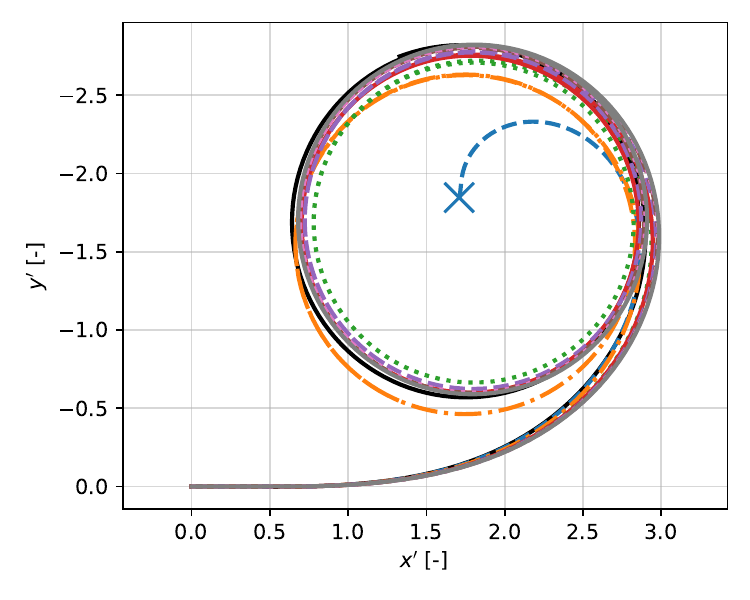}
        \caption{TC 35$^\circ$ PS manoeuvre (test manoeuvre)}
    \end{subfigure}

    \includegraphics[width=0.8\linewidth]{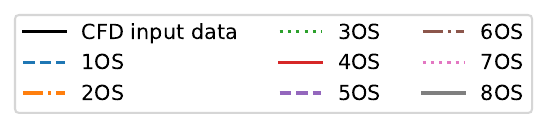}
    \caption{Predicted Trajectories of the input manoeuvre (a) and a test manoeuvre (b) in comparison with CFD simulation data for coefficients predicted with time series ending with 1 to 8 overshoots}
    \label{fig:trajets_diff_dataset_length}
\end{figure}

Using the predicted coefficients, both the input manoeuvre ZZ~$35^{\circ}/35^{\circ}$ PS and the test manoeuvre TC~$35^{\circ}$ PS are simulated. The resulting trajectories are shown in \autoref{fig:trajets_diff_dataset_length}. Numerical errors of the validation variables relative to the CFD data are reported in \autoref{tab:zz3535PS_os} and \ref{tab:TC35PS_os}.

\begin{table*}[htbp]
\centering
\caption{ZZ 35$^\circ$/35$^\circ$ (PS): IMO metrics and relative errors versus CFD for the reconstructed training manoeuvre, using models trained on a single manoeuvre truncated to 1--8 overshoots (OS).}
\label{tab:zz3535PS_os}
\footnotesize
\renewcommand{\arraystretch}{1.15}
\begin{tabular*}{\textwidth}{@{\extracolsep{\fill}} l
cc cc cc cc c}
\hline\hline
 & \multicolumn{2}{c}{Overshoot 1 [$^\circ$]} &
   \multicolumn{2}{c}{Overshoot 2 [$^\circ$]} &
   \multicolumn{2}{c}{$t_{\mathrm{exec2}}$ L/V [s]} &
   \multicolumn{2}{c}{$t_{\mathrm{check}}$ L/V [s]} &
   \multicolumn{1}{c}{Avg. MRAE [\%]} \\
\cline{2-9}
Caso & Pred. & Err. & Pred. & Err. & Pred. & Err. & Pred. & Err. &  \\
\hline
CFD (ref.) & \multicolumn{2}{c}{22.80} & \multicolumn{2}{c}{12.80} &
            \multicolumn{2}{c}{8.68}  & \multicolumn{2}{c}{1.10} & \\
\hline
1 OS & 22.80 & 0.0\% & n/a & n/a& n/a  & n/a& n/a& n/a& n/a\\
2 OS & 22.90 & 0.4\% & 12.70 & 0.5\% & 8.70 & 0.0\% & 1.10 & 0.0\% & 0.2 \\
3 OS & 22.63 & 0.7\% & 12.62 & 1.4\% & 8.66 & 0.2\% & 1.12 & 1.8\% & 1.0 \\
4 OS & 22.63 & 0.7\% & 12.51 & 2.3\% & 8.66 & 0.2\% & 1.11 & 0.9\% & 1.0 \\
5 OS & 22.66 & 0.6\% & 12.49 & 2.4\% & 8.66 & 0.2\% & 1.11 & 0.9\% & 1.0 \\
6 OS & 22.66 & 0.6\% & 12.56 & 1.9\% & 8.65 & 0.3\% & 1.12 & 1.8\% & 1.2 \\
7 OS & 22.69 & 0.5\% & 12.52 & 2.2\% & 8.66 & 0.2\% & 1.11 & 0.9\% & 1.0 \\
8 OS & 22.70 & 0.4\% & 12.49 & 2.4\% & 8.65 & 0.3\% & 1.11 & 0.9\% & 1.0 \\
\hline\hline
\end{tabular*}
\end{table*}

\begin{table*}[htbp]
\centering
\caption{Turning circle 35$^\circ$ (PS) predicted test manoeuvre: Turning-circle metrics (AD$_{90}$, TR$_{90}$, TD$_{180}$, SD) and relative errors versus CFD for models trained on ZZ training truncations of 1--8 overshoots (OS).}
\label{tab:TC35PS_os}
\footnotesize
\renewcommand{\arraystretch}{1.12}
\begin{tabular*}{\textwidth}{@{\extracolsep{\fill}} l cc cc cc cc c}
\hline\hline
 & \multicolumn{2}{c}{Advance 90 [-]} &
   \multicolumn{2}{c}{Transfer 90 [-]} &
   \multicolumn{2}{c}{Tactical Diameter 180 [-]} &
   \multicolumn{2}{c}{Steady Circle Diameter [-]} &
   \multicolumn{1}{c}{Avg. MRAE [\%]} \\
\cline{2-9}
Case & Pred. & Err. & Pred. & Err. & Pred. & Err. & Pred. & Err. &  \\
\hline
CFD (ref.) & \multicolumn{2}{c}{2.80} & \multicolumn{2}{c}{-1.11} &
             \multicolumn{2}{c}{2.73} & \multicolumn{2}{c}{2.09} & \\
\hline
1 OS & 3.80 & 35.7\% & -4.30 & 290.9\% & 10.40 & 285.2\% & 11.20 & 433.3\% & 261.3 \\
2 OS & 2.80 & 0.0\% & -1.20 & 9.1\% & 2.90 & 7.4\% & 2.30 & 9.5\% & 6.5 \\
3 OS & 2.84 & 1.4\% & -1.16 & 4.5\% & 2.90 & 6.2\% & 1.97 & 5.7\% & 1.6 \\
4 OS & 2.84 & 1.4\% & -1.17 & 5.4\% & 2.91 & 6.6\% & 2.08 & 1.0\% & 3.1 \\
5 OS & 2.84 & 1.4\% & -1.17 & 5.4\% & 2.91 & 6.6\% & 2.02 & 3.3\% & 2.5 \\
6 OS & 2.84 & 1.4\% & -1.17 & 5.4\% & 2.92 & 7.0\% & 2.00 & 4.3\% & 2.4 \\
7 OS & 2.84 & 1.4\% & -1.18 & 6.3\% & 2.92 & 7.0\% & 1.98 & 5.3\% & 2.4 \\
8 OS & 2.84 & 1.4\% & -1.18 & 6.3\% & 2.94 & 7.7\% & 1.95 & 6.7\% & 2.2 \\
\hline\hline
\end{tabular*}
\end{table*}

Taken together, \autoref{tab:zz3535PS_os} and \autoref{tab:TC35PS_os} show that the required input length depends on whether the case is in-sample ZZ~$35^{\circ}/35^{\circ}$ PS or out-of-sample ZZ~$35^{\circ}/35^{\circ}$ SB and TC~$35^{\circ}$. For the training manoeuvre ZZ~$35^{\circ}/35^{\circ}$ PS, the 1\,OS record does not provide a complete set of IMO scalar metrics (Overshoot~2, $t_{\mathrm{exec2}}$, and $t_{\mathrm{check}}$ remain undefined in \autoref{tab:zz3535PS_os}). From 2\,OS onwards, the average error remains low, reaching $0.2\%$ at 2\,OS and stabilising around $1.0\%$--$1.2\%$ for 3--8\,OS. Metric-wise (3--8\,OS), the largest deviations occur in Overshoot~2 (up to $2.4\%$) and $t_{\mathrm{check}}$ (up to $1.8\%$), while $t_{\mathrm{exec2}}$ remains within $0.2\%$--$0.3\%$.

Turning-circle generalisation remains more demanding. For TC~$35^{\circ}$ PS, the MRAE is computed as the arithmetic mean of the four geometric metric errors reported in \autoref{tab:TC35PS_os} (Advance, Transfer, Tactical Diameter, and Steady Diameter). Under this definition, the average error drops sharply from $261.3\%$ at 1\,OS to $6.5\%$ at 2\,OS and reaches its best performance at 4\,OS ($3.6\%$). Beyond 4\,OS, changes remain modest but the mean error increases to $4.2\%$--$5.5\%$ for 5--8\,OS, driven primarily by increasing errors in Transfer and Tactical Diameter. The 1\,OS case is dominated by very large deviations in Transfer ($290.9\%$), Tactical Diameter ($285.2\%$), and Steady Diameter ($433.3\%$), which indicates poor out-of-sample robustness under severe truncation.

Accordingly, 4\,OS is adopted as the baseline input length because it is the shortest setting that (i) avoids the incomplete IMO scalar metrics observed at 1\,OS for ZZ~$35^{\circ}/35^{\circ}$ PS, (ii) achieves the best average performance for TC~$35^{\circ}$ PS ($3.6\%$, \autoref{tab:TC35PS_os}) under the consistent averaging definition stated above, and (iii) keeps in-sample ZZ~$35^{\circ}/35^{\circ}$ PS errors low (average $1.0\%$, \autoref{tab:zz3535PS_os}). This choice also mitigates the instability observed under extreme truncation and contains CFD runtime, providing a balanced trade-off between fidelity, robustness, and computational cost for the subsequent analyses.

\subsection{Prediction Model Performance}
\label{subsec:model_performance}

This section evaluates the predictive performance of the identification models using simulated trajectories and manoeuvring metrics in non-dimensional form. A single ZZ~$35^{\circ}/35^{\circ}$ (PS) manoeuvre is used for training, and four estimators are compared under identical pre-processing and the same design matrix $\mathbf{A}$ with right-hand side $\mathbf{b}$ as defined in section~2.3: constrained LSQ, Lasso, Ridge, and SVR. The identified non-dimensional hydrodynamic coefficients for surge $X$, sway $Y$, and yaw $N$ are summarised in the bar plots of Figs.~18, 20, and 22, respectively. The vertical axis is logarithmic, and light pink shading highlights coefficients for which at least one model assigns an opposite sign or zero magnitude. Complementarily, Figs.~19, 21, and 23 decompose the total hydrodynamic force/moment into per-coefficient contributions during the ZZ~$35^{\circ}/35^{\circ}$ (PS) test, linking coefficient values to their dynamic effect.

\makeatletter
\@ifundefined{coefimgheight}{\newlength{\coefimgheight}}{}
\makeatother
\setlength{\coefimgheight}{0.25\textheight} 

\begin{figure*}[!ht] 
  \centering
  \begin{minipage}{0.88\textwidth}
    \centering
    \includegraphics[width=\textwidth,height=\coefimgheight,keepaspectratio]{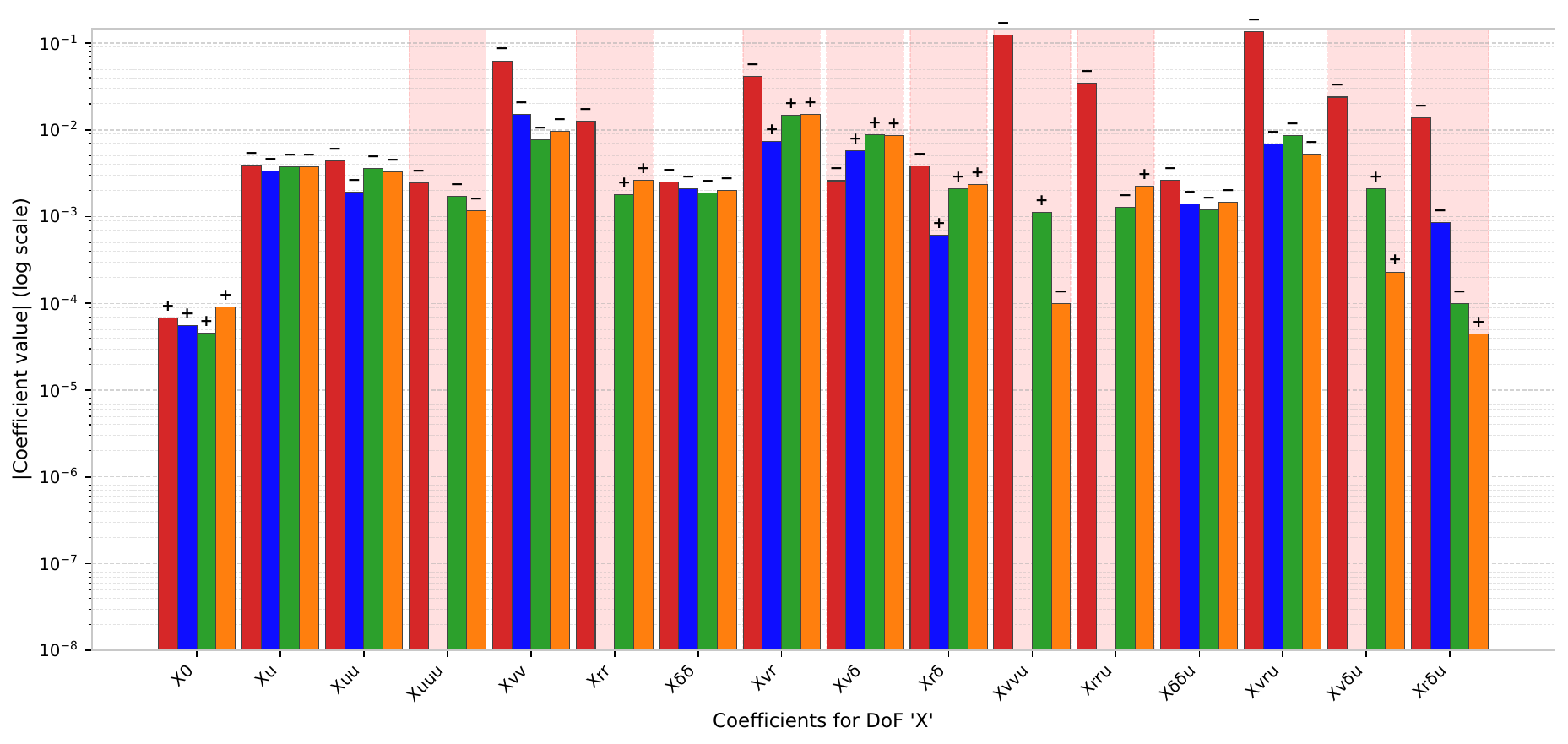}
    \includegraphics[width=\linewidth]{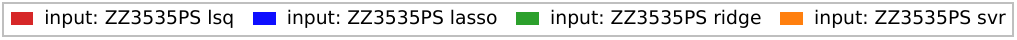}
  \caption{Non-dimensional surge-force coefficients (\(X\)) for different predictor models (LSQ, Lasso, Ridge, SVR).\newline  Pink shading highlights coefficients with inconsistent signs or zero magnitude. Vertical axis is logarithmic.}
 \label{fig:coeffs_X_full}
  \end{minipage}\hfill
  \begin{minipage}{0.9\textwidth}
    \centering
    \begin{tabular*}{\textwidth}[t]{cc}
        \includegraphics[width=0.5\linewidth, trim={10 10 10 10},clip]{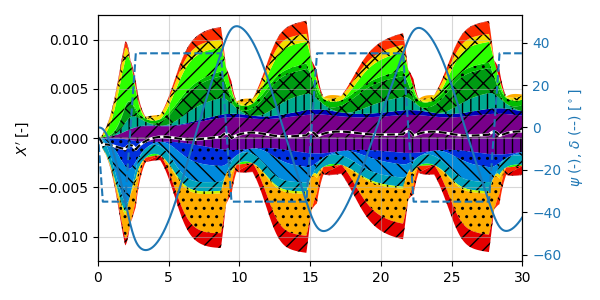} & \includegraphics[width=0.5\linewidth, trim={10 10 10 10},clip]{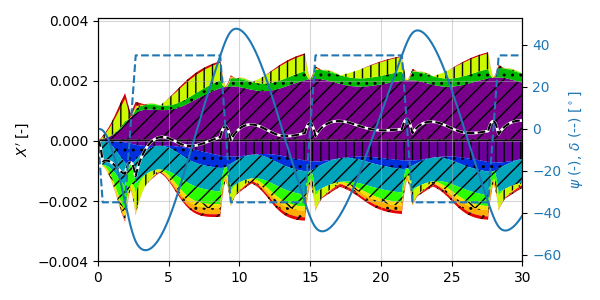} \\
        \begin{subfigure}[b]{0.45\textwidth}\caption{LSQ (larger vertical axis limits!)} \end{subfigure} & \begin{subfigure}[b]{0.45\textwidth}\caption{Lasso} \end{subfigure} \\
        \includegraphics[width=0.5\linewidth, trim={10 10 10 10},clip]{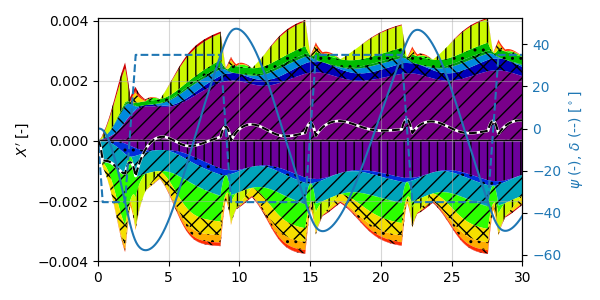
        } & \includegraphics[width=0.5\linewidth, trim={10 10 10 10},clip]{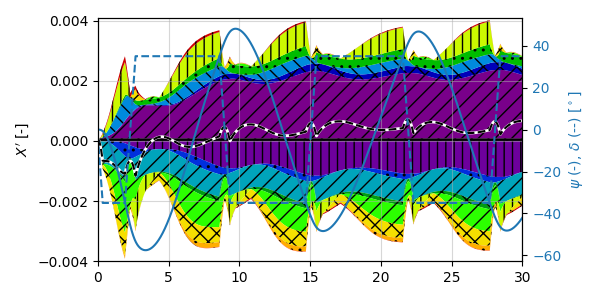} \\
        \begin{subfigure}[b]{0.45\textwidth}\caption{Ridge} \end{subfigure} & \begin{subfigure}[b]{0.45\textwidth}\caption{SVR} \end{subfigure} \\
    \end{tabular*}
    \raisebox{10pt}{\includegraphics[width=0.98\linewidth, trim={0 0 0 0},clip]{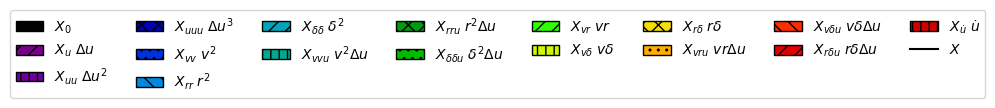}}
    \nextfloat
    \caption{Resulting non-dimensional force components of total the hydrodynamic force per hydrodynamic coefficient \newline  during the ZZ 35°/35° PS test manoeuvre for surge direction from different prediction methods}
    \label{fig:forceComps_X_zz3535ps_pred_methods}
  \end{minipage}
  
\end{figure*}

\subsubsection{Surge coefficients}

\autoref{fig:coeffs_X_full} summarizes the estimated surge force coefficients across the different regression methods. Constant and linear terms $(X_0, X_u)$ show minor variations and sign consistency. Quadratic and cubic surge dependencies ($X_{uu}$, $X_{uuu}$) from regularized models are reduced by up to ca. 60\%. Using Lasso regression, $X_{uuu}$ is driven to zero, attributing pure surge-dependent forces solely to linear and quadratic terms.

Pure sway and yaw dependencies ($X_{vv}$, $X_{rr}$) are predicted about one order of magnitude smaller from regularized methods indicating cancellation errors in LSQ predictions. 
Reductions of the rudder angle dependency ($X_{\delta \delta}$) from regularized methods are only 15\% to 25\% smaller compared to LSQ.

Predictions of higher-order coupling coefficients ($X_{vr}$, $X_{r\delta}$, $X_{vvu}$, $X_{rru}$ $\dots$) are generally significantly smaller from regularized methods compared to LSQ, typically more than one order of magnitude. One exception is the sway-rudder angle coupling term $X_{v \delta}$, for which LSQ resulted in the smallest magnitude. Otherwise, the larger coefficients resulted in unrealistically large force component contributions for LSQ in \autoref{fig:forceComps_X_zz3535ps_pred_methods} caused most probably by multicollinearity.

\begin{figure*}[!ht] 
  \centering
  \begin{minipage}{0.8\textwidth}
    \centering
    \includegraphics[width=\textwidth,height=\coefimgheight,keepaspectratio]{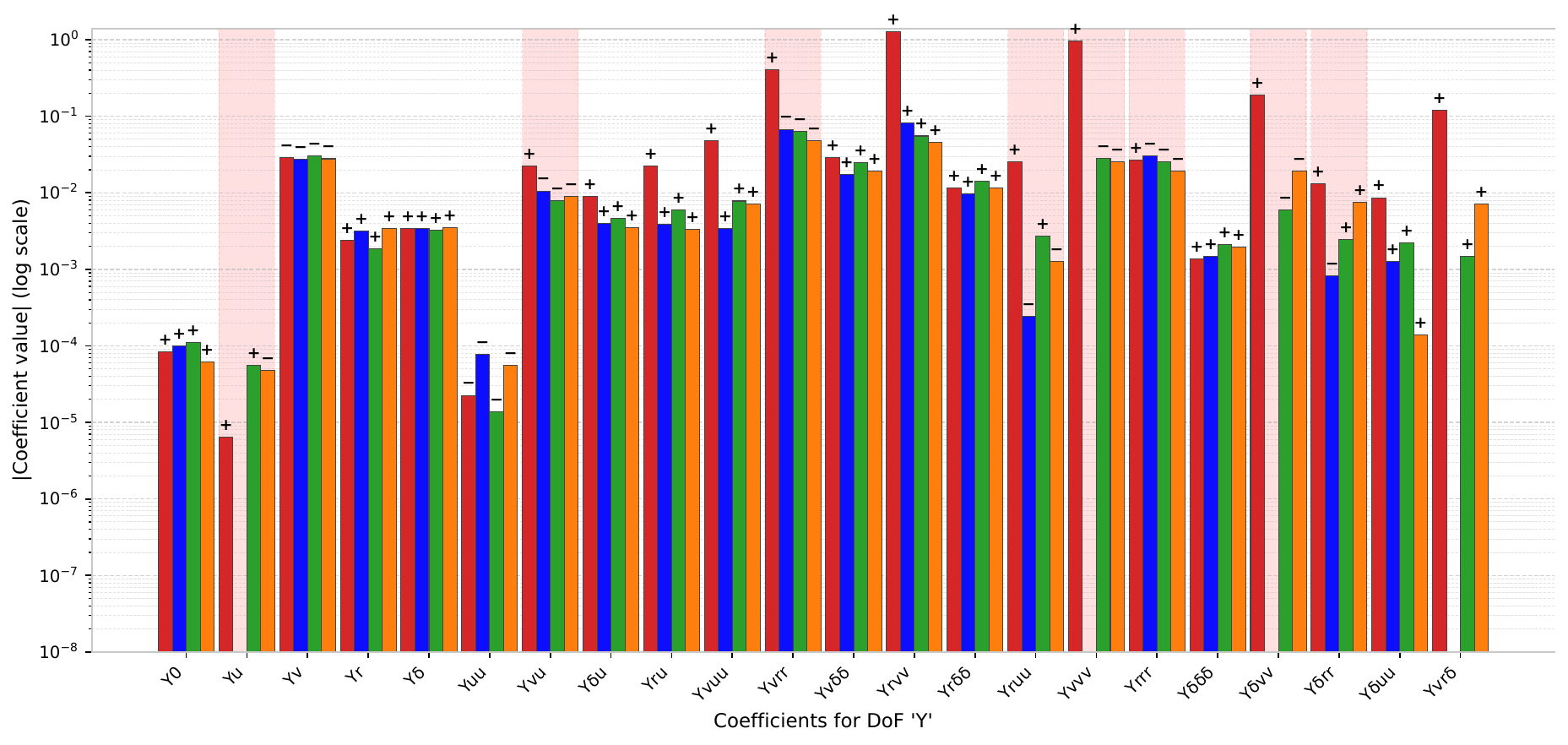}
    \includegraphics[width=\linewidth]{figs/results_modelperformance/legend_strip_coeffs_models.pdf}
 \caption{Non-dimensional sway-force coefficients (\(Y\)) for different predictor models (LSQ, Lasso, Ridge, SVR). \newline Pink shading highlights coefficients with inconsistent signs or zero magnitude. Vertical axis is logarithmic.}
 \label{fig:coeffs_Y_full}
  \end{minipage}\hfill
  \begin{minipage}{0.9\textwidth}
    \centering
    \begin{tabular*}{\textwidth}[t]{cc}
        \includegraphics[width=0.5\linewidth, trim={10 10 10 10},clip]{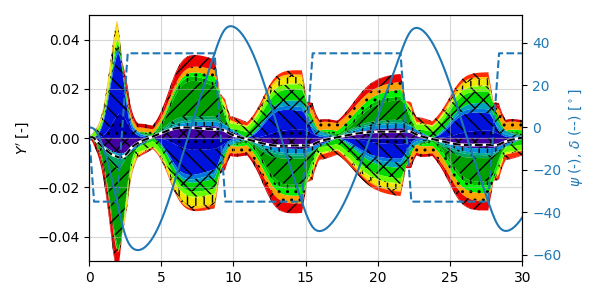} & \includegraphics[width=0.5\linewidth, trim={10 10 10 10},clip]{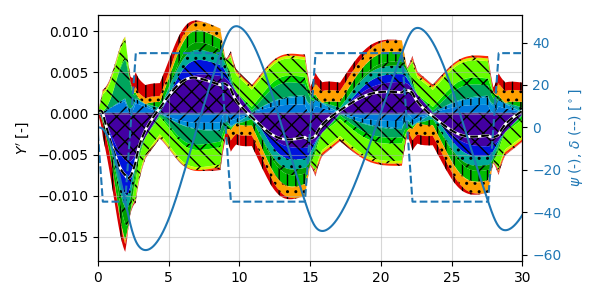} \\
        \begin{subfigure}[b]{0.45\textwidth}\caption{LSQ (larger vertical axis limits!)} \end{subfigure} & \begin{subfigure}[b]{0.45\textwidth}\caption{Lasso} \end{subfigure} \\
        \includegraphics[width=0.5\linewidth, trim={10 10 10 10},clip]{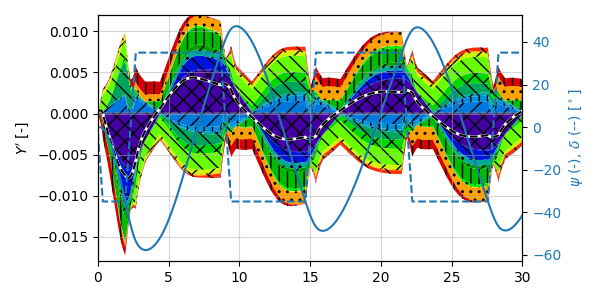
        } & \includegraphics[width=0.5\linewidth, trim={10 10 10 10},clip]{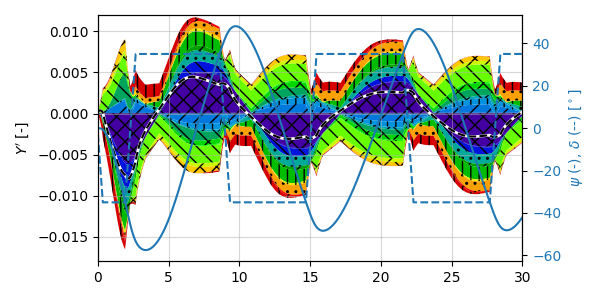} \\
        \begin{subfigure}[b]{0.45\textwidth}\caption{Ridge} \end{subfigure} & \begin{subfigure}[b]{0.45\textwidth}\caption{SVR} \end{subfigure} \\
    \end{tabular*}
    \raisebox{10pt}{\includegraphics[width=0.98\linewidth, trim={0 0 0 0},clip]{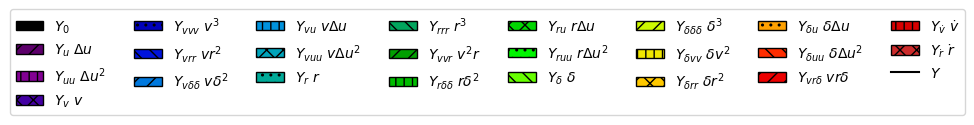}}
    \nextfloat
    \caption{Resulting non-dimensional force components of total the hydrodynamic force per hydrodynamic coefficient \newline during the ZZ 35°/35° PS test manoeuvre for surge direction from different prediction methods}
    \label{fig:forceComps_Y_zz3535ps_pred_methods}
  \end{minipage}
\end{figure*}


\subsubsection{Sway coefficients}

Estimated sway force coefficients are presented in \autoref{fig:coeffs_Y_full}. For the constant sway force $Y_0$ arising from asymmetry due to the propeller rotation direction of this single-screw vessel, differences across predictors remain limited ($<30\%$), and its contribution to the total sway force is relatively small compared to the damping and steering terms. Pure surge-dependent terms ($Y_u$, $Y_{uu}$) are negligible and do not notably influence the manoeuvre, owing to the hull's symmetry in the $x$-$z$ plane.



\begin{figure*}[!ht] 
  \centering
  \begin{minipage}{0.88\textwidth}
    \centering
    \includegraphics[width=\textwidth,height=\coefimgheight,keepaspectratio]{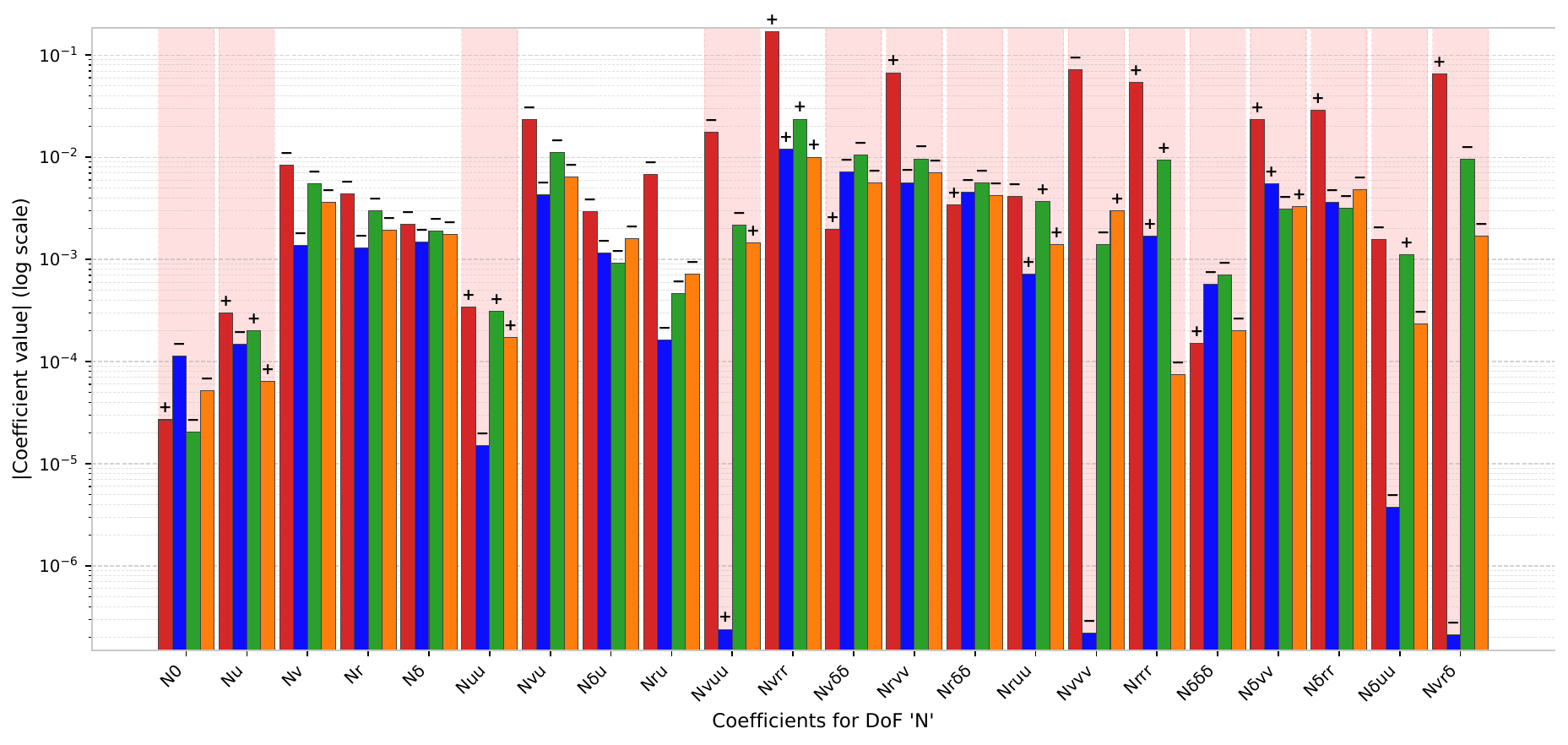}
    \includegraphics[width=\linewidth]{figs/results_modelperformance/legend_strip_coeffs_models.pdf}
 \caption{Non-dimensional yaw-moment coefficients (\(N\)) for different predictor models (LSQ, Lasso, Ridge, SVR). \newline  Pink shading highlights coefficients with inconsistent signs or zero magnitude. Vertical axis is logarithmic.}
 \label{fig:coeffs_N_full}
  \end{minipage}\hfill
  \begin{minipage}{0.9\textwidth}
    \centering
    \begin{tabular*}{\textwidth}[t]{cc}
        \includegraphics[width=0.5\linewidth, trim={0 10 0 10},clip]{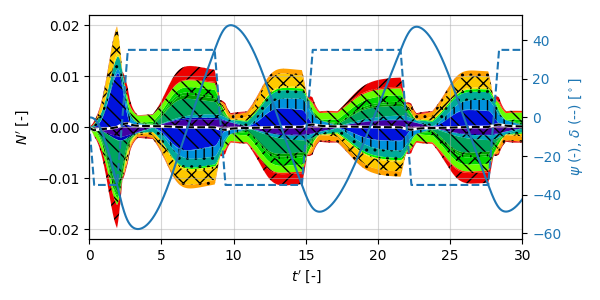} & \includegraphics[width=0.5\linewidth, trim={0 10 0 10},clip]{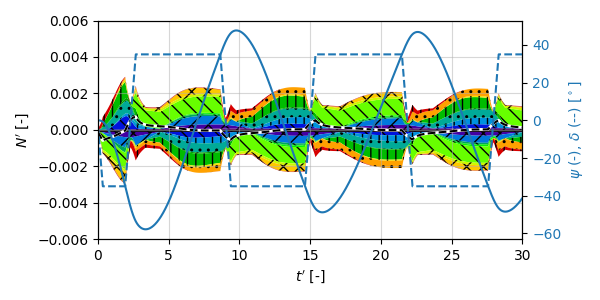} \\
        \begin{subfigure}[b]{0.45\textwidth}\caption{LSQ (larger vertical axis limits!)} \end{subfigure} & \begin{subfigure}[b]{0.45\textwidth}\caption{Lasso} \end{subfigure} \\
        \includegraphics[width=0.5\linewidth, trim={0 10 0 10},clip]{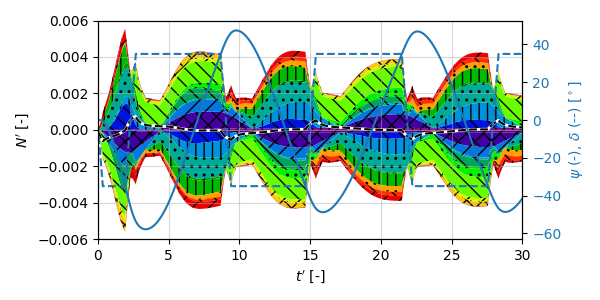
        } & \includegraphics[width=0.5\linewidth, trim={0 10 0 10},clip]{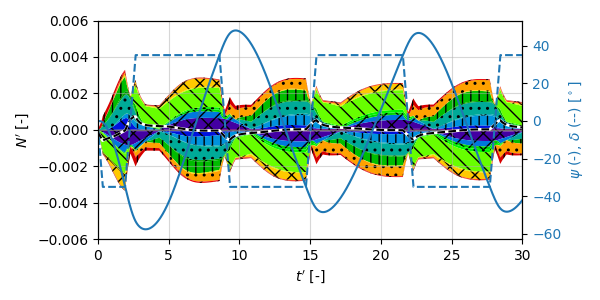} \\
        \begin{subfigure}[b]{0.45\textwidth}\caption{Ridge} \end{subfigure} & \begin{subfigure}[b]{0.45\textwidth}\caption{SVR} \end{subfigure} \\
    \end{tabular*}
    \raisebox{10pt}{\includegraphics[width=0.98\linewidth, trim={0 0 0 0},clip]{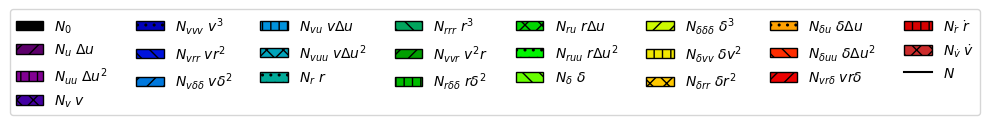}}
    \nextfloat
    \caption{Resulting non-dimensional force components of total the hydrodynamic force per hydrodynamic coefficient \newline during the ZZ 35°/35° PS test manoeuvre for yaw direction from different prediction methods}
    \label{fig:forceComps_N_zz3535ps_pred_methods}
  \end{minipage}
\end{figure*}

The linear sway-dependency coefficient $Y_v$ shows limited variance across predictors, with maximum differences of $<5\%$, and constitutes the most significant component of the sway force (see \autoref{fig:coeffs_Y_full}). In contrast, the cubic sway-dependency term $Y_{vvv}$ is substantially larger for LSQ than for regularized methods, leading to a prominent sway force contribution $Y_{vvv} v^3$ in \autoref{fig:forceComps_Y_zz3535ps_pred_methods}(a); regularized methods, however, attribute negligible sway force to this cubic term.

Pure yaw-dependent terms ($Y_r$, $Y_{rrr}$) exhibit similar values across all predictors, with differences up to $45\%$ for $Y_r$. The coefficient $Y_{rrr}$ even reverses sign relative to LSQ predictions, though absolute values show comparable differences to $Y_r$. The resulting force contributions $Y_r r$ and $Y_{rrr} r^3$ yield smaller --yet relevant-- sway force components than the sway-dependent terms (see \autoref{fig:coeffs_Y_full}).

Rudder-angle-dependent (steering) terms ($Y_\delta$, $Y_{\delta\delta\delta}$) agree reasonably well across estimators. Notably, the linear term $Y_\delta$ differs by $\approx 6\%$ for Ridge and $<1\%$ for Lasso and SVR an important aspect, as it forms a major component of the sway force $Y_\delta\, \delta$. The cubic terms $Y_{\delta\delta\delta}$ are $\approx 50\%$ higher for Ridge and SVR, attributing a larger portion of the steering force to higher-order dependencies.

Higher-order coupling terms ($Y_{vrr}$, $Y_{vvr}$, $Y_{\delta vv}$, $Y_{vr\delta}$, $\dots$) are generally more than one order of magnitude larger for LSQ than for regularized methods. Differences are smaller for steering-dominated couplings ($Y_{v\delta\delta}$, $Y_{r\delta\delta}$), typically remaining below $40\%$.

\subsubsection{Yaw coefficients}

Estimated yaw moment coefficients in \autoref{fig:coeffs_N_full} exhibit similar trends to the sway coefficients but to a greater extent. Differences are substantial even for constant and first-order coefficient estimates ($N_0$, $N_v$, $N_r$). The constant yaw moment $N_0$ from regularized methods reverses sign compared to LSQ, though its influence on the manoeuvre remains limited (see \autoref{fig:forceComps_N_zz3535ps_pred_methods}). Pure surge-dependent terms ($N_u$, $N_{uuu}$) have small influence on the manoeuvre across all estimators, however larger than expected given the hull's $x$-$z$ symmetry, conversely to the sway coefficient contributions, indicating over-prediction.

Differences in linear sway dependencies $N_v$ range from 35\% to 85\%, significantly larger than for the sway force coefficient $Y_v$. Only LSQ attributes a notable component to the cubic coefficient $N_{vvv}$.

For pure yaw dependencies, the linear term $N_r$ shows differences up to 71\%, despite $N_r\,r$ constituting the most relevant contribution to the total yaw moment over large portions of time. Again, LSQ predicts significantly larger cubic contributions than regularized methods.

The smallest disagreement is observed for rudder-angle-dependent (steering) terms ($N_\delta$, $N_{\delta\delta\delta}$), though still larger than for sway coefficients. Differences for $N_\delta$ reach up to 34\% compared to LSQ predictions. For the cubic term $N_{\delta\delta\delta}$, differences are larger including sign reversal, but only Lasso and Ridge attribute a notable (though small) portion $N_{\delta\delta\delta}\,\delta^3$ to the cubic dependency.

Following the trends observed for surge and sway coefficients, higher-order coupling terms ($N_{vrr}$, $N_{vvr}$, $N_{\delta vv}$, $N_{vr\delta}$, $N_{\delta rr}$, $\dots$) are typically one order of magnitude larger from LSQ than from regularized methods. Steering-dominant terms ($N_{v\delta\delta}$, $N_{r\delta\delta}$) form an exception, where regularized estimators predict larger magnitudes often with sign reversal.

\subsubsection{Prediction of test manoeuvres}

\begin{figure*}[!htbp]
  \centering

  \begin{subfigure}[b]{0.48\textwidth}
    \centering
    \includegraphics[width=\linewidth]{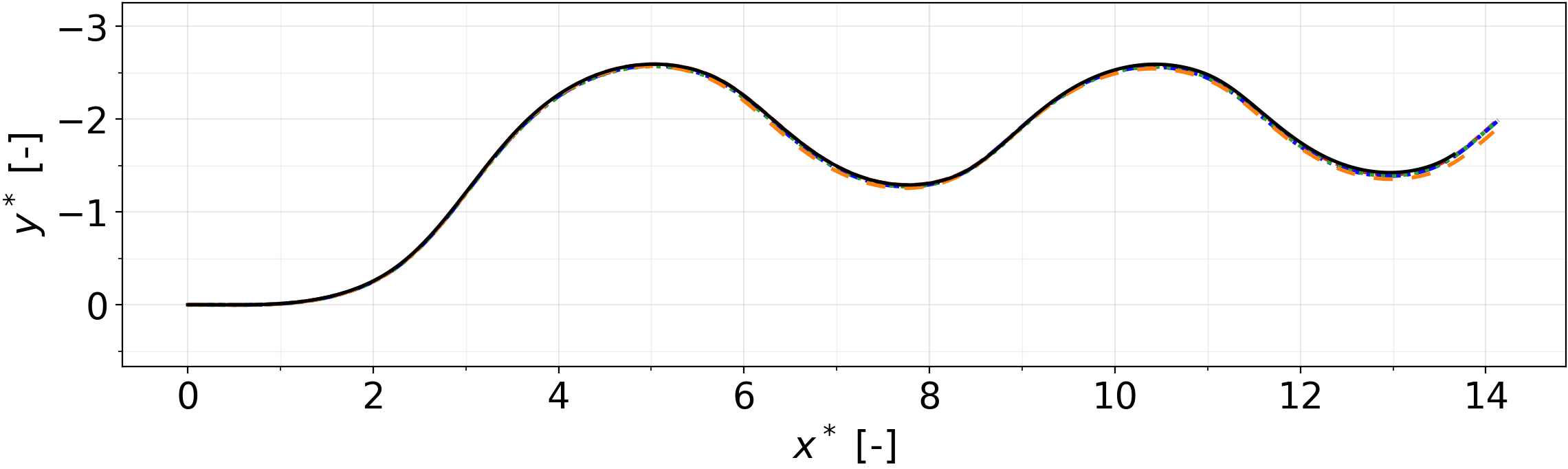}
    \caption{ZZ\,35$^\circ$/35$^\circ$ (PS)}
  \end{subfigure}\hfill
  \begin{subfigure}[b]{0.48\textwidth}
    \centering
    \includegraphics[width=\linewidth]{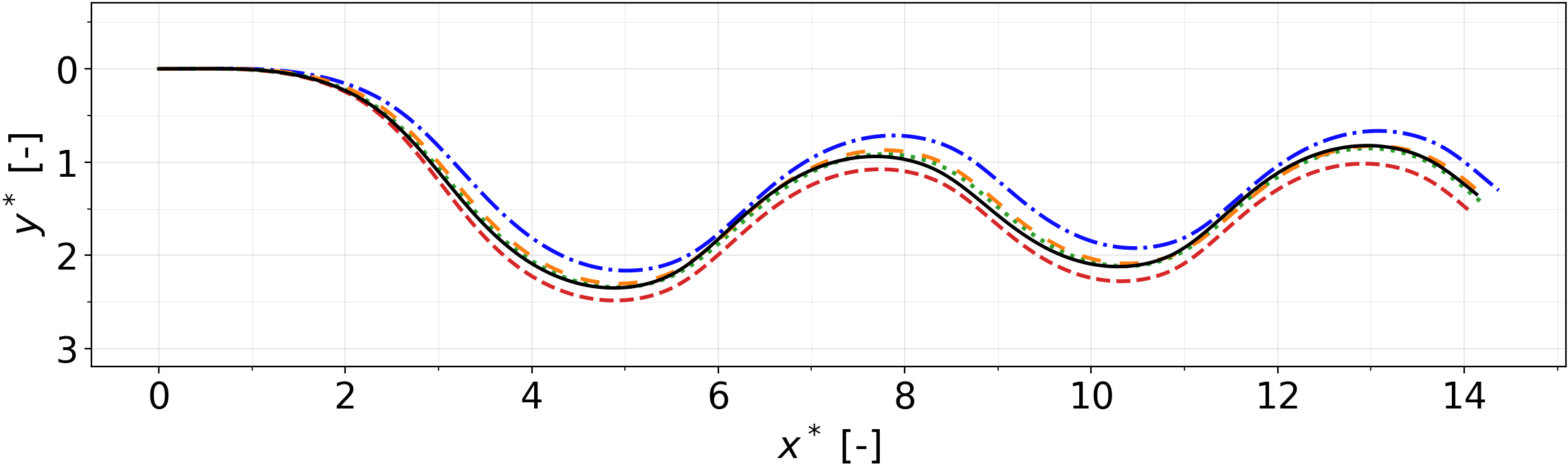}
    \caption{ZZ\,35$^\circ$/35$^\circ$ (SB)}
  \end{subfigure}

  \begin{subfigure}[b]{0.48\textwidth}
    \centering
    \includegraphics[width=\linewidth]{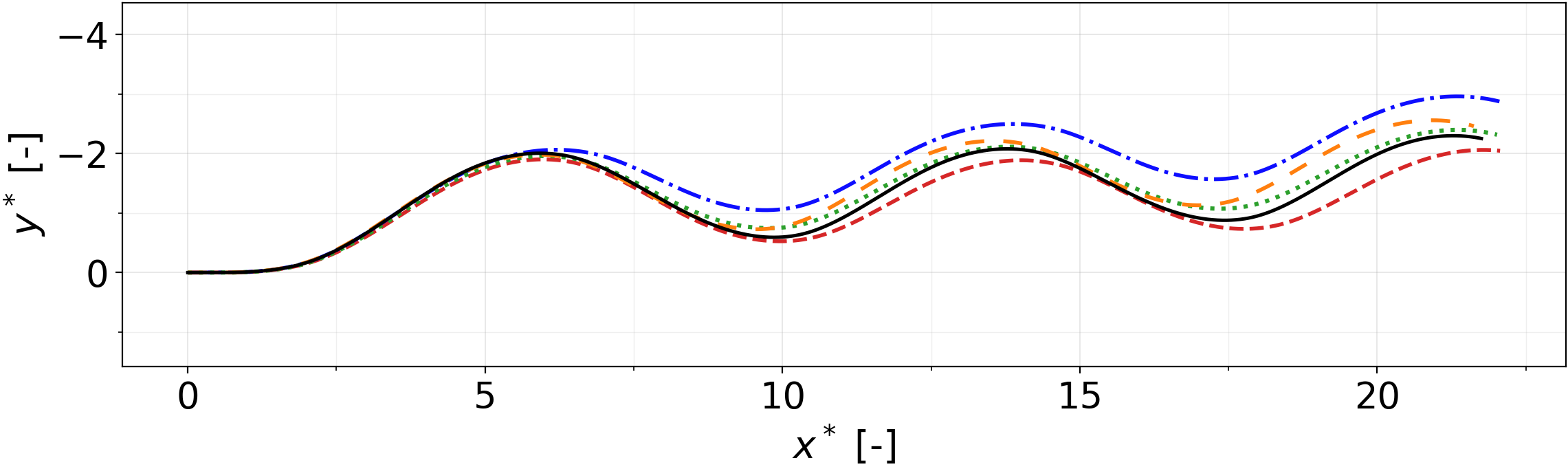}
    \caption{ZZ\,20$^\circ$/20$^\circ$(PS)}
  \end{subfigure}\hfill
  \begin{subfigure}[b]{0.48\textwidth}
    \centering
    \includegraphics[width=\linewidth]{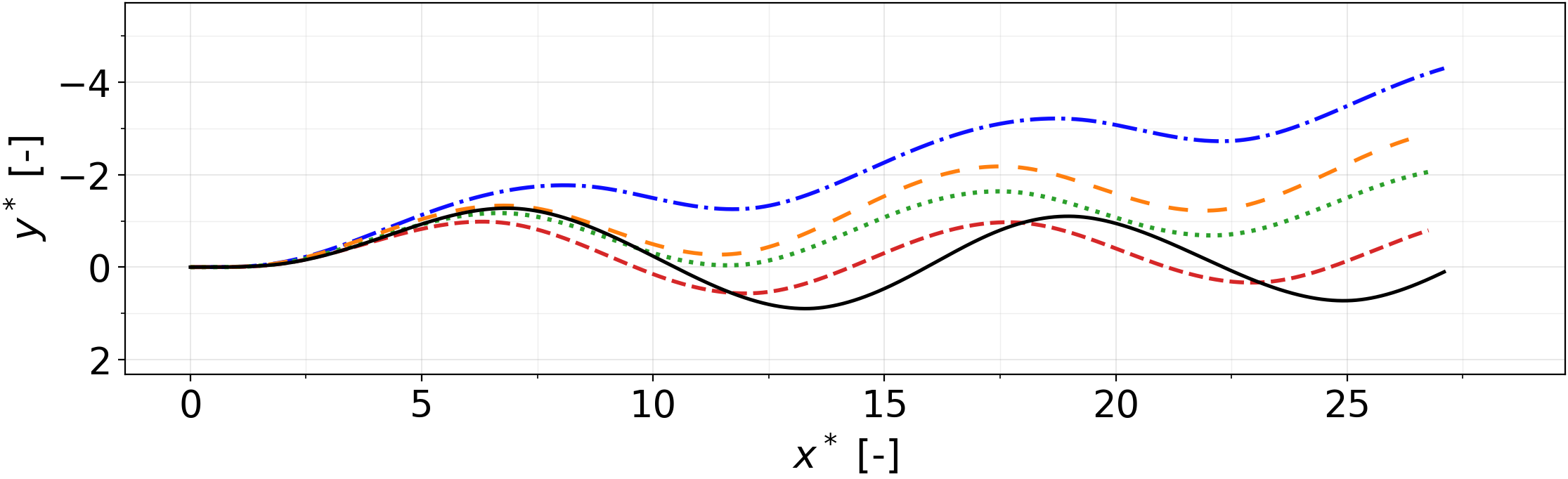}
    \caption{ZZ\,10$^\circ$/10$^\circ$(PS)}
  \end{subfigure}

  \begin{subfigure}[b]{0.48\textwidth}
    \centering
    \includegraphics[width=\linewidth]{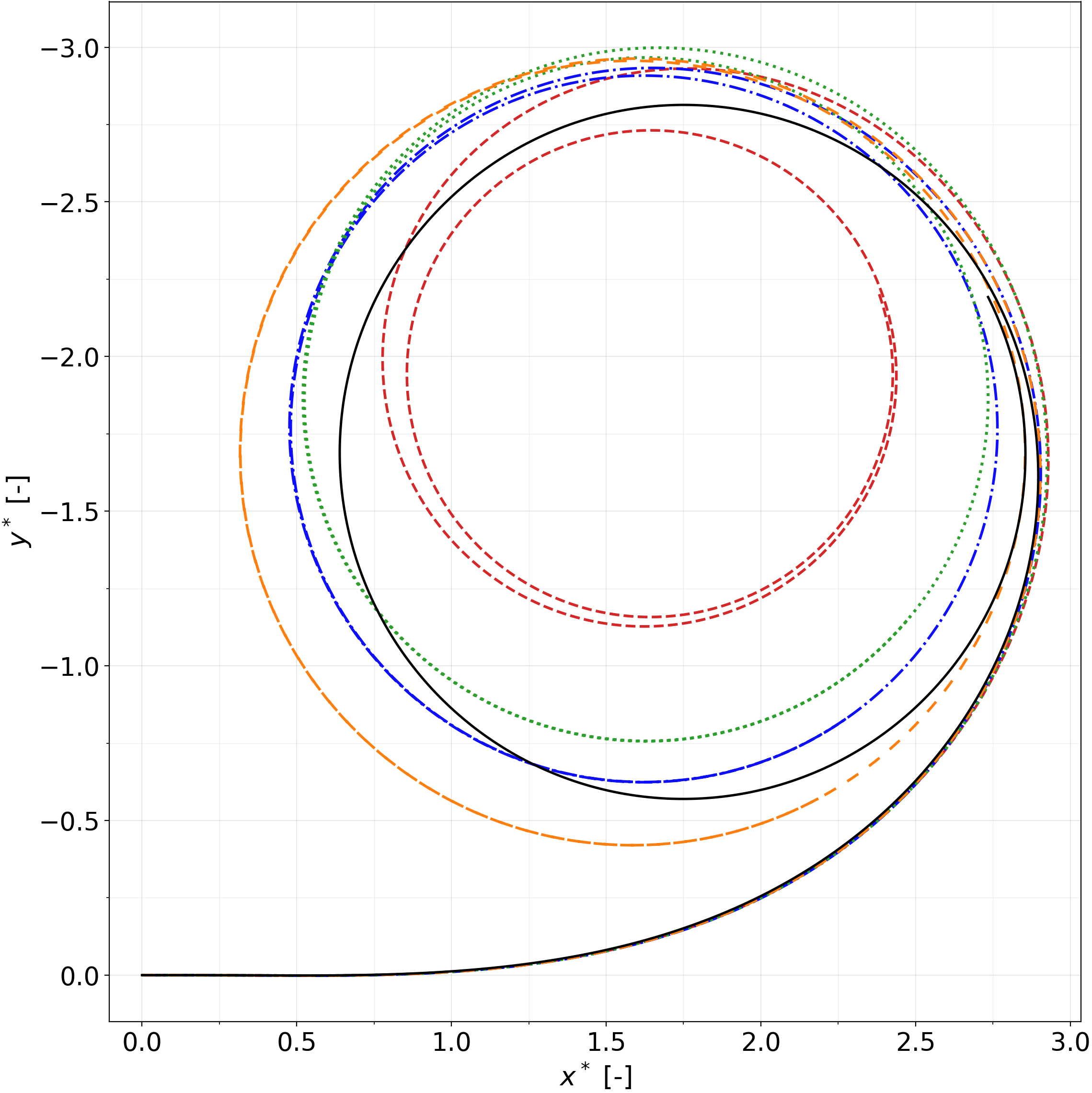}
    \caption{TC\,35$^\circ$ (PS)}
  \end{subfigure}\hfill
  \begin{subfigure}[b]{0.48\textwidth}
    \centering
    \includegraphics[width=\linewidth]{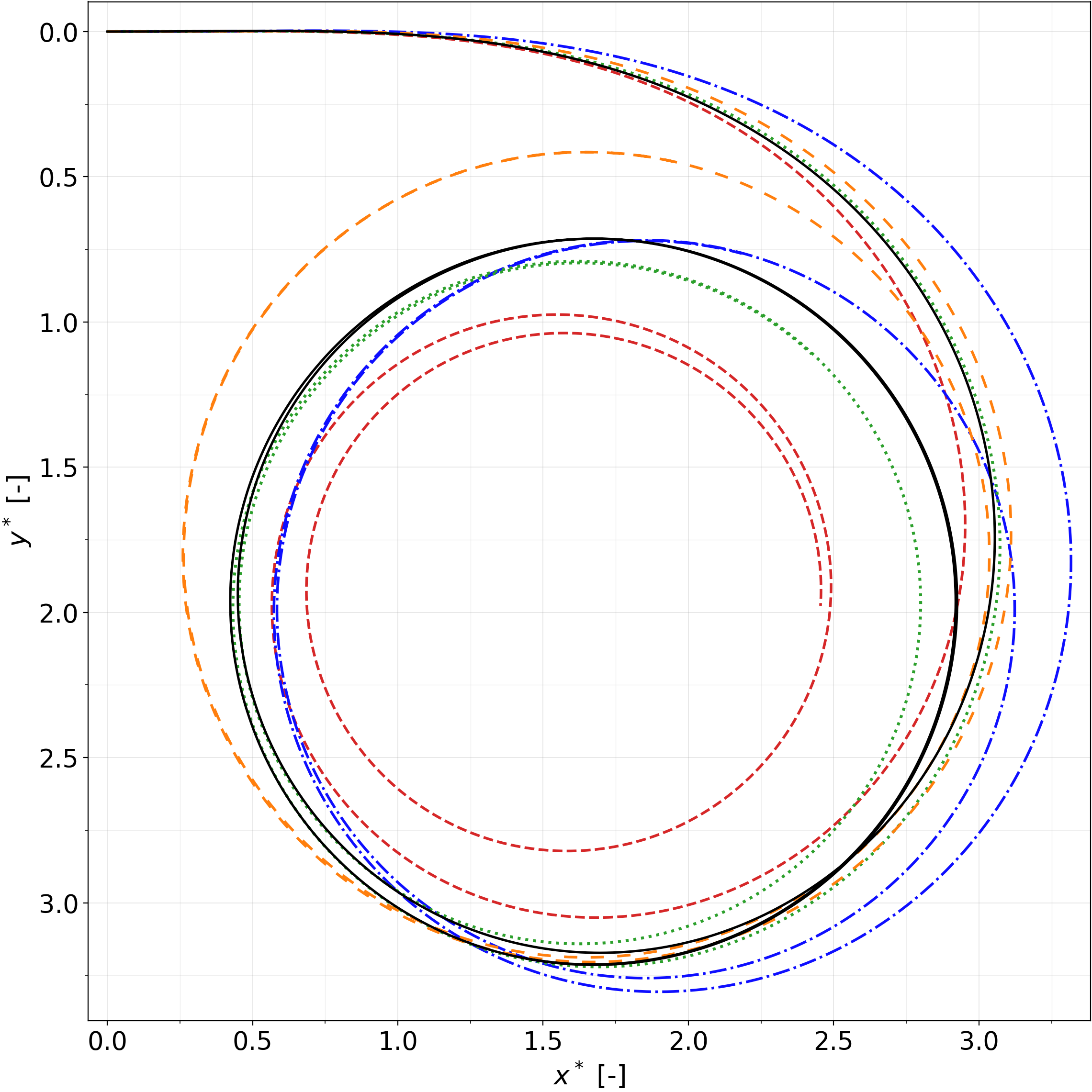}
    \caption{TC\,35$^\circ$ (SB)}
  \end{subfigure}

  \vspace{0.15em}
  \includegraphics[width=0.88\textwidth]{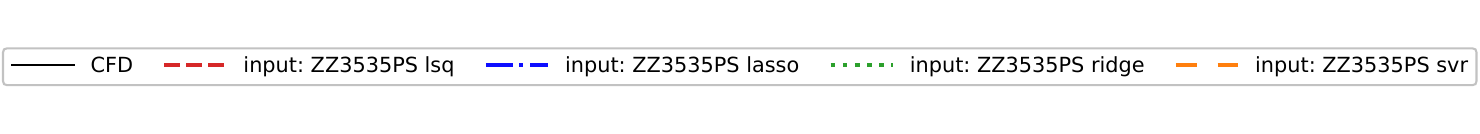}
  \vspace{-0.15em}

  \caption{Non-dimensional trajectories predicted by LSQ, Lasso, Ridge, and SVR models trained on a single ZZ $35^\circ/35^\circ$ (PS) manoeuvre using the large set of coefficients}
  \label{fig:traject_sevManoeuvr_Ridge001_PNA}
\end{figure*}

\begin{figure*}[!htbp]
  \centering

  \begin{subfigure}[b]{0.48\textwidth}
    \centering
    \includegraphics[width=\linewidth]{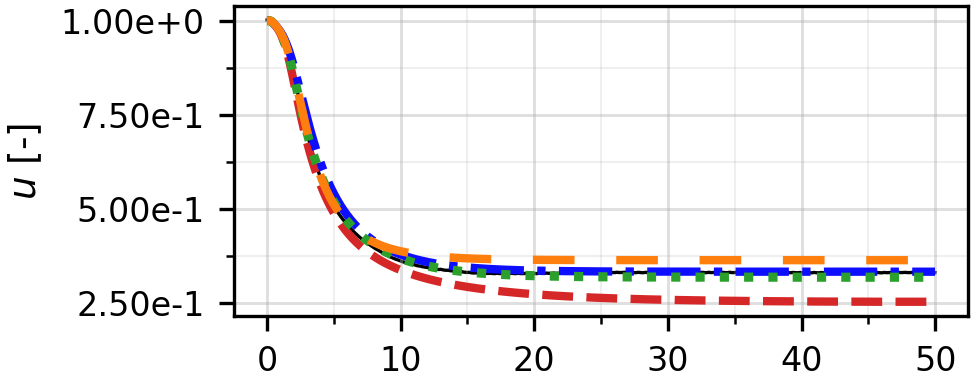}
    \caption{$u$}
  \end{subfigure}\hfill
  \begin{subfigure}[b]{0.48\textwidth}
    \centering
    \includegraphics[width=\linewidth]{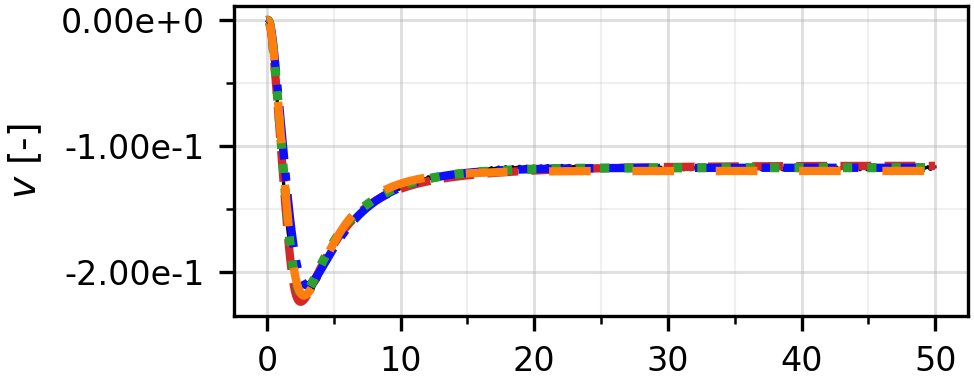}
    \caption{$v$}
  \end{subfigure}

  \begin{subfigure}[b]{0.48\textwidth}
    \centering
    \includegraphics[width=\linewidth]{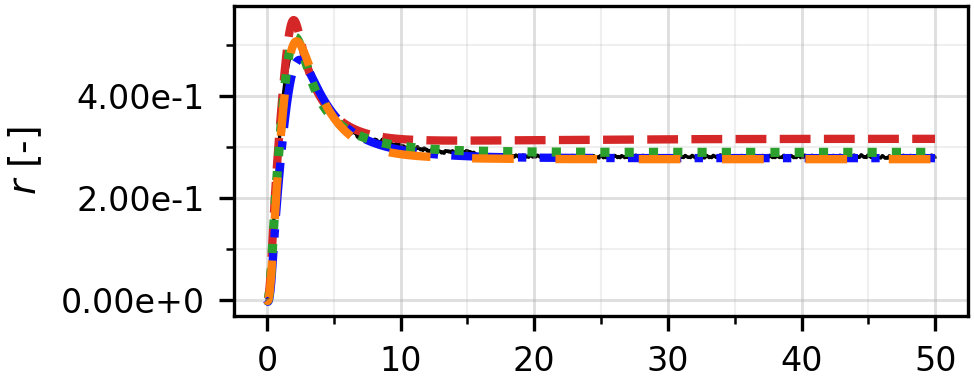}
    \caption{$r$}
  \end{subfigure}\hfill
  \begin{subfigure}[b]{0.48\textwidth}
    \centering
    \includegraphics[width=\linewidth]{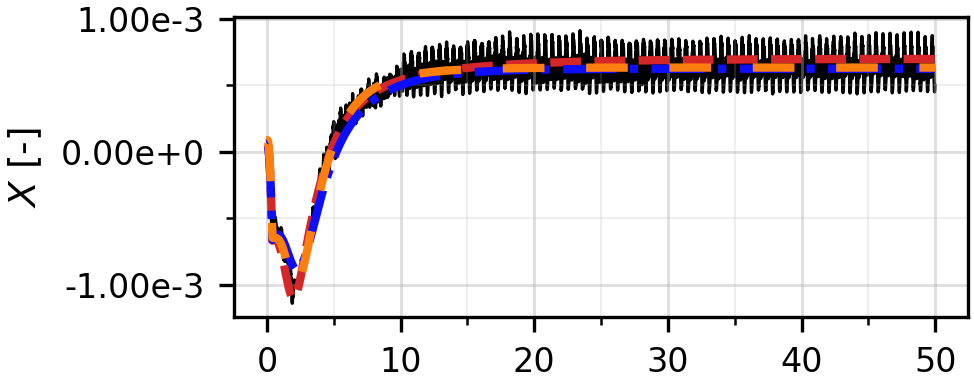}
    \caption{$X$}
  \end{subfigure}

  \begin{subfigure}[b]{0.48\textwidth}
    \centering
    \includegraphics[width=\linewidth]{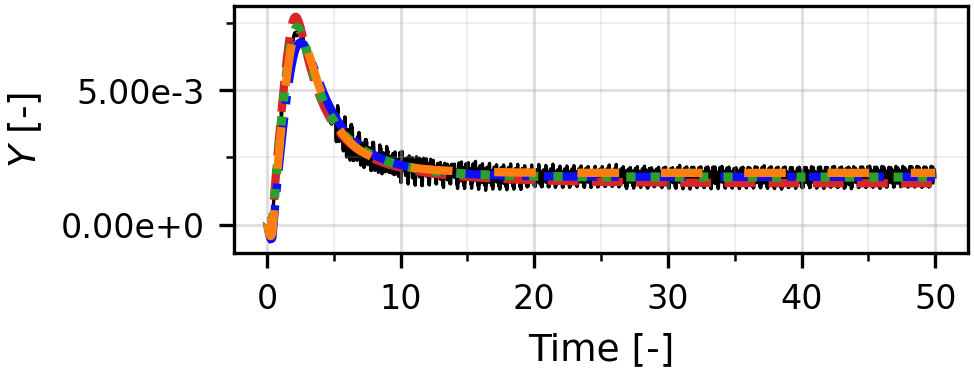}
    \caption{$Y$}
  \end{subfigure}\hfill
  \begin{subfigure}[b]{0.48\textwidth}
    \centering
    \includegraphics[width=\linewidth]{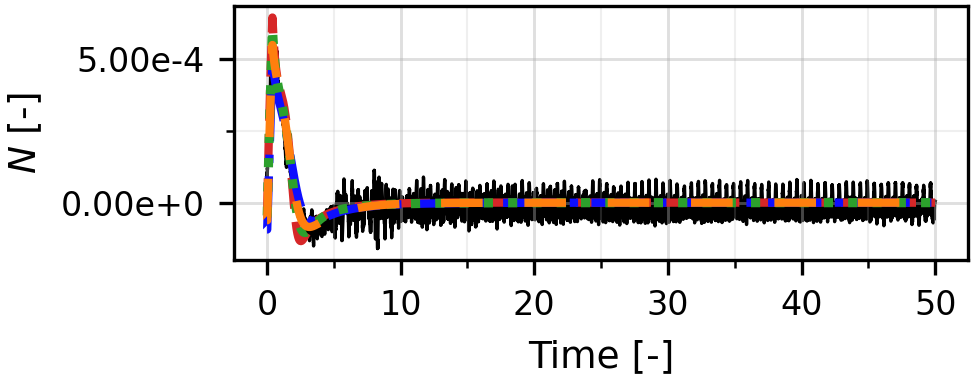}
    \caption{$N$}
  \end{subfigure}

  \vspace{0.15em}
  \includegraphics[width=0.88\textwidth]{figs/number_of_manoeuvres/legend_strip_Train_ZZ3535PS_Test_ZZ3535PS_models.pdf}  
  \vspace{-0.15em}

  \caption{Turning circle \(35^\circ\): non-dimensional time histories for \(u, v, r\) and \(X, Y, N\). Black lines denote CFD references; coloured lines are model predictions.}
  \label{fig:TC35PS_velocity_Manoevr_4_Models}
\end{figure*}

The following subsection presents the trajectories, motions, hydrodynamic forces, and metrics produced by all models using the ZZ \(35^\circ/35^\circ\) (PS) training data. The generated trajectories and selected time series are shown in \autoref{fig:traject_sevManoeuvr_Ridge001_PNA} and \autoref{fig:TC35PS_velocity_Manoevr_4_Models}, respectively; they are compared with the six CFD-generated manoeuvre trajectories (four zig-zag manoeuvres and two turning circles). The corresponding metrics and prediction errors are reported in \autoref{tab:train_ZZ3535PS}.


Each trajectory is computed using coefficients identified from a single ZZ $35^\circ/35^\circ$ (PS) manoeuvre, with the input time series truncated to four overshoots (4\,OS), and without further tuning. As shown in \autoref{fig:forceComps_X_zz3535ps_pred_methods}--\autoref{fig:forceComps_N_zz3535ps_pred_methods}, discrepancies reflect both differences in coefficient magnitudes and differences in the term structure used to reconstruct the total surge force $X$, sway force $Y$, and yaw moment $N$. LSQ distributes a substantial fraction of the reconstructed loads across higher-order, strongly coupled interactions, whereas the regularised estimators (Lasso, Ridge, SVR) suppress many weakly supported couplings and concentrate the response in a smaller, more interpretable subset of terms. These structural differences propagate to the predicted trajectories and the resulting IMO manoeuvring metrics.

\autoref{fig:traject_sevManoeuvr_Ridge001_PNA} shows analogous behaviour for the zig-zag manoeuvres. The training trajectory (ZZ~$35^{\circ}/35^{\circ}$ (PS)) is closely reproduced by all models, as indicated in \autoref{fig:traject_sevManoeuvr_Ridge001_PNA}(a) and \autoref{tab:train_ZZ3535PS}: the maximum in-sample MRAE is $3.6\%$ at $t_{\mathrm{check}}$ (SVR), while all remaining metric errors are below $2.7\%$. Force--moment decompositions in \autoref{fig:forceComps_X_zz3535ps_pred_methods}--\autoref{fig:forceComps_N_zz3535ps_pred_methods} confirm that each estimator reconstructs $X$, $Y$, and $N$ using a distinct active subset of hydrodynamic terms. Subfigures (b)--(d) in \autoref{fig:traject_sevManoeuvr_Ridge001_PNA} show that this single-manoeuvre identification does not fully constrain the dynamics: when rudder amplitude, overshoot level, or reversal timing differ from the training case, predictions depart from the CFD reference.


For ZZ~$10^{\circ}/10^{\circ}$ (PS), errors increase substantially. At the first overshoot, LSQ yields the largest error ($59.6\%$), whereas the regularised estimators reduce it to $46.3\%$ (Ridge), $39.6\%$ (Lasso), and $33.9\%$ (SVR). At the second overshoot, Lasso exhibits the largest error ($69.7\%$), while LSQ gives the smallest ($24.6\%$), with Ridge and SVR remaining intermediate ($41.9\%$ and $34.4\%$). Across the remaining zig-zag cases, LSQ can match individual overshoot metrics, but Ridge and SVR provide the most consistent performance across both overshoot angles and timing. This trend reflects extrapolation: off-training manoeuvres involve smaller heading excursions and different rudder histories, activating coupled surge--sway--yaw--rudder dynamics that are not fully excited in the single ZZ~$35^{\circ}/35^{\circ}$ (PS) identification run; Ridge and SVR remain comparatively stable, whereas Lasso can degrade when suppressed higher-order couplings become relevant.


Differences are more pronounced for turning-circle manoeuvres (\autoref{fig:traject_sevManoeuvr_Ridge001_PNA}(e)--(f)). This is only partially reflected in \autoref{tab:train_ZZ3535PS}: advance, transfer, and tactical diameter capture the early and intermediate turning response, whereas the steady turning diameter isolates long-horizon balance effects between resistance and yaw moment. Accordingly, LSQ remains competitive up to the tactical diameter (e.g.\ TD~$180^{\circ}$ error $4.0\%$) but diverges in the steady regime, producing a steady-diameter error of $29.7\%$ for TC~$35^{\circ}$ (PS). Ridge maintains a near-CFD steady diameter (error $1.0\%$), with Lasso at $2.4\%$ and SVR at $14.4\%$. \autoref{fig:TC35PS_velocity_Manoevr_4_Models} shows that LSQ predicts a larger steady-phase speed loss and yaw-rate deviation, whereas the regularised models track the CFD trends more closely, consistent with their more compact force--moment reconstructions.

\begin{table*}[htbp]
\centering
\caption{IMO manoeuvring metrics for zig-zag and turning-circle test cases, reported as CFD reference values and model predictions (relative errors in parentheses) for models trained on the ZZ $35^\circ/35^\circ$ (PS) manoeuvre.}
\label{tab:train_ZZ3535PS}
\footnotesize
\renewcommand{\arraystretch}{1.18}
\begin{tabular*}{\textwidth}{@{\extracolsep{\fill}} l l r r r r r}
\hline\hline
Manoeuvre & Metric & \multicolumn{1}{c}{CFD} & \multicolumn{1}{c}{LSQ} & \multicolumn{1}{c}{LASSO} & \multicolumn{1}{c}{Ridge} & \multicolumn{1}{c}{SVR} \\
\hline
ZZ 35--35 (PS) & Overshoot 1 [$^\circ$] & 22.80 & 22.92 (0.5\%) & 22.89 (0.4\%) & 22.63 (0.8\%) & 22.61 (0.8\%) \\
 & Overshoot 2 [$^\circ$] & 12.80 & 12.71 (0.7\%) & 12.80 (0.0\%) & 12.51 (2.3\%) & 13.00 (1.6\%) \\
 & $t_{\mathrm{exec2}}$ L/V [s] & 8.68 & 8.70 (0.2\%) & 8.66 (0.2\%) & 8.66 (0.2\%) & 8.60 (0.9\%) \\
 & $t_{\mathrm{check}}$ L/V [s] & 1.10 & 1.11 (0.9\%) & 1.13 (2.7\%) & 1.11 (0.9\%) & 1.14 (3.6\%) \\
\hline
ZZ 35--35 (SB) & Overshoot 1 [$^\circ$] & 20.86 & 22.88 (9.7\%) & 18.26 (12.5\%) & 20.60 (1.3\%) & 20.24 (3.0\%) \\
 & Overshoot 2 [$^\circ$] & 15.27 & 14.25 (6.7\%) & 15.14 (0.9\%) & 14.45 (5.4\%) & 14.79 (3.1\%) \\
 & $t_{\mathrm{exec2}}$ L/V [s] & 2.00 & 1.94 (3.0\%) & 2.25 (12.5\%) & 2.03 (1.5\%) & 2.08 (4.0\%) \\
 & $t_{\mathrm{check}}$ L/V [s] & 1.34 & 1.39 (3.7\%) & 1.22 (9.0\%) & 1.31 (2.2\%) & 1.30 (3.0\%) \\
\hline
ZZ 20--20 (PS) & Overshoot 1 [$^\circ$] & 16.85 & 15.12 (10.3\%) & 16.58 (1.6\%) & 15.86 (5.9\%) & 16.65 (1.2\%) \\
 & Overshoot 2 [$^\circ$] & 13.31 & 12.56 (5.6\%) & 9.40 (29.4\%) & 11.31 (15.0\%) & 13.02 (2.2\%) \\
 & $t_{\mathrm{exec2}}$ L/V [s] & 7.51 & 7.42 (1.2\%) & 7.92 (5.5\%) & 7.55 (0.5\%) & 7.37 (1.9\%) \\
 & $t_{\mathrm{check}}$ L/V [s] & 1.54 & 1.37 (11.0\%) & 1.07 (30.5\%) & 1.22 (20.8\%) & 1.31 (14.9\%) \\
\hline
ZZ 10--10 (PS) & Overshoot 1 [$^\circ$] & 20.48 & 8.28 (59.6\%) & 12.36 (39.6\%) & 10.99 (46.3\%) & 13.53 (33.9\%) \\
 & Overshoot 2 [$^\circ$] & 16.03 & 12.09 (24.6\%) & 4.85 (69.7\%) & 9.31 (41.9\%) & 10.51 (34.4\%) \\
 & $t_{\mathrm{exec2}}$ L/V [s] & 7.31 & 6.75 (7.7\%) & 9.01 (23.3\%) & 7.21 (1.4\%) & 7.30 (0.1\%) \\
 & $t_{\mathrm{check}}$ L/V [s] & 3.19 & 2.24 (29.8\%) & 0.97 (69.6\%) & 1.77 (44.5\%) & 1.70 (46.7\%) \\
\hline
TC 35$^\circ$ (PS) & AD $90^\circ$  [AD/L] & 2.80 & 2.84 (1.4\%) & 2.82 (0.7\%) & 2.84 (1.4\%) & 2.82 (0.7\%) \\
 & TR $90^\circ$  [TR/L] & -1.11 & -1.16 (4.5\%) & -1.13 (1.8\%) & -1.17 (5.4\%) & -1.13 (1.8\%) \\
 & TD $180^\circ$ [TD/L] & 2.73 & 2.84 (4.0\%) & 2.85 (4.4\%) & 2.91 (6.6\%) & 2.88 (5.5\%) \\
 & SD [SD/L] & 2.09 & 1.47 (29.7\%) & 2.14 (2.4\%) & 2.07 (1.0\%) & 2.39 (14.4\%) \\
\hline
TC 35$^\circ$ (SB) & Advance 90 [AD/L] & 2.97 & 2.87 (3.4\%) & 3.24 (9.1\%) & 2.99 (0.7\%) & 3.03 (2.0\%) \\
 & Transfer 90 [TR/L] & 1.24 & 1.19 (4.0\%) & 1.33 (7.3\%) & 1.27 (2.4\%) & 1.23 (0.8\%) \\
 & TD $180^\circ$ [TD/L] & 3.09 & 2.97 (3.9\%) & 3.23 (4.5\%) & 3.14 (1.6\%) & 3.13 (1.3\%) \\
 & SD [SD/L] & 2.35 & 1.73 (26.4\%) & 2.42 (3.0\%) & 2.24 (4.7\%) & 2.64 (12.3\%) \\
\hline\hline
\end{tabular*}
\end{table*}

Taken together, \autoref{fig:coeffs_X_full}--\autoref{fig:TC35PS_velocity_Manoevr_4_Models} and \autoref{tab:train_ZZ3535PS} show that all models reproduce the identification manoeuvre ZZ~$35^{\circ}/35^{\circ}$ (PS) with small in-sample metric discrepancies, whereas off-training predictions become manoeuvre and estimator dependent. The largest differences emerge for low-amplitude zig-zags (e.g.\ ZZ~$10^{\circ}/10^{\circ}$ (PS)) and for turning circles, where the steady turning diameter separates models more clearly than early-turn metrics. These trends are consistent with the estimator-specific force--moment reconstructions of $X$, $Y$, and $N$ reported above.

\subsection{Identification using multiple input manoeuvres}
\label{subsec:several_manoeuvre_sets_used}

This section extends the identification procedure by including more than one input manoeuvre in the training set. The objective is two-fold: (i) to improve predictive capability outside the specific condition used for identification in section~\ref{subsec:model_performance}, and (ii) to mitigate ill-conditioning due to multicollinearity in the regression by increasing the dynamic excitation of the model inputs.

Two hydrodynamic coefficient sets are considered. The first is a reduced set (32 parameters) based on the formulation in \citet{Mucha2017}, which has less structural complexity but is attractive for fast identification. The second is the full coefficient set from \citet{lewis_PNA_1988}, which retains a larger number of non-linear and coupled terms. Using both sets allows the separation of the two effects: limitations due to model complexity (reduced set) and limitations due to parameter estimation under multicollinearity (full set).

Three training configurations are considered:
\begin{enumerate}
  \item ZZ 35$^\circ$/35$^\circ$ PS + TC 35$^\circ$ PS: The turning circle data is included because a zig-zag input alone does not reproduce the sustained turning behaviour to port and starboard; the turning-circle manoeuvre provides the missing steady-turn information.
  \item ZZ 35$^\circ$/35$^\circ$ (PS) + ZZ 10$^\circ$/10$^\circ$ (PS): A smaller angle zig-zag is included because ZZ 35$^\circ$/35$^\circ$ (PS) does not extrapolate reliably to much smaller amplitudes; the ZZ 10$^\circ$/10$^\circ$ (PS) manoeuvre constrains the small-amplitude response.
  \item ZZ 35$^\circ$/35$^\circ$ (PS) + ZZ 10$^\circ$/10$^\circ$ (PS) + TC 35$^\circ$ (PS): Combination of three manoeuvres to provide extreme amplitudes (maximum and minimum ranges) coverage and broader dynamic excitation.
\end{enumerate}

The comparison is carried out for both coefficient sets and for two identification methods: Ordinary least squares and Ridge regression. Ridge regression is preferred over Lasso and SVR in this section because it showed reliable performance in section \ref{subsec:model_performance} while remaining computationally inexpensive relative to SVR, as is shown in \autoref{tab:comp-time-model-training-minimal-noitalic} and because it avoids aggressive coefficient cancellation, which in Lasso can remove physically relevant non-linear couplings needed to sustain turning or to reproduce low-amplitude zig-zag overshoots.

\begin{figure*}[!Hb]
  \centering
  \begin{subfigure}[b]{0.48\textwidth}
    \centering
    \includegraphics[width=\linewidth]{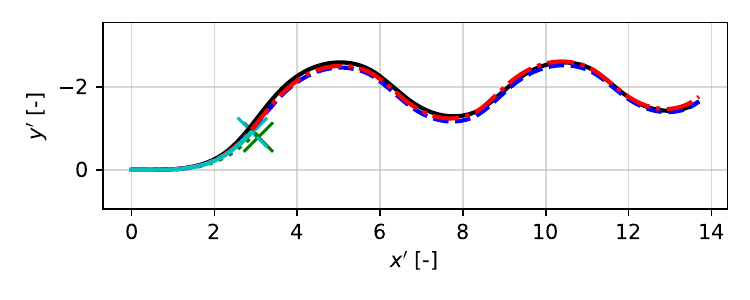}
    \caption{ZZ\,35$^{\circ}$/35$^{\circ}$ PS}
  \end{subfigure}\hfill
  \begin{subfigure}[b]{0.48\textwidth}
    \centering
    \includegraphics[width=\linewidth]{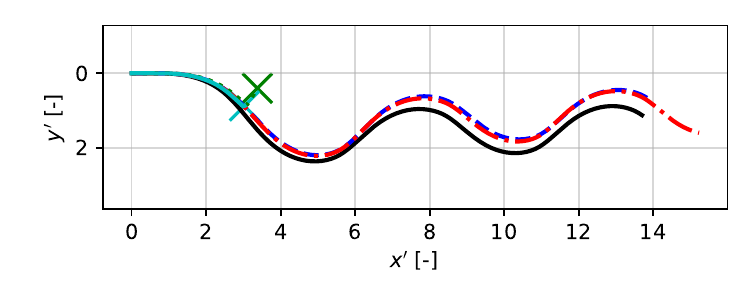}
    \caption{ZZ\,35$^{\circ}$/35$^{\circ}$ SB}
  \end{subfigure}

  \begin{subfigure}[b]{0.48\textwidth}
    \centering
    \includegraphics[width=\linewidth]{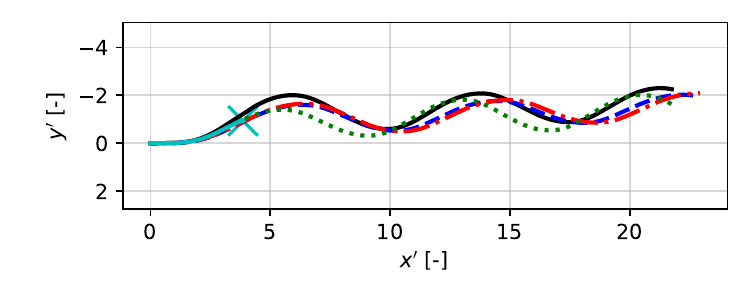}
    \caption{ZZ\,20$^{\circ}$/20$^{\circ}$ PS}
  \end{subfigure}\hfill
  \begin{subfigure}[b]{0.48\textwidth}
    \centering
    \includegraphics[width=\linewidth]{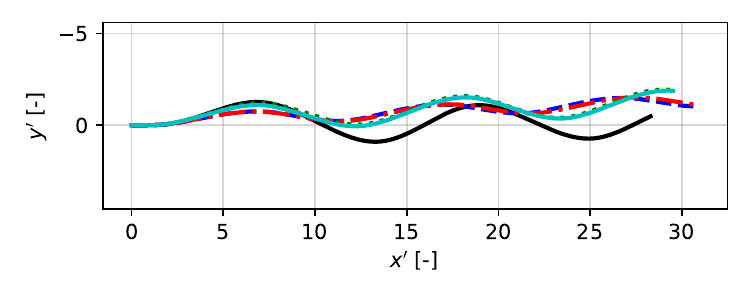}
    \caption{ZZ\,10$^{\circ}$/10$^{\circ}$ PS}
  \end{subfigure}

  \begin{subfigure}[b]{0.48\textwidth}
    \centering
    \includegraphics[width=\linewidth]{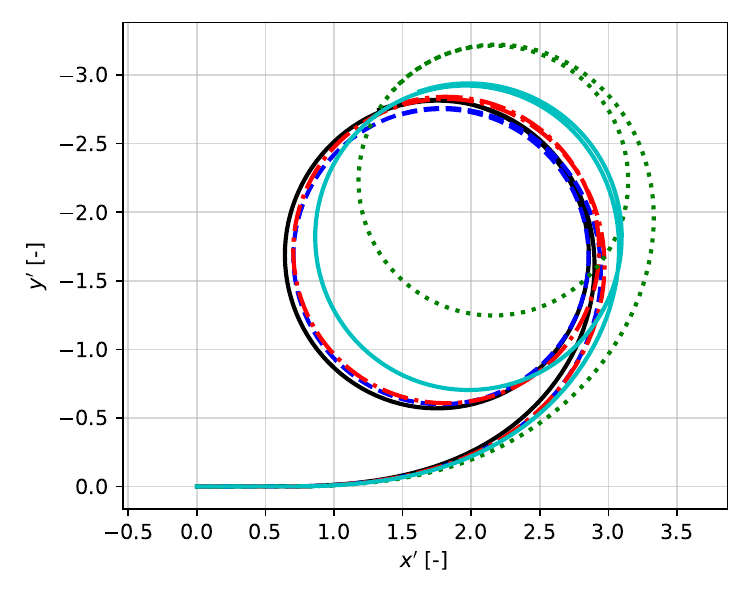}
    \caption{TC\,35$^\circ$ PS}
  \end{subfigure}\hfill
  \begin{subfigure}[b]{0.48\textwidth}
    \centering
    \includegraphics[width=\linewidth]{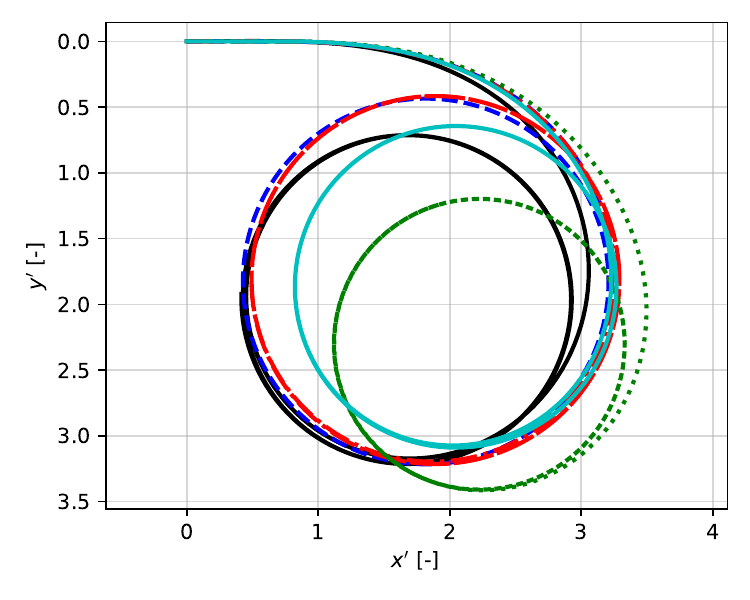}
    \caption{TC\,35$^\circ$ SB}
  \end{subfigure}
  \includegraphics[width=0.8\linewidth]{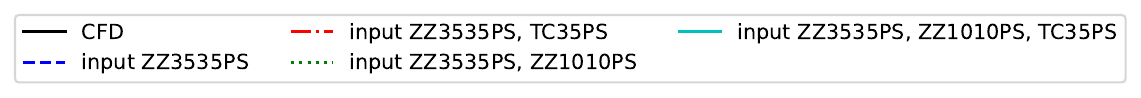}
  \caption{Predicted trajectories of input manoeuvre and test manoeuvres with predicted hydrodynamic coefficients using LSQ with several input manoeuvres in combination with the smaller set of coefficients from \citet{Mucha2017}; Black: CFD; coloured: model.}
  \label{fig:traject_sevManoeuvr_LSQ_Mucha}
\end{figure*}

The results obtained using only a single input manoeuvre and the reduced hydrodynamic coefficient set identified via LSQ were promising initially (see section \ref{subsec:set_of_coeffs}). With ZZ~35$^\circ$/35$^\circ$ (PS) as the sole identification input, the model was able to reproduce that input manoeuvre and predict the TC~35$^\circ$ (PS) test manoeuvre with high accuracy. Up to this point, however, only one input manoeuvre had been used for identification, and only one test manouevre was predicted. \autoref{fig:traject_sevManoeuvr_LSQ_Mucha} extends this analysis by comparing least-squares predictions against a wider set of CFD manoeuvres, including TC~35$^\circ$ (PS) and (SB), ZZ~35$^\circ$/35$^\circ$ (SB), ZZ~20$^\circ$/20$^\circ$ (PS), and ZZ~10$^\circ$/10$^\circ$ (PS). 

The dashed blue curves correspond to the predictions using only ZZ~35$^\circ$/35$^\circ$ (PS) as input data. For ZZ~35$^\circ$/35$^\circ$ and for TC~35$^\circ$ in both port (PS) and starboard (SB) directions, this single-input case shows satisfactory qualitative agreement with the CFD reference, and similarly good agreement is obtained for ZZ~20$^\circ$/20$^\circ$ (PS). In contrast, ZZ~10$^\circ$/10$^\circ$ (PS) deviates markedly from CFD under the same identification. 

Adding TC~35$^\circ$ (PS) to the input data (red dash-dotted curves) produces only marginal changes in the predicted trajectories and does not resolve the mismatch at low rudder amplitudes, which is consistent with the fact that a steady turning circle does not sufficiently constrain the small-amplitude zig-zag response. 

\noindent
\begin{minipage}{0.97\linewidth}
  \centering

  {%
    \captionsetup{labelfont=bf}
    \captionof{table}{Total computation time by model and training data (all DOFs combined).}
    \label{tab:comp-time-model-training-minimal-noitalic}
  }

  \setlength{\tabcolsep}{8pt}
  \renewcommand{\arraystretch}{1.25}

  \begin{tabular}{lrrrr}
    \hline\hline
    Model & \multicolumn{4}{c}{Training data* -- Time [s]} \\
    \cline{2-5}
          & 1 & 2 & 3 & 4 \\
    \hline
    LSQ   &   0.071   &   0.125   &   0.144   &   0.183   \\
    Lasso &   1.524   &   2.941   &   3.636   &   4.694   \\
    Ridge &   0.071   &   0.696   &   0.787   &   2.058   \\
    SVR   & 107.397   & 247.777   & 310.268   & 401.801   \\
    \hline\hline
  \end{tabular}
  \begin{flushleft}
  \footnotesize
  * Notes: Training data mapping:\\
  1 $\rightarrow$ ZZ3535PS\\
  2 $\rightarrow$ ZZ3535PS + TC35PS\\
  3 $\rightarrow$ ZZ3535PS + ZZ1010PS\\
  4 $\rightarrow$ ZZ3535PS + ZZ1010PS + TC35PS\\
  \end{flushleft}
\end{minipage}
\vspace{0.5\baselineskip} 

A possible strategy is to include ZZ~10$^\circ$/10$^\circ$ (PS) in the identification data to improve the prediction of both ZZ~10$^\circ$/10$^\circ$ and ZZ~20$^\circ$/20$^\circ$. However, when this additional input manoeuvre is included, the system-based predictions for both port and starboard turning circles resulted in large deviations of the predicted trajectories. (cyan and green curves in \autoref{fig:traject_sevManoeuvr_LSQ_Mucha}). For larger-amplitude zig-zag manoeuvres (ZZ~35$^\circ$/35$^\circ$, ZZ~20$^\circ$/20$^\circ$) including the input manoeuvre ZZ~35$^\circ$/35$^\circ$ (PS), the predictions result in numerical overflow, marked by a cross marker in the Figures. This loss of stability is attributed to the limited complexity of the reduced coefficient set: when multiple manoeuvres with very different amplitude ranges are combined, the model cannot represent all regimes simultaneously.

\begin{figure*}[ht!]
    \centering
    \begin{subfigure}[b]{0.49\textwidth}
        \centering
        \includegraphics[width=1\linewidth]{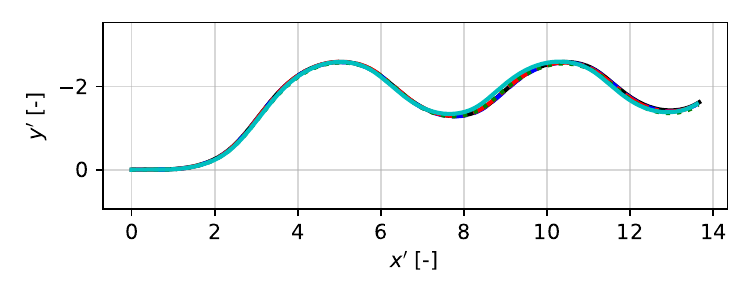}
        \caption{ZZ 35$^{\circ}$/35$^{\circ}$ PS}
    \end{subfigure}%
    ~
    \begin{subfigure}[b]{0.49\textwidth}
        \centering
        \includegraphics[width=1\linewidth]{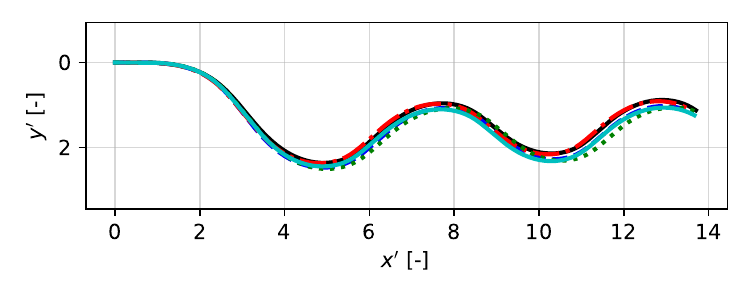}
        \caption{ZZ 35$^{\circ}$/35$^{\circ}$ SB}
    \end{subfigure}

    \begin{subfigure}[b]{0.49\textwidth}
        \centering
        \includegraphics[width=1\linewidth]{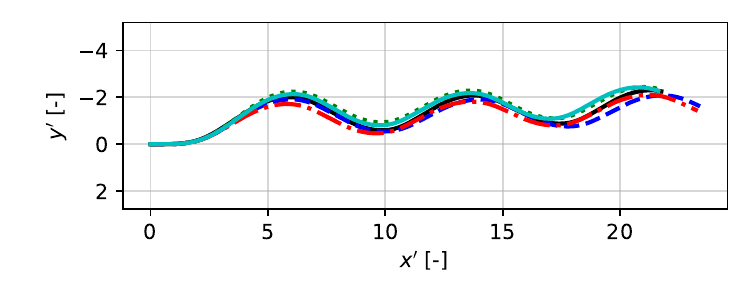}
        \caption{ZZ 20$^{\circ}$/20$^{\circ}$ PS}
    \end{subfigure}%
    ~
    \begin{subfigure}[b]{0.49\textwidth}
        \centering
        \includegraphics[width=1\linewidth]{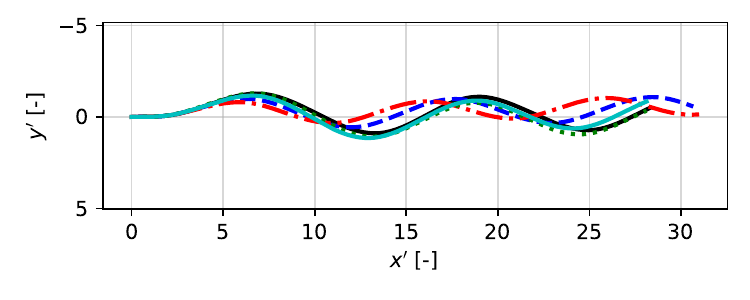}
        \caption{ZZ 10$^{\circ}$/10$^{\circ}$ PS}
    \end{subfigure}

    \begin{subfigure}[b]{0.49\textwidth}
        \centering
        \includegraphics[width=1\linewidth]{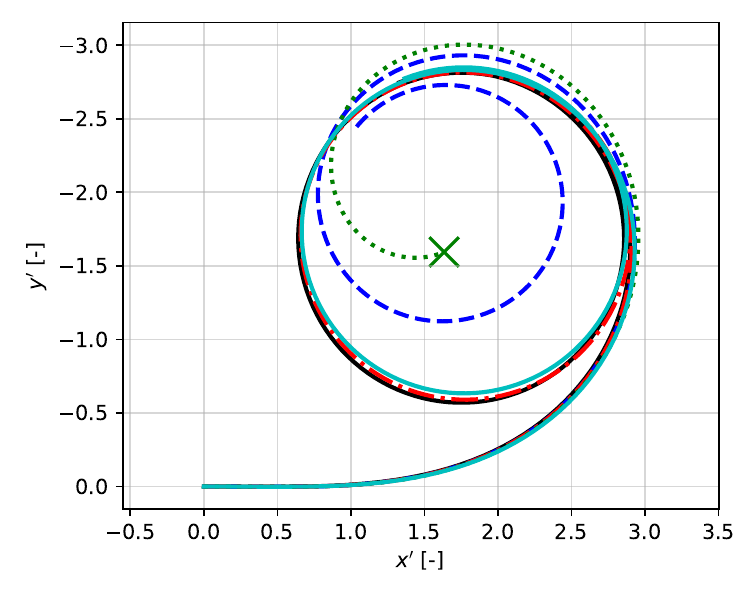}
        \caption{TC 35$^\circ$ PS}
    \end{subfigure}%
    ~
    \begin{subfigure}[b]{0.49\textwidth}
        \centering
        \includegraphics[width=1\linewidth]{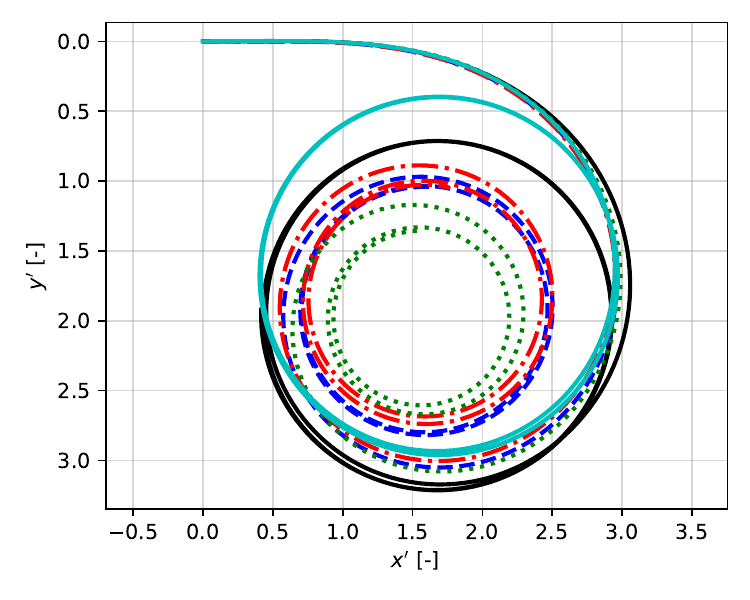}
        \caption{TC 35$^\circ$ SB}
    \end{subfigure}%
    
    \includegraphics[width=0.8\linewidth]{figs/number_of_manoeuvres/051a_legend.pdf}
    \caption{As \autoref{fig:traject_sevManoeuvr_LSQ_Mucha} but with the full set of hydrodynamic coefficients from \cite{lewis_PNA_1988}; Black: CFD; coloured: model.}
    \label{fig:traject_sevManoevr_LSQ_PNA}
\end{figure*}

To improve the versatility of the identified model, the full set of hydrodynamic coefficients is used next in combination with least-squares identification. The resulting predicted trajectories are shown in \autoref{fig:traject_sevManoevr_LSQ_PNA}. 
Using only one input manoeuvre for identification (dashed blue curves) leads to noticeable prediction errors, particularly for TC~35$^\circ$ (PS) and (SB), which is consistent with the multicollinearity effects observed already in section~\ref{subsec:set_of_coeffs}. 
Adding a second input manoeuvre reduces these errors, but the outcome depends on which manoeuvre is included. Incorporating TC~35$^\circ$ (PS) in the identification data (red dash-dotted curves) improves the predicted turning-circle trajectories especially for the port-side (PS) case but slightly degrades the ZZ~10$^\circ$/10$^\circ$ (PS) prediction. Conversely, including ZZ~10$^\circ$/10$^\circ$ (PS) in the identification data (green dashed curves) improves the low-amplitude zig-zag predictions (ZZ~10$^\circ$/10$^\circ$ and ZZ~20$^\circ$/20$^\circ$), while worsening the sustained turning response; in particular, free-run predictions for TC~35$^\circ$ (PS) become numerically unstable. It can be noted that for both options, the identified hydrodynamic model is able to accurately reproduce these input manoeuvres, which was not generally the case for the reduced set of coefficients. 

The best overall predictive capability is achieved when all these three manoeuvres are used together for identification, i.e. ZZ~35$^\circ$/35$^\circ$ (PS), ZZ~10$^\circ$/10$^\circ$ (PS), and TC~35$^\circ$ (PS) (cyan solid curves). In that case, the model reproduces most zig-zag and turning-circle trajectories with good qualitative agreement in comparison with CFD reference data. A persistent limitation remains in TC~35$^\circ$ (SB), where the predicted turning-circle trajectory still deviates from CFD. This could indicate that, even with the larger coefficient set and multiple input manoeuvres, some asymmetric turning behaviour between port (PS) and starboard (SB) is not fully captured, which is consistent with residual simplifications in the adopted Abkowitz-type modelling framework.

\begin{figure*}[ht!]
   \centering
   \begin{subfigure}[b]{0.49\textwidth}
       \centering
       \includegraphics[width=1\linewidth]{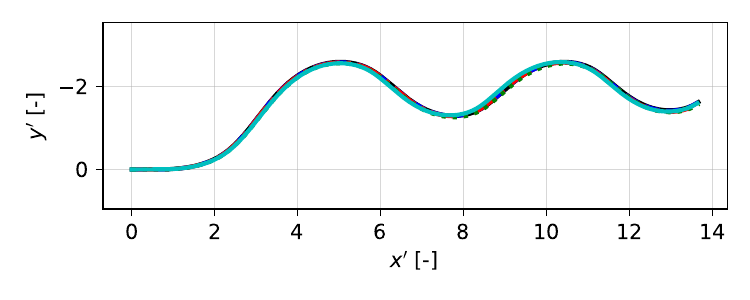}
       \caption{ZZ 35$^{\circ}$/35$^{\circ}$ PS (input data)}
   \end{subfigure}%
   ~
   \begin{subfigure}[b]{0.49\textwidth}
       \centering
       \includegraphics[width=1\linewidth]{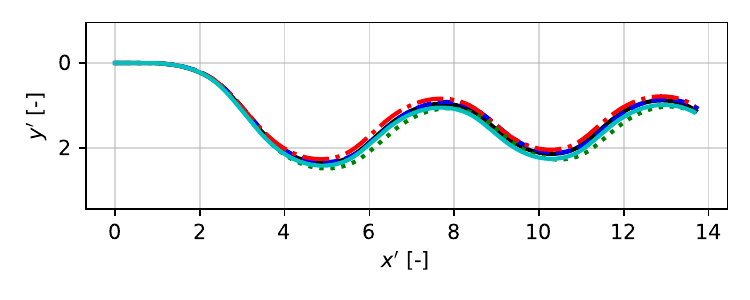}
       \caption{ZZ 35$^{\circ}$/35$^{\circ}$ SB}
   \end{subfigure}

   \begin{subfigure}[b]{0.49\textwidth}
       \centering
       \includegraphics[width=1\linewidth]{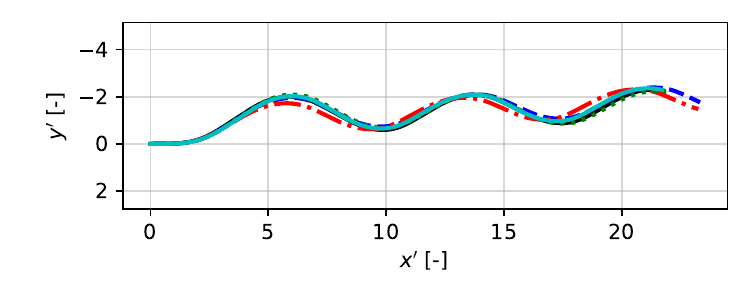}
       \caption{ZZ 20$^{\circ}$/20$^{\circ}$ PS}
   \end{subfigure}%
    ~
   \begin{subfigure}[b]{0.49\textwidth}
       \centering
       \includegraphics[width=1\linewidth]{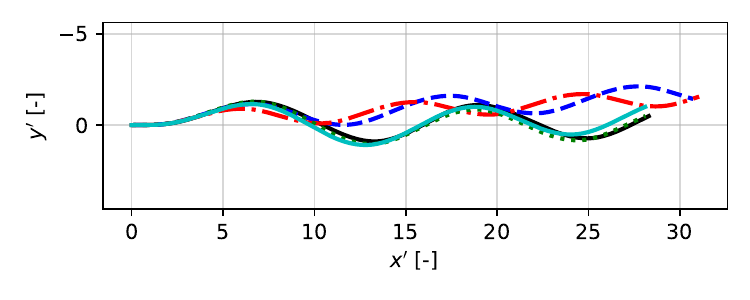}
       \caption{ZZ 10$^{\circ}$/10$^{\circ}$ PS}
   \end{subfigure}%

   \begin{subfigure}[b]{0.49\textwidth}
       \centering
       \includegraphics[width=1\linewidth]{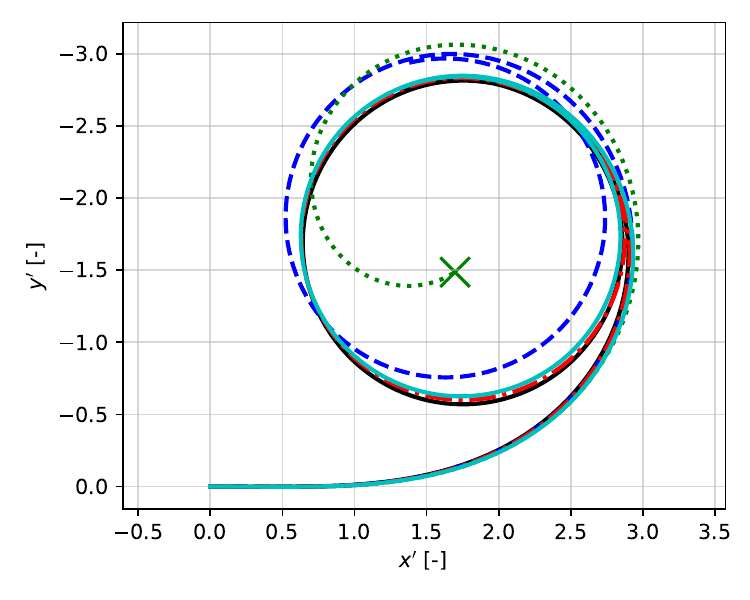}
       \caption{TC 35$^\circ$ PS}
   \end{subfigure}%
   ~
   \begin{subfigure}[b]{0.49\textwidth}
       \centering
       \includegraphics[width=1\linewidth]{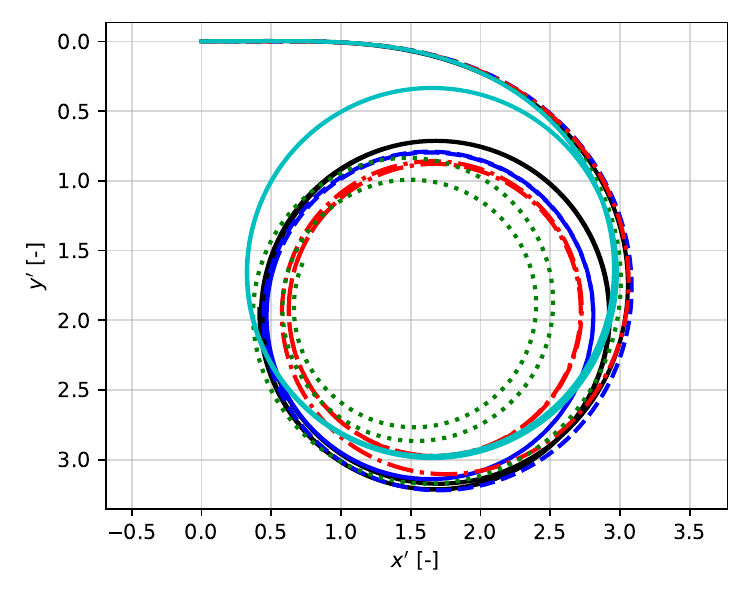}
       \caption{TC 35$^\circ$ SB}
   \end{subfigure}

   \includegraphics[width=0.8\linewidth]{figs/number_of_manoeuvres/051a_legend.pdf}
   \caption{As \autoref{fig:traject_sevManoevr_LSQ_PNA} but with hydrodynamic coefficient identification using Ridge regression with hyperparameter $\lambda = 0.001$}
   \label{fig:traject_sevManoevr_Ridge001_PNA}
\end{figure*}

Finally, the effect of using multiple input manoeuvres was assessed with the Ridge regression method instead of LSQ for parameter identification. Ridge introduces an $\ell_{2}$ regularisation term that inherently mitigates multicollinearity in the regression, as discussed in section~\ref{subsec:model_performance}, and therefore provided more reliable coefficient estimates even when only one manoeuvre is available for identification. The regularisation parameter was selected to $\lambda = 10^{-3}$ after comparing $\lambda = 10^{-1}$, $10^{-2}$, and $10^{-3}$; the smallest value offered the best overall agreement with the CFD reference across all manoeuvres without over-smoothing the dynamics. 
The predicted trajectories are shown in \autoref{fig:traject_sevManoevr_Ridge001_PNA}. 

In the single-input case (dashed blue curves in Fig. \autoref{fig:traject_sevManoevr_Ridge001_PNA}, Ridge achieves performance comparable to LSQ with the reduced set of hydrodynamic coefficients in \autoref{fig:traject_sevManoeuvr_LSQ_Mucha}: It reproduces the training zig-zag manoeuvre ZZ~35$^\circ$/35$^\circ$ (PS), delivers acceptable predictions for ZZ~20$^\circ$/20$^\circ$ (PS), and gives turning-circle trajectories for TC~35$^\circ$ (PS) that are slightly less accurate than LSQ with the reduced set, but already improved for TC~35$^\circ$ (SB). For the low angle zig-zag manouevre (ZZ~10$^\circ$/10$^\circ$ (PS)) deviations from reference data remain large.
When additional manoeuvres are included in the Ridge identification, the trends resemble those observed for LSQ with the full coefficient set (\autoref{fig:traject_sevManoevr_LSQ_PNA}), but with two notable differences. First, Ridge delivers useful predictions with only one or two input manoeuvres, whereas LSQ required three input manoeuvres to reach a similar level of overall agreement. Second, the degradation outside the training regime is generally less severe under Ridge: adding TC~35$^\circ$ (PS) data (red dash-dotted curves) improves the turning-circle behaviour at the cost of a slight deterioration in the small-amplitude zig-zag response, whereas adding ZZ~10$^\circ$/10$^\circ$ (PS) data (green dashed curves) improves ZZ~10$^\circ$/10$^\circ$ and ZZ~20$^\circ$/20$^\circ$ at the cost of a weaker sustained turn, in case of TC~35$^\circ$ (PS) leading to numerical instability. Using all three manoeuvres simultaneously for identification (cyan solid curves) results in the best global predictive capability The resulting Ridge-based predictions are close to those obtained with LSQ under the same three-manoeuvre input set. A residual asymmetry remains in TC~35$^\circ$ (SB), consistent with the limitations already discussed for LSQ and attributed to modelling simplifications in the Abkowitz-type framework rather than to the regression method.

\section{Discussion}
\label{sec:discussion}

The results in section~\ref{sec:results} highlight interactions between four key elements of model-based system identification for the KVLCC2 tanker: (i) fidelity and limitations of the CFD-generated training data used as reference, (ii) composition and conditioning of the hydrodynamic coefficient set, (iii) richness of the identification data in terms of excitation level, time-series length, and manoeuvre diversity, and (iv) the estimator used for coefficient prediction. These are discussed below in terms of robustness, extrapolation to unseen manoeuvres, and physical interpretability, with emphasis on reproducing both transient zig-zag and steady turning-circle performance.

\subsection{Synthetic data generation for training and validation}
\label{subsec:cfd_input_data_discussion}

The CFD input data stem from URANS simulations optimised for computation efficiency, using coarse grids, large time steps, and a virtual disk propeller model. Despite these reductions in model fidelity, the results agree well with recent literature benchmarks (see section~\ref{subsec:cfd_sim_results}), confirming suitability for hydrodynamic model identification. The presented identification procedures are data-source agnostic and could equivalently employ physical model or full-scale measurements, provided the data are prepared as outlined in Section~\ref{subsec:data_prep}.

\subsection{Model structure, multicollinearity, and identifiability}
\label{subsec:model_structure}

Dimensionality reduction of the hydrodynamic coefficient set diminishes multicollinearity, improves conditioning and enhances generalisation. While both reduced and full set reproduce the identification manoeuvre, only the reduced set maintains reliable out-of-sample predictions, as excessive cancellation occurs among overparameterised higher-order terms. Thus, reproducing the training manoeuvre alone does not guarantee predictive performance. However, when several input manoeuvres are included for identification, the reduced set becomes too restrictive and produces unphysical responses and numerical divergence, whereas the larger set benefits from the increased excitation and improves accordingly.

Regarding the length of the input manoeuvre, coefficient convergence improves markedly after four overshoots in a large-amplitude zig–zag; dominant linear and quadratic terms stabilise, while weak higher-order terms fluctuate without affecting predicted forces. Beyond this, additional data contribute little to predictions.

Different estimators show distinct trade-offs. Unconstrained least squares overfits weakly excited terms, due to multicollinearity, leading to unrealistic decay of surge speed and overestimation of yaw rate. Lasso regularisation stabilises fits but may suppress physically relevant couplings. Ridge regression and linear SVR effectively attenuate ill-conditioned terms while preserving nonlinear structure; Ridge achieves similar robustness at lower computational cost.


\subsection{Practical implications and limitations}
\label{subsec:implications}

Single-manoeuvre identification reproduces training behaviour but transfers poorly to other manoeuvres. Combining a large and a low-amplitude zig–zag with a turning circle enables balanced excitation of transient and quasi-steady conditions but requires sufficient model complexity permitting the application of strongly reduced sets of coefficients.

Ridge regression with mild regularisation ($\lambda =10^{-3}$) maintains robust performance even with limited input and approaches multi-manoeuvre least-squares accuracy with lower cost than SVR and better accuracy than Lasso regression. Remaining discrepancies, notably the port–starboard asymmetry in turning circles, originate from limitations in the CFD training data due to simplified propulsion modelling.

Practical guidelines emerge: (i) at least four overshoots of a large-amplitude zig–zag are required for stable, transferable coefficients; (ii) training should include manoeuvres that excite both transient and steady yaw conditions, such as ZZ\,35$^\circ$/35$^\circ$, ZZ\,10$^\circ$/10$^\circ$, and TC\,35$^\circ$; and (iii) Ridge regression provides the best compromise between accuracy, robustness, and cost, while Lasso suits interpretability and least squares should be reserved for reference.


\section{Conclusion}
\label{sec:conclusion}

This study establishes practical guidelines for Abkowitz-type manoeuvring model identification from free-running manoeuvring data, addressing multicollinearity, excitation sufficiency, and estimator robustness.


A large-angle zig–zag with at least four overshoots provides an effective minimum dataset for consistent sign stability and transferable predictions. Additional manoeuvres, notably ZZ\,10$^\circ$/10$^\circ$ and TC\,35$^\circ$, extend validity across small-amplitude and steady-turn regimes, while longer datasets yield diminishing returns.


Among tested estimators, Ridge regression best balances fidelity, robustness, and cost. It mitigates multicollinearity, preserves the nonlinear structure necessary for manoeuvre transfer, and supports stable turning predictions. SVR attains comparable performance but at higher computational expense; Lasso aids in model reduction and interpretability, though excessive sparsity can suppress key couplings. Least squares remains the least robust baseline.


Multicollinearity is inherent in polynomial Abkowitz models and should be mitigated through excitation diversity and regularisation rather than coefficient pruning. Recommended practice combines the above three manoeuvres, applies a quadratic penalty (Ridge) within a consistently scaled QP framework, and monitors identifiability using VIFs, condition numbers, and per-term stability. VIFs signal collinearity but should not drive automatic term removal, as this may eliminate necessary couplings, especially those involving surge perturbation $\Delta u$.


An effective minimum training set thus comprises manoeuvres such as ZZ\,35$^\circ$/35$^\circ$, ZZ\,10$^\circ$/10$^\circ$, and TC\,35$^\circ$. If only two manoeuvres are possible, the large zig–zag and the turning circle should be prioritised.


Future work should refine the treatment of multicollinearity using singular-value-based diagnostics. Regularisation strategies can then be targeted more precisely, while maintaining physically motivated term retention to improve steady-turn predictions and low-amplitude manoeuvre fidelity. 

Furthermore, extensions of the prediction model to four (or more) degrees of freedom, e.g. including roll motions for slender ships at large Froude numbers, is relatively straightforward and could be investigated in future studies.


\section{Acknowledgements}


This research was supported by the German Federal Ministry of Transport under the IHATEC ErosVSP project, grant number 19H22010C. The authors gratefully acknowledge this funding, without which the presented work would not have been possible.

%


\printcredits

\bibliographystyle{cas-model2-names}

\bibliography{cas-refs}


%

\clearpage

\end{document}